\documentclass{jfm}
\usepackage{graphicx}
\usepackage{epstopdf, epsfig}
\usepackage{xcolor}

\shorttitle{Turbulent channel flow of rigid oblate particles}
\shortauthor{M. Niazi Ardekani, P. Costa, W.-P. Breugem, F. Picano and L. Brandt}

\title{Drag reduction in turbulent channel flow laden with finite-size oblate spheroids}

\author{M. Niazi Ardekani\aff{1}
  \corresp{\email{mehd@mech.kth.se}},
  P. Costa\aff{2}, W.-P. Breugem\aff{2}, \\ F. Picano\aff{3}
 \and L.  Brandt\aff{1}}

\affiliation{\aff{1} Linn\'e Flow Centre and SeRC (Swedish e-Science Research Centre), KTH Mechanics, \\ S-100 44 Stockholm, Sweden
\aff{2}Laboratory for Aero $\&$ Hydrodynamics, Delft University of Technology, Delft, The Netherlands
\aff{3} Department of Industrial Engineering, University of Padova, \\ Via Venezia 1, 35131 Padua, Italy}

\begin{document}

\maketitle

\begin{abstract}
We study suspensions of oblate rigid particles in a viscous fluid for different values of the particle volume fractions.
Direct numerical simulations have been performed using a direct-forcing immersed boundary method to account for the dispersed phase, combined with a soft-sphere collision model and lubrication corrections for short-range particle-particle and particle-wall interactions. 
With respect to the single phase flow, we show that in flows laden with oblate spheroids the drag is reduced and the turbulent fluctuations attenuated.
In particular, the turbulence activity decreases to lower values than those obtained by only accounting for the effective suspension viscosity.
To explain the observed drag reduction we consider the particle dynamics and the interactions of the particles with the turbulent velocity field and show that the particle wall layer, previously observed and found to be responsible for the increased dissipation in suspensions of spheres, disappears in the case of oblate particles.
These rotate significantly slower than spheres near the wall and tend to stay with their major axes parallel to the wall, which leads to a decrease of the Reynolds stresses and turbulence production and so to the overall drag reduction. 

\end{abstract}


\section{Introduction}\label{sec:Introduction}

Suspensions of solid particles in fluids can be found in many environmental and industrial applications. Sediment transport in estuaries \citep{Mehta2014}, blood flow in 
the human body, pyroclastic flows from volcanos and pulp fibers in paper making \citep{Lundell2011} are among the examples of flows that deserve further investigations. 
The presence of solid rigid particles alters the global transport and rheological properties of the mixture in complex (and often unpredictable) ways. Many efforts have therefore been devoted to quantify the effects of the particles in these flows, starting from the simpler case of monodisperse rigid neutrally-buoyant spherical particles. 
The first studies of suspensions under laminar 
conditions can be traced back to Einstein \citep{Einstein1906,Einstein1911} who analytically derived an expression for the effective viscosity $\nu_e$ of a suspension of rigid 
spheres in the dilute and viscous limit: $\nu_e / \nu = 1 + (5/2)\phi$, where $\phi$ is the volume fraction and $\nu$ is the kinematic viscosity of the suspending fluid. A quadratic correction, accounting for particle-particle interactions was later proposed for higher volume fractions \citep{Batchelor1970,Batchelor1972}. 
The rheology 
of dense suspensions is usually characterized by semi-empirical formulas for the effective viscosity \citep{Stickel2005,Guazzelli2011}. 

Inertial effects, yet in laminar flows, are shown to induce significant modifications of the suspension microstructure and to create a local anisotropy 
responsible for shear-thickening \citep{Kulkarni2008,Picano2013},  thus to a change of the macroscopic suspension dynamics. 
Shear-thickening and particle migration towards 
regions of low shear had been observed in several previous studies for dense suspensions at low Reynolds number \citep{Hampton1997,Brown2009,Yeo2011}. 
The  highly inertial regime was considered in the pioneering work of 
\cite{Bagnold1954} who showed that shearing closely spaced particles 
 induces an effective viscosity that increases linearly with the shear rate, 
resulting in a normal or dispersive stress in addition to the shear stress \citep{Hunt2002}. 
Recently,
\cite{Lashgari2014, Lashgari2016} documented the existence of three different regimes when changing the volume fraction $\phi$ of neutrally-buoyant spherical particles and the Reynolds number $Re$: 
a laminar-like regime at low $Re$ and low to intermediate $\phi$ where the viscous stress dominates dissipation, a turbulent-like regime at high Reynolds number and low to intermediate $\phi$ where the turbulent Reynolds stress plays the main role in the momentum transfer across the channel and a 
third regime at higher $\phi$, denoted as inertial shear-thickening, characterised by a significant enhancement of the wall shear stress due to the particle-induced stresses. 

When the Reynolds number is sufficiently high, the flow becomes turbulent, exhibiting chaotic and multi-scale dynamics. The presence of the finite-size particles (particles comparable to or larger than than the smallest hydrodynamic scales of the flow) can change the turbulent structures at or below the particle size \citep{Naso2010,Homann2013}.  
These interactions modulate the whole process by inducing non-trivial effects on the turbulence, see e.g.\  the studies in homogenous isotropic turbulence by \cite{Lucci2010} and \cite{Fornari2016}, the latter including sedimentation. 
The first simulations of finite-size particles in a turbulent channel flow, as those discussed here,
were performed by \cite{Pan1996}, revealing that turbulent fluctuations and stresses increase in the presence of the solid phase.  
\cite{Matas2003,Loisel2013,Yu2013,Lashgari2015} considered the turbulence onset in suspensions of neutrally-buoyant spherical particles and reported a decrease of the critical Reynolds number for transition to turbulence in the semi-dilute 
regime. The simulations by \cite{Shao2012} revealed a decrease of the fluid streamwise velocity fluctuations due to an attenuation of the large-scale streamwise vortices in a turbulent channel 
flow. \cite{Kidanemariam2013} considered heavy 
finite-size particles and showed accumulation in the near-wall low-speed streaks at low $\phi$. 
\cite{Picano2015} studied dense suspensions of neutrally-buoyant particles  in turbulent 
channel flow up to volume fraction $\phi = 20 \%$. These authors showed that the velocity fluctuation intensities and the Reynolds shear stress gently increase with $\phi$ and then 
sharply decrease at $\phi = 20 \%$, even though the overall drag still increases. They attributed the drag increase to the enhancement of turbulence activity for $
\phi \le 10\% $ and then to the particle-induced stresses that govern the dynamics at high $\phi$. \cite{Costa2016} showed that the turbulent drag of a suspension 
of spherical particles is always higher than what predicted by only accounting for the effective suspension viscosity. This is attributed to the formation of a particle-wall 
layer, a layer of particles forming near the wall. \cite{Fornari2015} investigated the role of fluid and particle inertia, also in the semi-dilute regime, and show that the excluded volume is responsible for the turbulence modulations, while the particle inertia is negligible for solid to fluid density ratios below 10.

The dynamics of suspension in the presence of non-spherical particles are less understood \citep{Prosperetti2015life}, the majority of the previous studies in the turbulent regime dealing with the point-like spheroid particles.
Most of these investigations further assume dilute conditions and neglect the feedback on the flow. 
In this so-called one-way coupling regime, spherical particles display accumulation near the wall (turbophoresis) and preferentially sampling of low-speed regions
\citep[see e.g.][]{Sardina2011,Sardina2012}. 
 Turbulent channel flow of 
non-spherical particles has been investigated by several authors \citep{Zhang2001,Mortensen2008,Marchioli2010,Challabotla2015}, with focus on the particle  dynamics. 
\cite{Challabotla2015b} investigated the rotational motion of inertia-free spheroids in turbulent channel flow 
using the equations by \cite{Jeffery1922} 
for the particle rotation. These authors showed that oblate spheroids preferentially align their symmetry axes normal to the wall, whereas prolates are preferentially
parallel to the wall. 
The mean 
particle rotation was also reported to reduce when increasing the particle aspect ratio. Far from the wall, where the mean shear vanishes, this preferential alignment 
disappear and the behaviour is similar to that observed in homogeneous and isotropic turbulence \citep{Voth2015}.
 \cite{Kulick1994} studied 
small particles at higher concentrations, modelling feedback on the flow (two-way coupling),  and reported that
the fluid turbulence is attenuated by the addition of particles, while the turbulence anisotropy  increases. This effect was reported to increase with the particle Stokes number, 
particle mass loading and distance to the wall. \cite{Paschkewitz2004,Gillissen2008} showed drag reduction in suspensions of rigid fibres, similarly to what was reported for dilute polymer solutions \citep{Ptasinski2003,Dubief2004}.  

Despite these previous efforts, the turbulent flow of finite-size non-spherical particles is still unexplored. 
This is therefore the object of the present study.
In particular, we consider turbulent channel flow of finite-size oblate spheroids at volume fractions up to $\phi = 15 \%$. Aspect ratio (ratio of polar over equatorial radius) $\mathcal{AR} = 1/3$ is chosen for the particles to depart adequately from sphericity, where the effect of shape is more noticeable.  We show that, unlike spherical particles,
oblate particles cause drag reduction as the volume fraction $\phi$ increases within the investigated range. We attribute the drag 
reduction to the absence of a particle-wall layer and to an attenuation of the near-wall turbulence, explained  by the particle preferential orientation and reduced rotation near the wall.

The paper is organised as follows. The governing equations and the flow geometry are introduced in \S\ref{sec:Methodology}, followed by the results of the numerical simulations in  section \S\ref{sec:Results}. The main conclusions and final remarks are drawn in \S\ref{sec:Final_remarks}. Results for laminar flow are reported in appendix~\ref{app:Laminar_Simulations} as comparison. 
     

\section{Methodology}\label{sec:Methodology}
\vspace{10pt}

Several approaches for performing interface-resolved direct numerical simulations (DNS) of particle-laden flows have been proposed in recent years. Among these methods, \textit{force coupling} \citep{Lomholt2003}, \textit{front tracking} \citep{Unverdi1992}, \textit{Physalis} \citep{Zhang2005,Sierakowski2016}, algorithms based on the \textit{lattice Boltzmann} method for resolving the fluid phase \citep{Ladd1994,Ladd19942} and the \textit{Immersed boundary} method (IBM) \citep{Peskin1972} as used here. Several algorithms for IBM have been proposed \citep{Mittal2005,Uhlmann2005,Kempe2012JCP,Breugem2012} since the original work by \citep{Peskin1972}. 
The possibility of exploiting efficient computational algorithms for solving the Navier-Stokes equations on a Cartesian grid has made IBM a popular tool to investigate particle suspensions. The IBM algorithm proposed by \cite{Breugem2012} has been recently extended to ellipsoidal particles by \cite{Ardekani2016}, using lubrication, friction and collision models for the short-range particle interactions. In this work we use the same numerical model to simulate dense suspensions of oblate spheroidal particles in turbulent plane channel flow.

\subsection{Governing equations}
\vspace{8pt}

The incompressible Navier-Stokes equations describe the flow field in the Eulerian phase: 
\begin{eqnarray}
\label{eq:NS1}  
\rho_f (\frac{\partial \textbf{u}}{\partial t} + \textbf{u} \cdot \nabla \textbf{u} ) &=&  -\nabla p - \nabla p_e + \mu_{f} \nabla^2 \textbf{u} + \rho_f \textbf{f} \, , \\ [8pt]
\nabla \cdot \textbf{u} &=& \, 0 \, .
\label{eq:NS2} 
\end{eqnarray}
where $\textbf{u}$ is the fluid velocity, $p_e$ is the contribution to the total pressure from a constant pressure gradient that drives the flow,  $p$ is the modified pressure (the total pressure minus $p_e$ and the contribution from the hydrostatic pressure) and $\rho_f$ and $ \mu_{f}$ are the density and dynamic viscosity of the fluid. The extra term $\textbf{f}$ on the right hand side of equation (\ref{eq:NS1}) is the IBM force field, active in the immediate vicinity of a particle surface to enforce no-slip and no-penetration boundary conditions. 

The motion of rigid spheroidal particles are described by Newton-Euler Lagrangian equations,
\begin{eqnarray}
\label{eq:NewtonEuler1}  
\rho_p V_p \frac{ \mathrm{d} \textbf{U}_{p}}{\mathrm{d} t} &=& \oint_{\partial {S}_p}  \pmb{\tau} \cdot  \textbf{n} \mathrm{d}A - V_p \nabla p_e + \left( \rho_p - \rho_f \right)V_p\textbf{g} + \textbf{F}_c , \, \\ [8pt] 
\frac{ \mathrm{d} \left( \textbf{I}_p \, \pmb{\omega}_{p} \right) }{\mathrm{d} t} &=& \, \oint_{\partial {S}_p} \textbf{r} \times \left( \pmb{\tau} \cdot  \textbf{n} \right) \mathrm{d}A + \textbf{T}_c  \, , 
\label{eq:NewtonEuler2}
\end{eqnarray} 
where $\textbf{U}_p$ and  $\pmb{\omega}_{p}$ are the particle translational and the angular velocity. $\rho_p$, $V_p$ and $\textbf{I}_p$ are the mass density, volume and moment-of-inertia tensor of a spheroidal particle. $\textbf{r}$ indicates the position vector with respect to the center of the spheroid and $\textbf{n}$ is the outward unit normal vector at the particle surface $\partial {S}_p$ 
where the stress tensor $\boldsymbol{\tau} = -p\textbf{I} + \mu_f \left( \nabla \textbf{u} + \nabla \textbf{u}^T \right)$, acting on the surface of the particle is integrated. The force terms $- \rho_f V_p\textbf{g}$ and $V_p \nabla p_e$ account for the hydrostatic pressure and a constant pressure gradient $\nabla p_e$ with $\textbf{g}$ the gravitational acceleration. $\textbf{F}_c$ and $\textbf{T}_c$ are the force and torque resulting from particle-particle (particle-wall) collisions.  

\subsection{Numerical algorithm}    
\vspace{8pt}

The flow field is resolved on a uniform ($ \Delta x = \Delta y = \Delta z$), staggered, Cartesian grid while particles are represented by a set of Lagrangian points, uniformly distributed on the surface of each particle. The number of Lagrangian grid points $N_L$ on the surface of each particle is defined such that the Lagrangian grid volume $\Delta V_l$ becomes equal to the volume of the Eulerian mesh ${\Delta x}^3$. 

Taking into account the inertia of the fictitious fluid phase inside the particle volumes, \cite{Breugem2012} showed that  equations (\ref{eq:NewtonEuler1}) and (\ref{eq:NewtonEuler2}) can be rewritten as:
\begin{eqnarray}
\label{eq:NewtonEulerWim1}  
\rho_p V_p \frac{ \mathrm{d} \textbf{U}_{p}}{\mathrm{d} t} &\approx&  -\rho_f \sum\limits_{l=1}^{N_L} \textbf{F}_l \Delta V_l + \rho_f  \frac{ \mathrm{d}}{\mathrm{d} t} \left( \int_{V_p} \textbf{u} \mathrm{d} V  \right)  + \left( \rho_p - \rho_f \right)V_p\textbf{g} + \textbf{F}_c , \, \\ [8pt] 
\frac{ \mathrm{d} \left( \textbf{I}_p \, \pmb{\omega}_{p} \right) }{\mathrm{d} t} &\approx& -\rho_f \sum\limits_{l=1}^{N_L} \textbf{r}_l \times \textbf{F}_l \Delta V_l + \rho_f  \frac{ \mathrm{d}}{\mathrm{d} t} \left( \int_{V_p} \textbf{r} \times \textbf{u} \mathrm{d} V  \right) + \textbf{T}_c  \, .
\label{eq:NewtonEulerWim2}  
\end{eqnarray}
The point force $F_l$ is calculated at each Lagrangian point using the difference between the particle surface velocity ($\textbf{U}_p + \pmb{\omega}_p \times \textbf{r}$) and the interpolated first prediction velocity at the same point. The first prediction velocity is obtained by advancing equation (\ref{eq:NS1}) in time without considering the force field $\textbf{f}$. 

The forces, $\textbf{F}_l$, integrate to the force field $\textbf{f}$  using the regularized Dirac delta function $\delta_d$ of \cite{Roma1999}: 
 \begin{equation}
\label{eq:Fl}  
\textbf{f}_{\, ijk} = \sum\limits_{l=1}^{N_L} \textbf{F}_l \delta_d \left( \textbf{x}_{ijk} - \textbf{X}_l \right) \Delta V_l \,  \, 
\end{equation}
with $\textbf{x}_{ijk}$ and $\textbf{X}_l$ referring to an Eulerian and a Lagrangian grid cell. This smooth delta function essentially replaces the sharp interface with a thin porous shell of width  $3\Delta x$; it preserves the total force and torque on the particle provided that the Eulerian grid is uniform. An iterative algorithm is employed to calculate the force field $\textbf{f}$, allowing for a better estimate of no-slip and no-penetration boundary conditions. \citep{Breugem2012}. Equations (\ref{eq:NewtonEulerWim1}--\ref{eq:NewtonEulerWim2}) for the particle motion and (\ref{eq:NS1}--\ref{eq:NS2}) for the flow are integrated in time using an explicit low-storage Runge-Kutta method with the pressure-correction scheme used in \cite{Breugem2012} to project the velocity field in the divergence-free space.

When the distance between particles (or a particle and a wall) are smaller than one Eulerian grid size, the lubrication force is under-predicted by the IBM. To compensate for this inaccuracy and to avoid computationally expensive grid refinements, a lubrication model based on the asymptotic analytical expression for the normal lubrication force between 
unequal spheres \citep{Jeffrey1982} is used; here we approximate the two spheroidal particles with two spheres with same mass and radius corresponding to the local 
curvature at the points of contact. Using these approximating spheres, a soft-sphere collision model with Coulomb friction takes over the interaction when the particles touch. 
The restitution coefficients used for normal and tangential collisions are $0.97$ and $0.1$, with Coulomb friction coefficient set to $0.15$.
More details about the models and validations can be found in \cite{Ardekani2016,Costa2015}. 

\subsection{Flow geometry}
\vspace{8pt}

\begin{table}
  \begin{center}
\def~{\hphantom{0}}
  \begin{tabular}{lccccccc}
   & &  \multicolumn{4}{c}{Oblate - $\mathcal{AR} = 1/3$}  & &  \multicolumn{1}{c}{Sphere} \\
   \cline{3-6} \cline{8-8} \\
       \,\,\,\,\,\,\,\,\,  \,Case & $\phi=0\% \,\,\,\,\,\,\,\,\,\,$  & $\phi=5\%$ & $\phi=7.9\%$  & $\phi=10\% \,\,\,$  & \,\,\, $\phi=15\%$  & \,\,\,\, \,\,\,\,&  $\phi=10\%$  \\[3pt]
       $ \,\,\,\,\,\,\,\,\,\,\,\, \, \,  \,N_p$  & $0 \,\,\,\,\,\,\,\,\,\,$ & $2500$ & $3965$ & $5000$ & $7500$ && $5000$ \\[3pt]
      $ \,\,\,\,\,\,\,\,\,\,\,\, \, \,Re_b$  &  \,\,\,\,\,\,\, &  \,\,\,\,\,\,\, &  $5600$\\[3pt]
       $\, L_x \times L_y \times L_z$  &  \,\,\,\,\,\,\, &  \,\,\,\,\,\,\, &  $6h \times 2h \times 3h$\\[3pt]
       $N_x \times N_y \times N_z$  &  \,\,\,\,\,\,\, &  \,\,\,\,\,\,\, &  $1728 \times 576 \times 864$\\[3pt]
       $ \,\,\,\,\,\,\,\,\,\,\,\, \,  {\Delta x}^{\,\,+}$  & $0.625 \,\,\,\,\,\,\,\,\,\,$ & $0.623$ & $0.615$ & $0.608$ & $0.599$ && $0.684$ \\[3pt]    
       $ \,\,\,\,\,\,\,\,\,\,\,\, \, {D_{eq}}^+$  & $- \,\,\,\,\,\,\,\,\,\,$ & $19.94$ & $19.67$ & $19.44$ & $19.17$ && $21.89$ \\[3pt]                 
  \end{tabular}
  \caption{Summary of the different simulations cases. $N_p$ indicates the number of particles with equivalent diameter $D_{eq} = h/9$; $N_x$, $N_y$ and $N_z$ are the number of grid cells in each direction. ${\Delta x}^{\,\,+}$ and ${D_{eq}}^+$ are the Eulerian grid size and the particle equivalent diameter, given in viscous wall units for each simulation.} 
 \label{tab:cases}
 \end{center}
\end{table}

We study a pressure-driven plane channel flow in a computational domain of size $L_x=6h$, $L_y=2h$ and $L_z=3h$ in the streamwise, wall-normal and spanwise directions, where $h$ is half the channel height. The bulk velocity $U_b$ is fixed to guarantee a constant bulk Reynolds number $Re_b = 2hU_b/\nu = 5600$ corresponding to a friction Reynolds number $Re_\tau = U_*h/\nu =180$ for the single phase case with $\nu$, the kinematic viscosity of the fluid phase and $U_* = \sqrt{\tau_w/\rho_f}$, the friction velocity, calculated with the shear stress $\tau_w$ at the wall. Periodic boundary conditions are imposed for both fluid and particles in the streamwise, $x$, and spanwise, $z$, directions while the no-slip and no-penetration boundary conditions are employed at the walls. We note here that simulations in larger domains would be quite expensive computationally and have not been performed for the case of oblate particles. Nevertheless, the domain used in this study is larger than the minimal-unit channels adopted to identify the physical mechanisms underlying self-sustaining turbulence in Newtonian fluids \citep{Hamilton1995} as well as polymer suspensions where drag reduction is also observed \citep{Xi2010}. In addition, as also reported in \cite{Picano2015}, the presence of particles tends to disrupt long flow structures so that we believe the main conclusions from the simulations, focusing on drag reduction and particle dynamics, would not be different in longer domains.

We consider non-Brownian neutrally-buoyant rigid spheroidal particles with aspect ratio $\mathcal{AR} = 1/3$ (ratio of polar over equatorial radius). The particle equivalent diameter $D_{eq}$, i.e.\ the diameter of a sphere with the same volume, is $h/D_{eq} = 9$ to compare with the results of \cite{Picano2015} for spheres with diameter $D=D_{eq}$. The corresponding maximum and minimum diameters of the oblate particle, used here are $D_1=h/6.24$ and $D_2=h/18.72$. The particle Reynolds number based on the local shear, $Re_p \equiv \dot{\gamma} D_{eq}^2 / 4\nu$, ranges between approximately $95$ close to the walls and $0$ at the centerline of the channel, with $\dot{\gamma}$ approximated by the wall-normal gradient of the mean velocity.

We perform simulations at four different volume fractions $\phi = 5$; $7.9$; $10$; $15 \%$, corresponding to $7500$ particles at  $\phi=15\%$; we also consider the case of spheres at $\phi = 10\%$ and the unladen case for a direct comparison. 
 We reproduced the results at  $\phi = 10\%$ in \cite{Picano2015} with the collision model discussed in \cite{Ardekani2016}, accounting also for friction between the particles (and the wall), and at higher resolution, the same used for the oblate particles.
The results show a $4\%$ difference in the friction Reynolds number $Re_\tau$, at twice the grid resolution.

  The simulations are performed over a uniform Cartesian grid with the resolution of 32 grids per equivalent diameter $D_{eq}$ and $N_L = 3720$ and $N_L = 3219$ Lagrangian points on the surface of oblate and spherical particles, respectively. A summary of the simulated cases is given in table~\ref{tab:cases}. 
The simulations start from the laminar Poiseuille flow with random distribution of the particle position and orientation. The noise introduced by the presence of the particles triggers rapidly the transition to a fully turbulent state, after which the statistics are collected for about $16$ large-eddy turnover times $h/U_*$.


\section{Results}\label{sec:Results}
\vspace{10pt}

\begin{figure}
   \centering
   \includegraphics[width=0.495\textwidth]{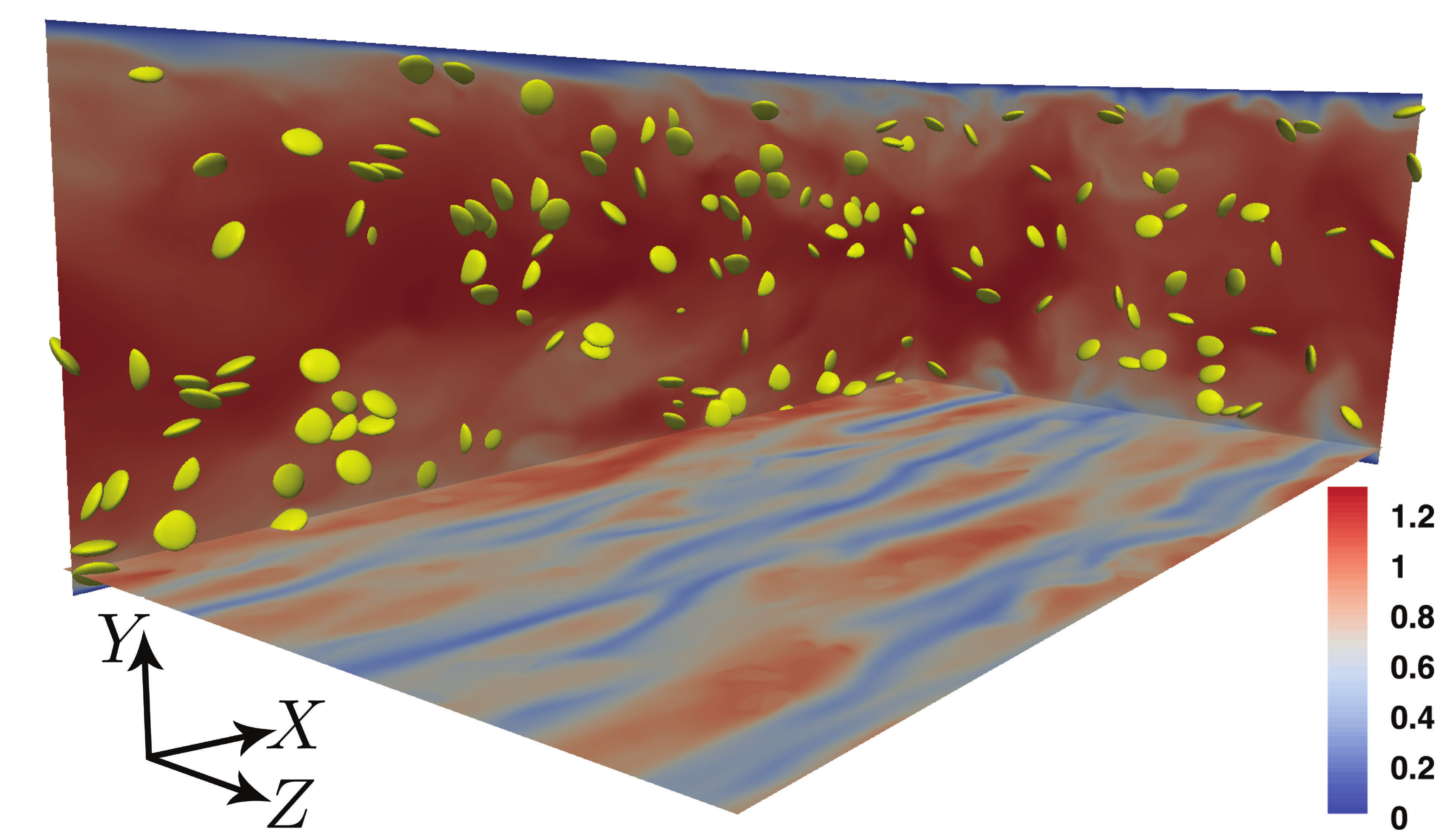}
   \includegraphics[width=0.495\textwidth]{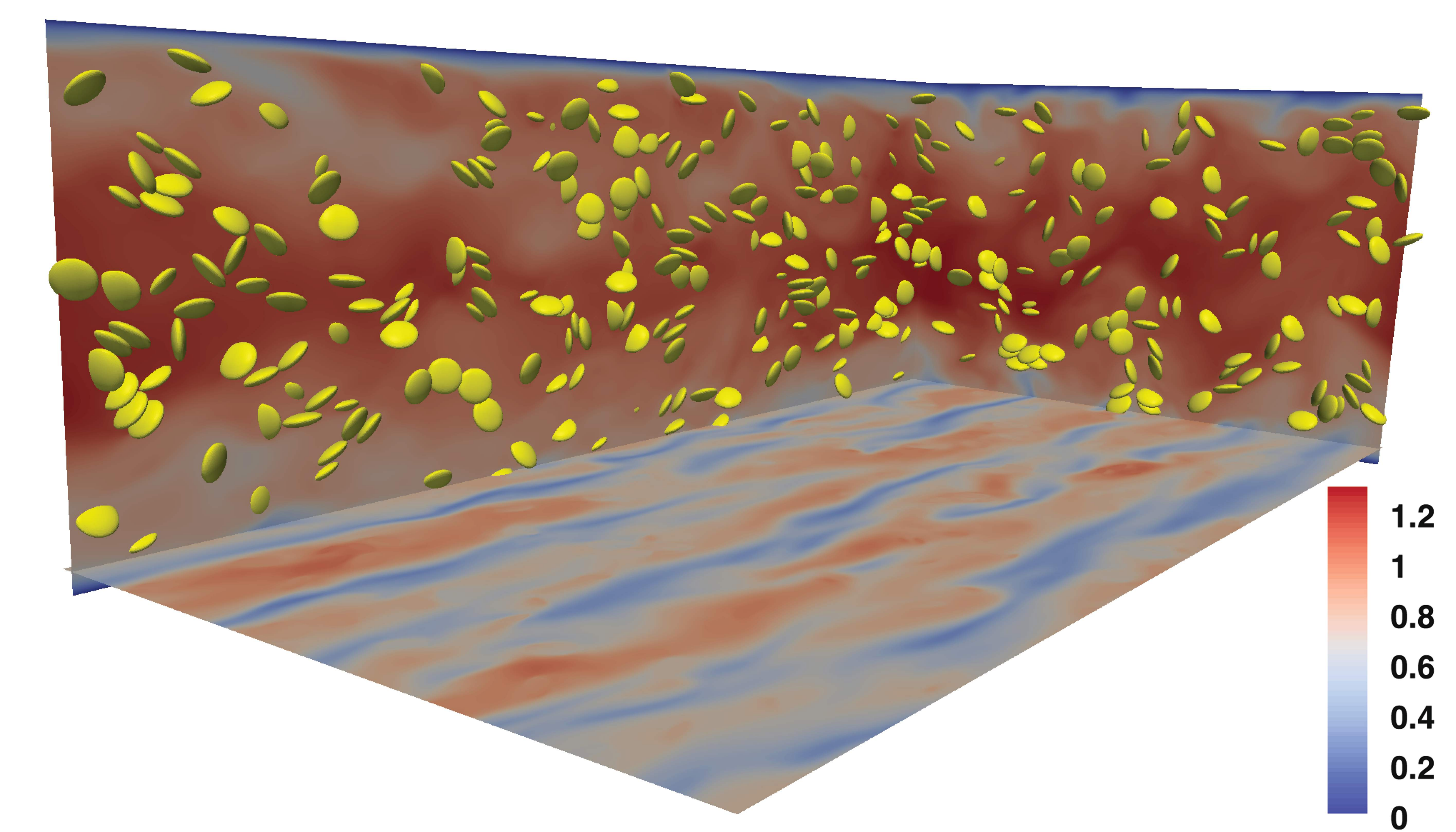} 
   \put(-395,110){\footnotesize $(a)$}
   \put(-200,110){\footnotesize $(b)$} \\ [5pt]
   \includegraphics[width=0.495\textwidth]{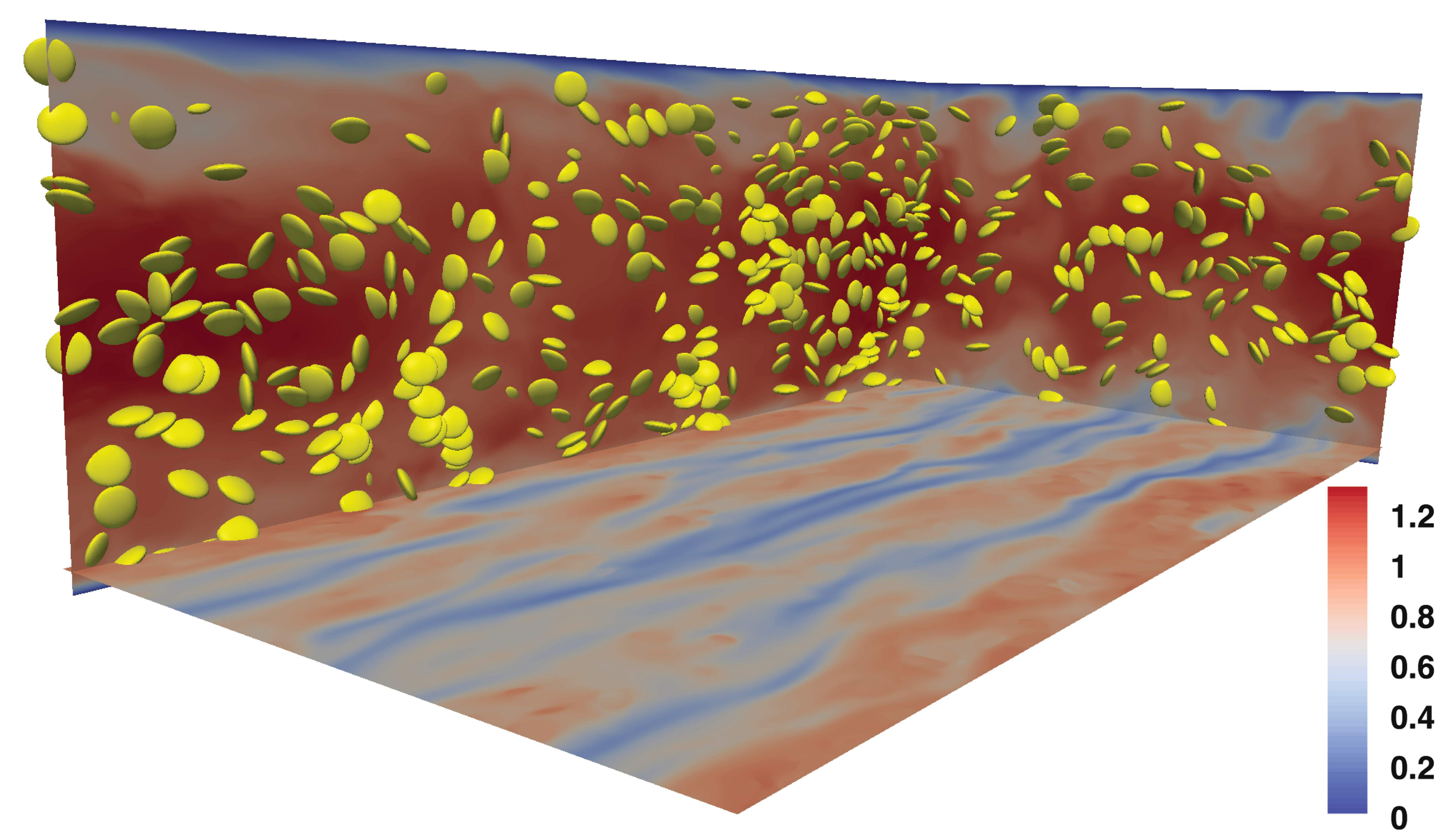}
   \includegraphics[width=0.495\textwidth]{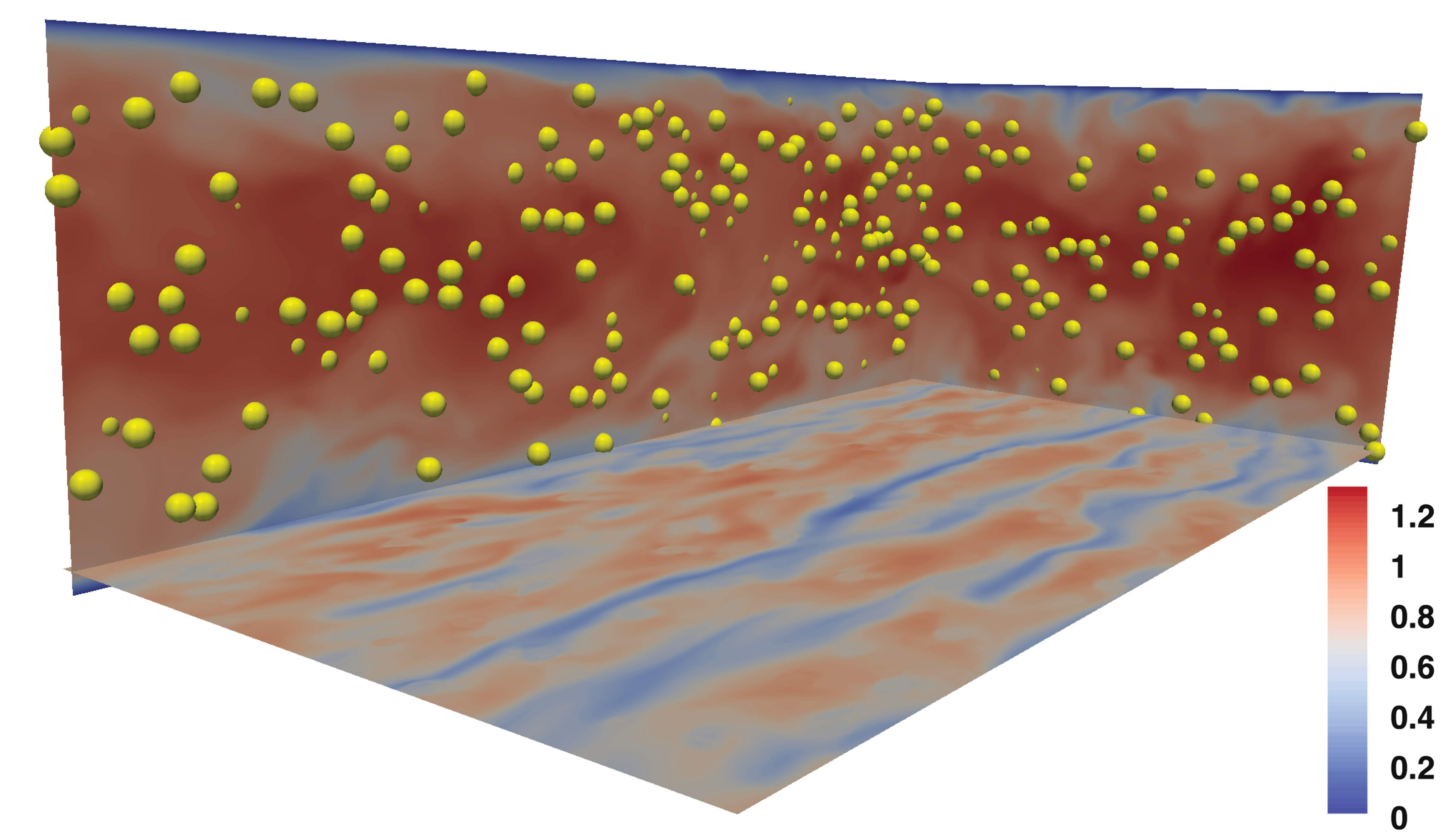} 
   \put(-395,110){\footnotesize $(c)$}
   \put(-200,110){\footnotesize $(d)$} \\ [5pt]
  \caption{Instantaneous snapshots of streamwise velocity $u$ in the presence of particles on orthogonal planes $xz$, $yz$ and $xy$ for $(a)$ oblates at $\phi = 5\%$, $(b)$ oblates at $\phi = 10\%$, $(c)$  oblates at $\phi = 15\%$ and $(d)$ spheres at $\phi = 10\%$. For clarity, just a fraction of particles, lying within the $xz$ and $yz$ planes are displayed.}
\label{Snapshots}
\end{figure}

We first display snapshots of the fluid flow and particles, see figure~\ref{Snapshots} where the instantaneous streamwise velocity $u$ is depicted on different horizontal and vertical planes for the cases with oblates at $\phi = 5\%$, $\phi = 10\%$ and $\phi = 15\%$ and for spheres at $\phi = 10\%$. For clarity, just a fraction of the particles (those lying within the visualized $xz$ and $yz$ planes) are displayed. The streamwise low-speed streaks, characteristics of wall-bounded turbulence, can be observed near the wall for all cases; being however more noisy in the flow laden with spheres.     

\subsection{Fluid phase statistics: drag reduction}

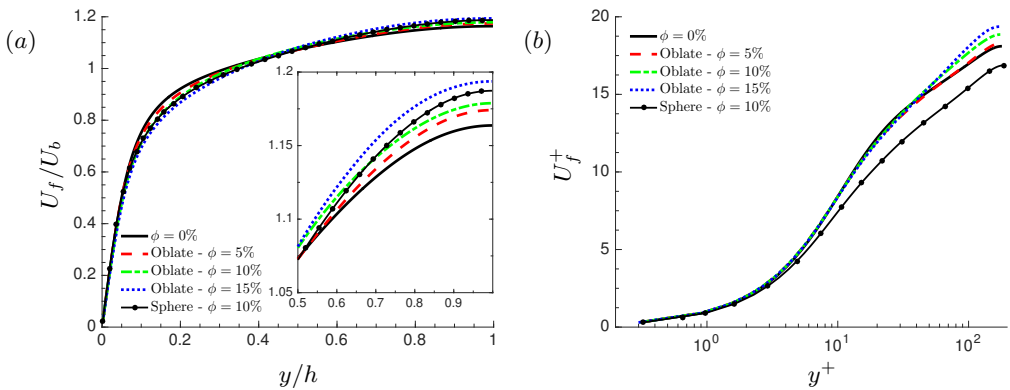
\begin{figure}
  \centering
   \includegraphics[width=0.495\textwidth]{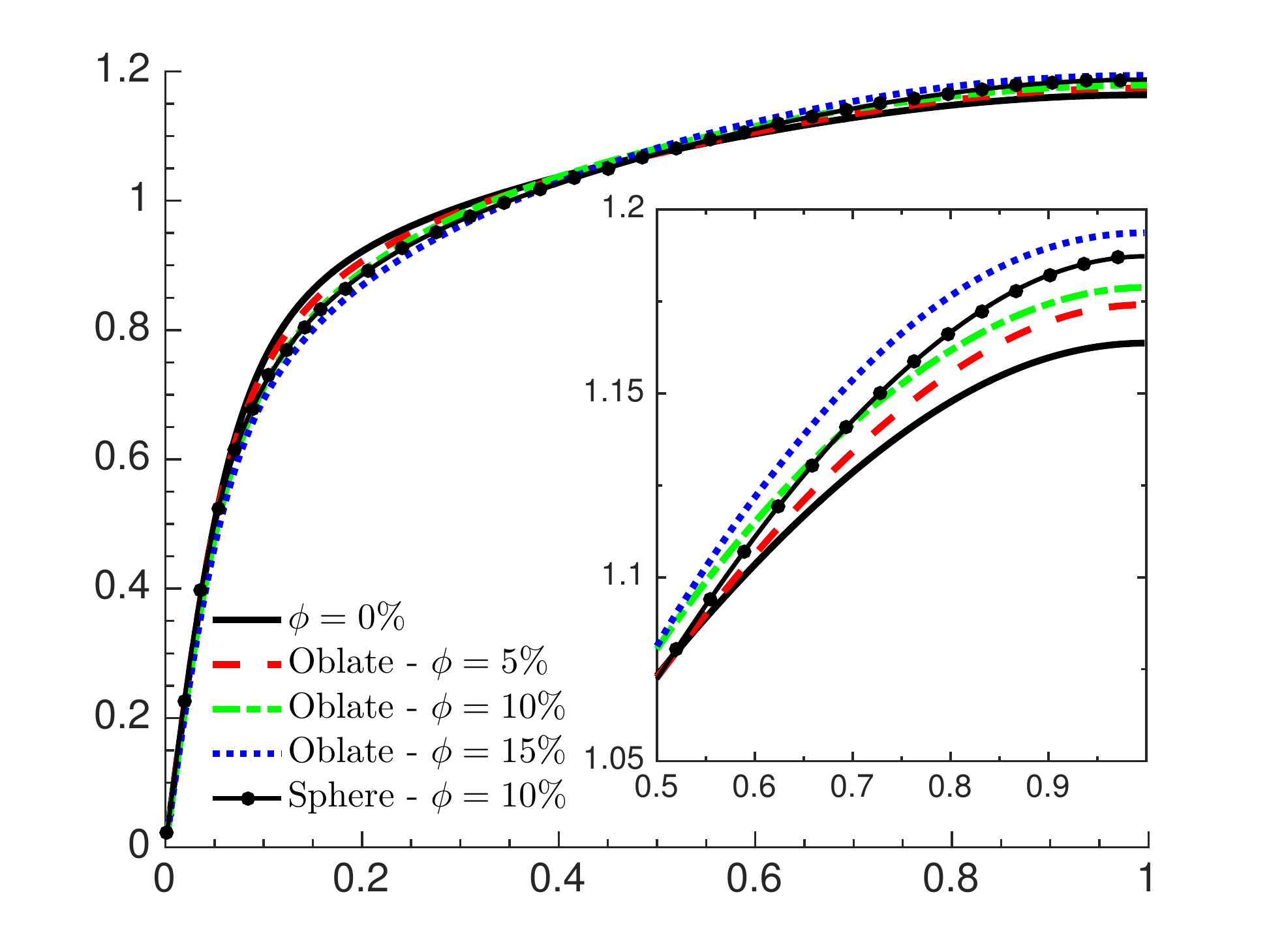}
   \includegraphics[width=0.495\textwidth]{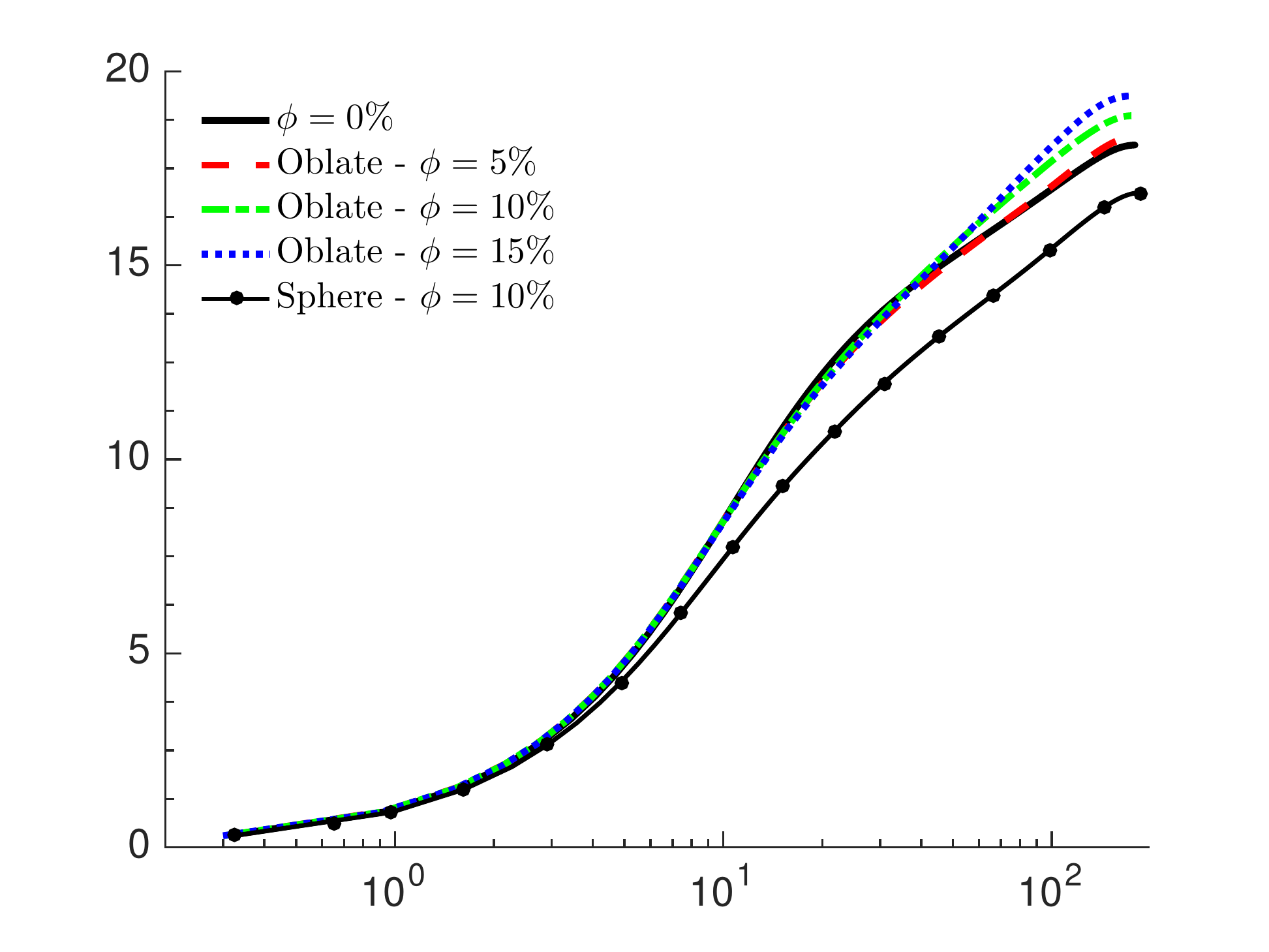}
   \put(-382,60){\rotatebox{90}{$U_f / U_b$}}
   \put(-190,70){\rotatebox{90}{$U^+_f$}}
   \put(-93,-5){{$y^+$}}
   \put(-292,-5){{$y / h$}}
   \put(-395,120){\footnotesize $(a)$}
   \put(-200,120){\footnotesize $(b)$}
   \vspace{5pt}
  \caption{Mean fluid velocity profiles $U_f$ in the streamwise direction for the different cases under investigation. The data are scaled with: $(a)$ outer units: $U_f / U_b$ versus $y / h$ and $(b)$ inner units: $U^+_f \equiv U_f / U_*$ versus $y^+ \equiv yU_* / \nu$, depicted in semi-logarithmic scale. The inset in panel $(a)$ displays the increase of the mean velocity at the core of the channel. }
\label{fig:mean}
\end{figure}
\begin{table}
  \begin{center}
\def~{\hphantom{0}}
  \begin{tabular}{lccccccc}
   & &  \multicolumn{4}{c}{Oblate - $\mathcal{AR} = 1/3$}  & &  \multicolumn{1}{c}{Sphere} \\
   \cline{3-6} \cline{8-8} \\
       \,\,\,\,\,\,\,\,\,  \,Case & $\,\,\,\, \phi=0\% \,\,\,\,\,\,\,\,\,\,$  & $\phi=5\% \,\,$ & $\phi=7.9\% \,\,$  & $\phi=10\% \,\,$  & $\phi=15\%$  & \,\,\,\, \,\,\,\,&  $\phi=10\%$  \\[3pt]
$ \,\,\,\,\,\,\,\,\,\,\,\,\,  Re_\tau$  & $\,\,\,\, 180 \,\,\,\,\,\,\,\,\,\,$ & $179.5 \,\,$ & $177 \,\,$ & $175 \,\,$ & $172.5$ && $197$ \\[3pt]
$ \,\,\,\,\,\,\,\,\,\,\,\, \, \,  \, \kappa$  & $ \,\,\,\, 0.4 \,\,\,\,\,\,\,\,\,\,$ & $0.38 \,\,$ & $0.36 \,\,$ & $0.34 \,\,$ & $0.28$ && $0.32$ \\[3pt]
$ \,\,\,\,\,\,\,\,\,\,\,\, \, \,  \, \beta$  & $\,\,\,\, 5.5 \,\,\,\,\,\,\,\,\,\,$ & $4.9 \,\,$ & $4.6 \,\,$ & $4 \,\,$ & $1.6$ && $0.8$ \\[3pt]
  \end{tabular}
  \caption{The friction Reynolds number $Re_\tau$ from all simulations at fixed bulk Reynolds number $Re_b =5600$ and the von K\'arm\'an constant $\kappa$ and the additive constant $\beta$ of the logarithmic law, calculated in the range of $50 \, <  \, y^+  < \, 150$.}
 \label{tab:kbeta}
 \end{center}
\end{table}

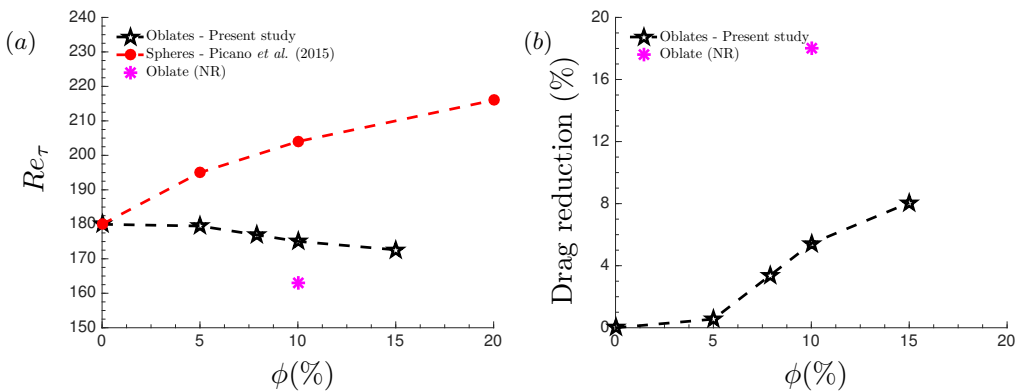
\begin{figure}
   \centering
   \includegraphics[width=0.495\textwidth]{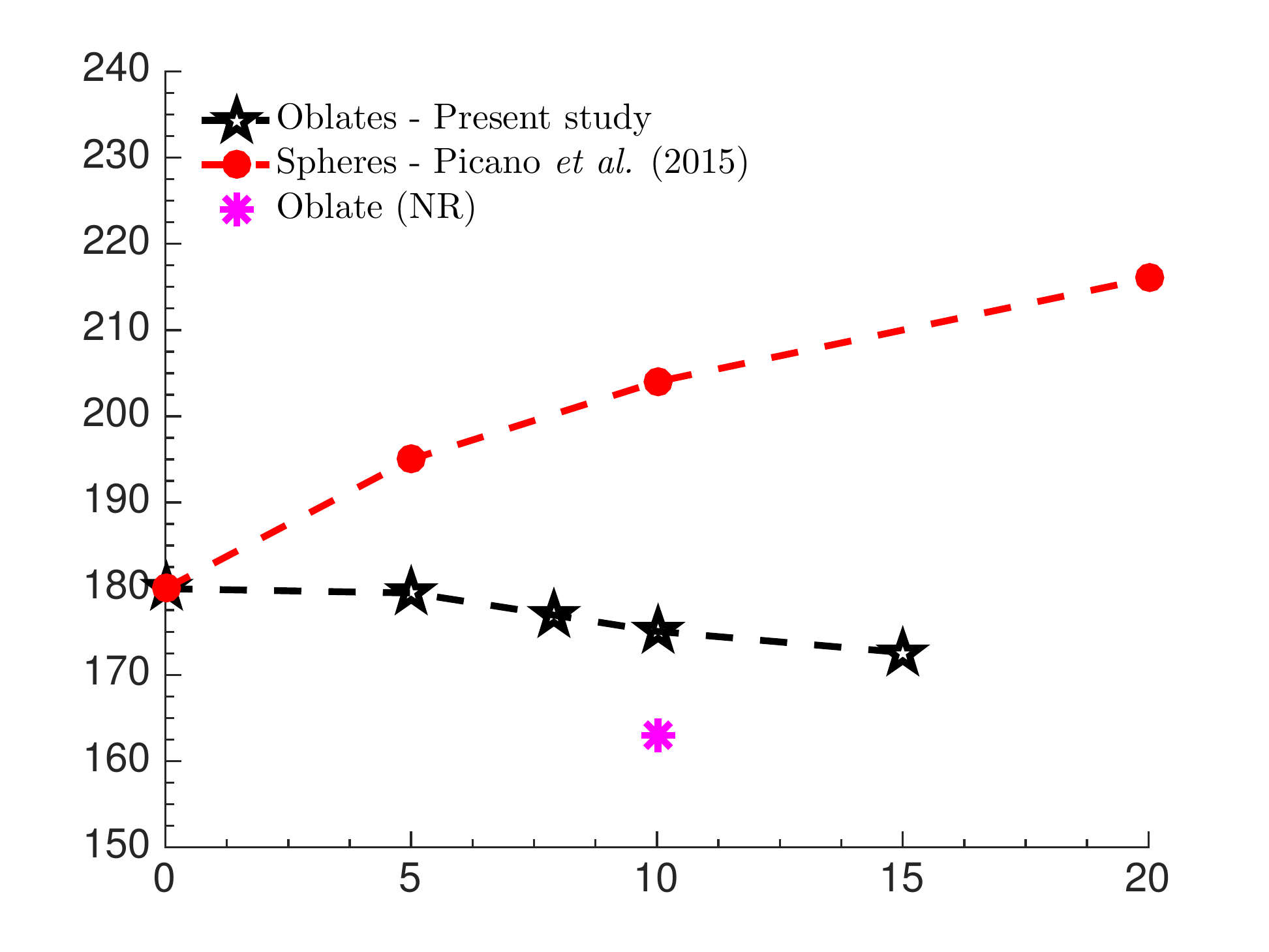}
   \includegraphics[width=0.495\textwidth]{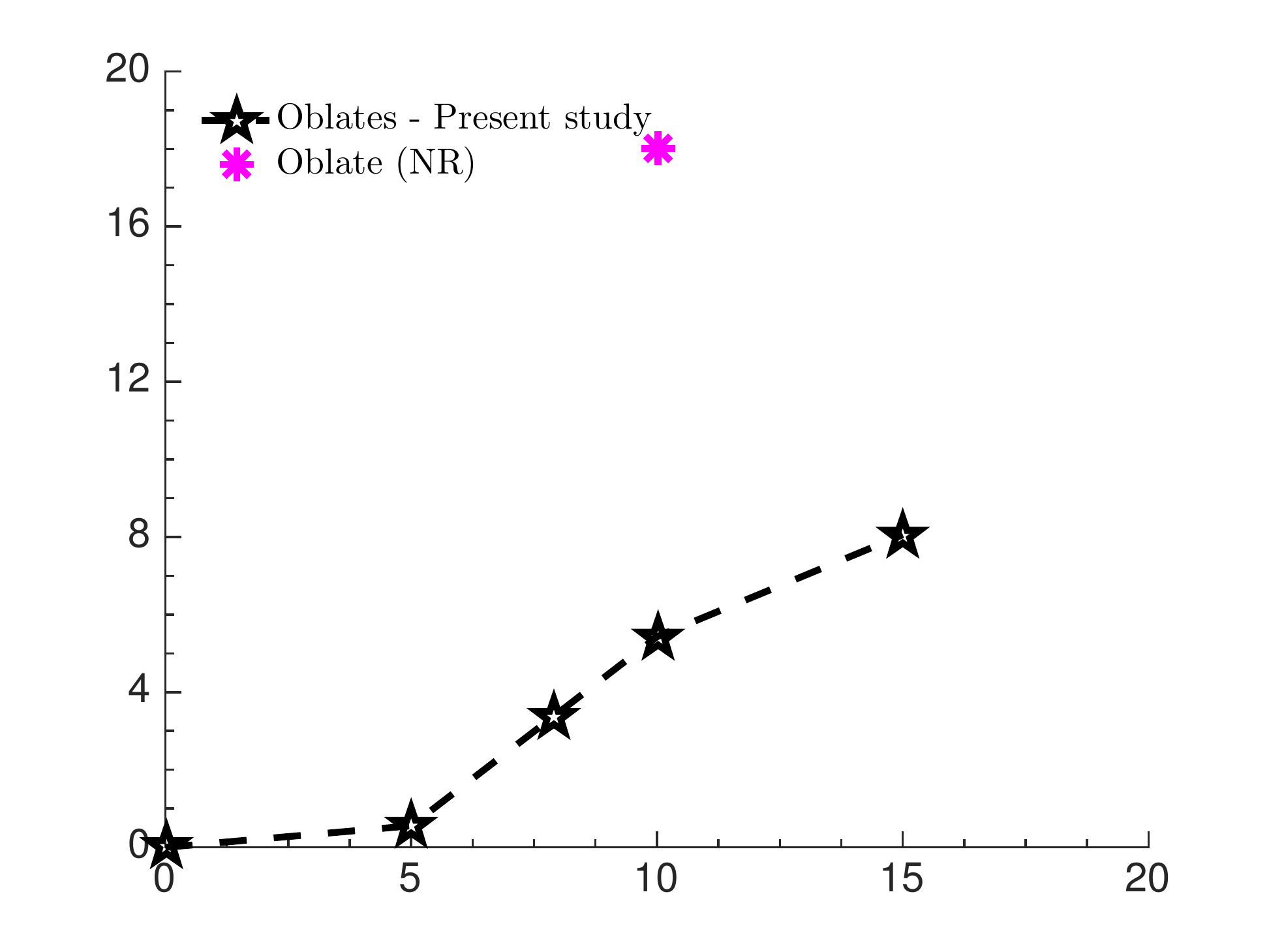}  
   \put(-395,120){\footnotesize $(a)$}
   \put(-200,120){\footnotesize $(b)$}
   \put(-388,63){\rotatebox{90}{ \large $Re_\tau$}}
   \put(-190,21){\rotatebox{90}{ \large Drag reduction $(\%)$}}   
   \put(-298,-5){{ \large $\phi (\%)$}}
   \put(-105,-5){{ \large $\phi (\%)$}}   
  \caption{$(a)$ Friction Reynolds number, $Re_\tau$, for oblate particles and spherical particles from \cite{Picano2015} and $(b)$ the drag reduction percentage for oblate particles versus volume fraction $\phi$. The case, denoted Oblate(NR, no rotation) will be addressed later is this section.}
\label{fig:Drag}
\end{figure}

For the statistics presented hereafter, the fluid velocity is considered only when outside the particles, whereas 
values pertaining the rigid body motion of the solid phase are used for the particle statistics. To distinguish between phases we use a phase indicator function that determines whether a grid point is located inside the fluid or solid phase. 
The  mean fluid velocity profiles, scaled in outer units $U_f / U_b$  and inner units $U^+_f \equiv U_f / U_*$, are depicted in figure~\ref{fig:mean} versus $y / h$ and $y^+ \equiv yU_* / \nu$ for the different cases. The results in figure~\ref{fig:mean}$(a)$ show that the mean velocity at the core of the channel increases when increasing the solid volume fraction, see inset, whereas it decreases around $y/h \approx 0.2$, that is at the beginning of the log-layer ($y^+ \approx 35$) for the studied Reynolds number. As the simulations assume constant bulk velocity $U_b$, this is a first indication of reduced drag.
The mean fluid velocity profile  for spheres at $\phi = 10\%$
deviates from the unladen case more than for oblates at same volume fraction. 
Displaying data in inner scaling, see figure~\ref{fig:mean}$(b)$, we identify a region ($50 \, <  \, y^+  < \, 150$) where the mean profiles follow a log-law of the type
\begin{equation} \label{eq:loglaw}
U_f^+ = \frac{1}{\kappa}\ln(y^+) + \beta .
\end{equation}
The values of the the effective von K\'arm\'an and additive constants, $\kappa$ and $\beta$, that best fit our results are reported in table~\ref{tab:kbeta} together with  the friction Reynolds number $Re_\tau$. Note that a reduced $\kappa$ is a sign of drag reduction, while a reduced $\beta$ indicates the opposite \citep{Virk1975}. 
The data in the table show that $\kappa$ and $\beta$ decrease with increasing the volume fraction $\phi$ for oblate particles. 
The largest reduction of $\beta$ is however found for spheres, while the value of $\kappa$  is close to that for oblates at $\phi=10\%$. 
The combination of the two effects results in drag reduction for oblate particles and a drag enhancement for the spheres as confirmed by the friction Reynolds number $Re_\tau$, shown in figure~\ref{fig:Drag}$(a)$ versus $\phi$ for oblate and  spherical particles.
For oblate particles, the effect of the solid phase is small for  
$\phi \lesssim 5\%$;  the total drag clearly decreases below the value of single phase flow for larger volume fractions:  
$Re_\tau=172.5$  for $\phi = 15\%$, corresponding to a drag reduction of $8.2 \%$ with respect to the single phase case ($(\tau_w |_{\phi = 15\%} - \tau_w |_{\phi = 0\%}) / \tau_w |_{\phi = 0\%}$). 
The drag reduction percentage is depicted versus the volume fraction  in figure~\ref{fig:Drag}$(b)$. 
The data in \cite{Picano2015} for perfect spheres, conversely, shows a clear increase of the friction Reynolds number 
when increasing the volume fraction.

\begin{figure}
   \centering
   \includegraphics[width=0.495\textwidth]{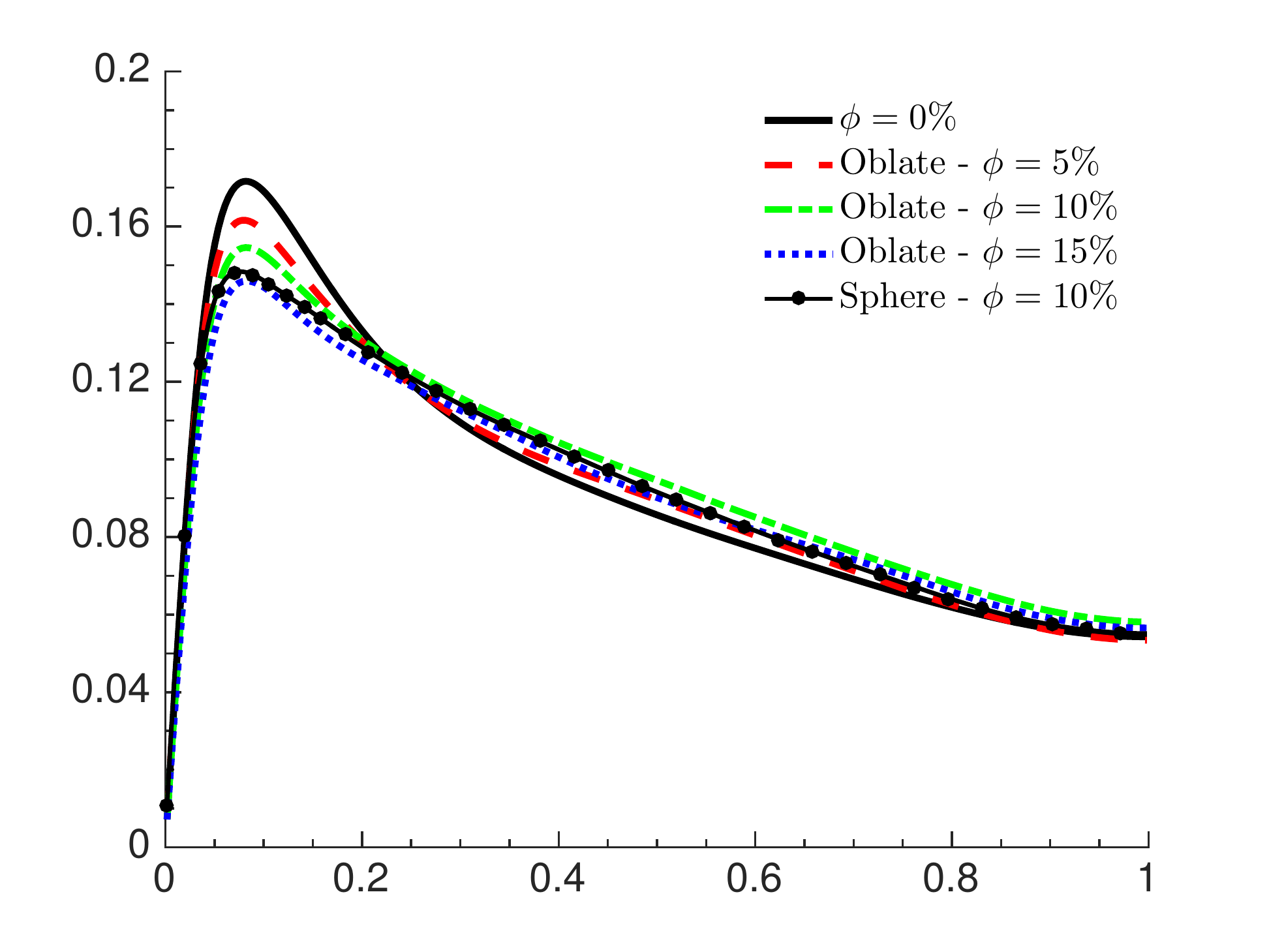}
   \includegraphics[width=0.495\textwidth]{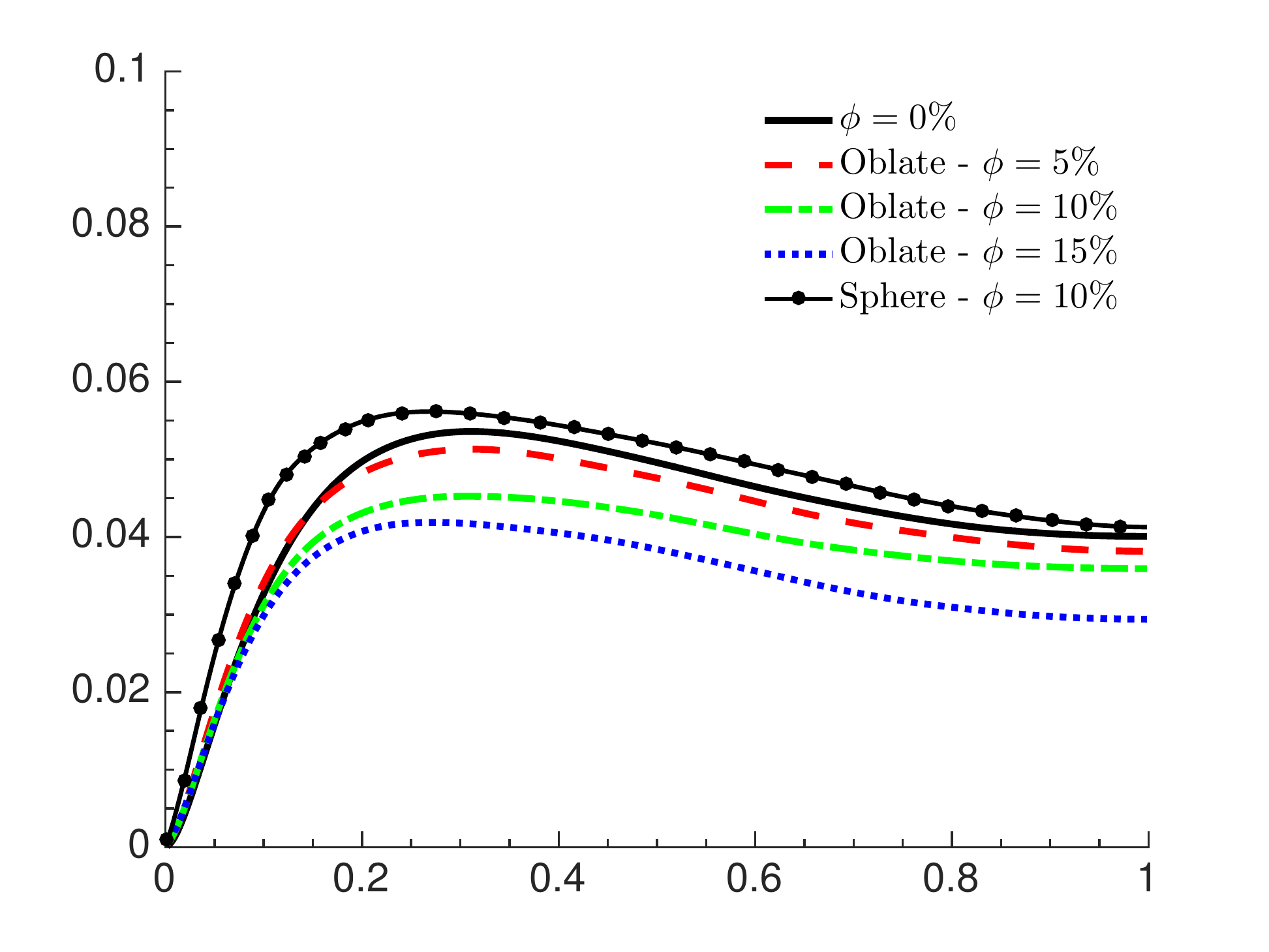} 
   \put(-292,-5){{$y / h$}}   
   \put(-98,-5){{$y / h$}}   
   \put(-388,60){\rotatebox{90}{${u^\prime_f}_{rms}$}}
   \put(-195,60){\rotatebox{90}{${v^\prime_f}_{rms}$}}   
   \put(-395,120){\footnotesize $(a)$}
   \put(-200,120){\footnotesize $(b)$}\\
   \includegraphics[width=0.495\textwidth]{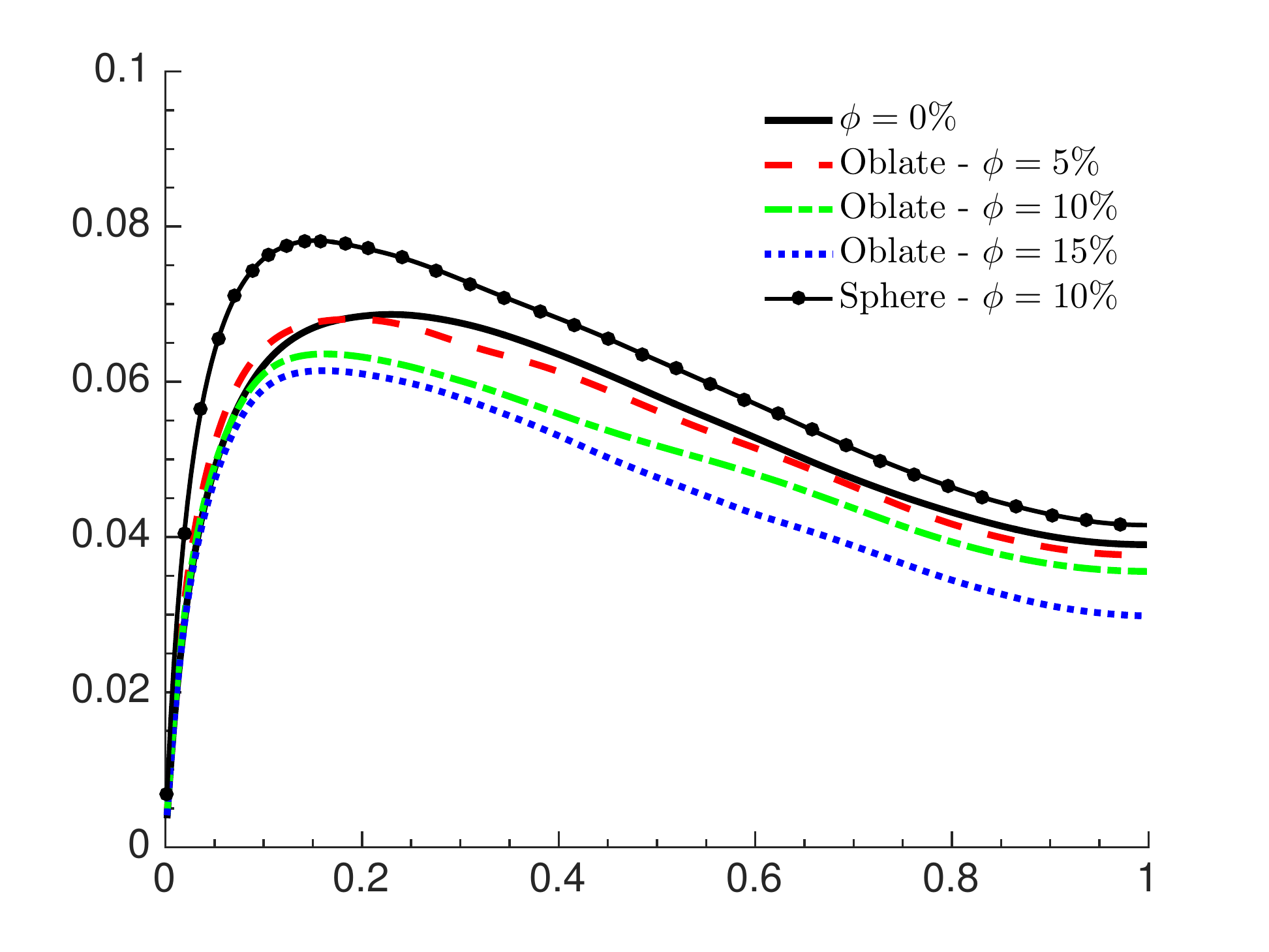}
   \includegraphics[width=0.495\textwidth]{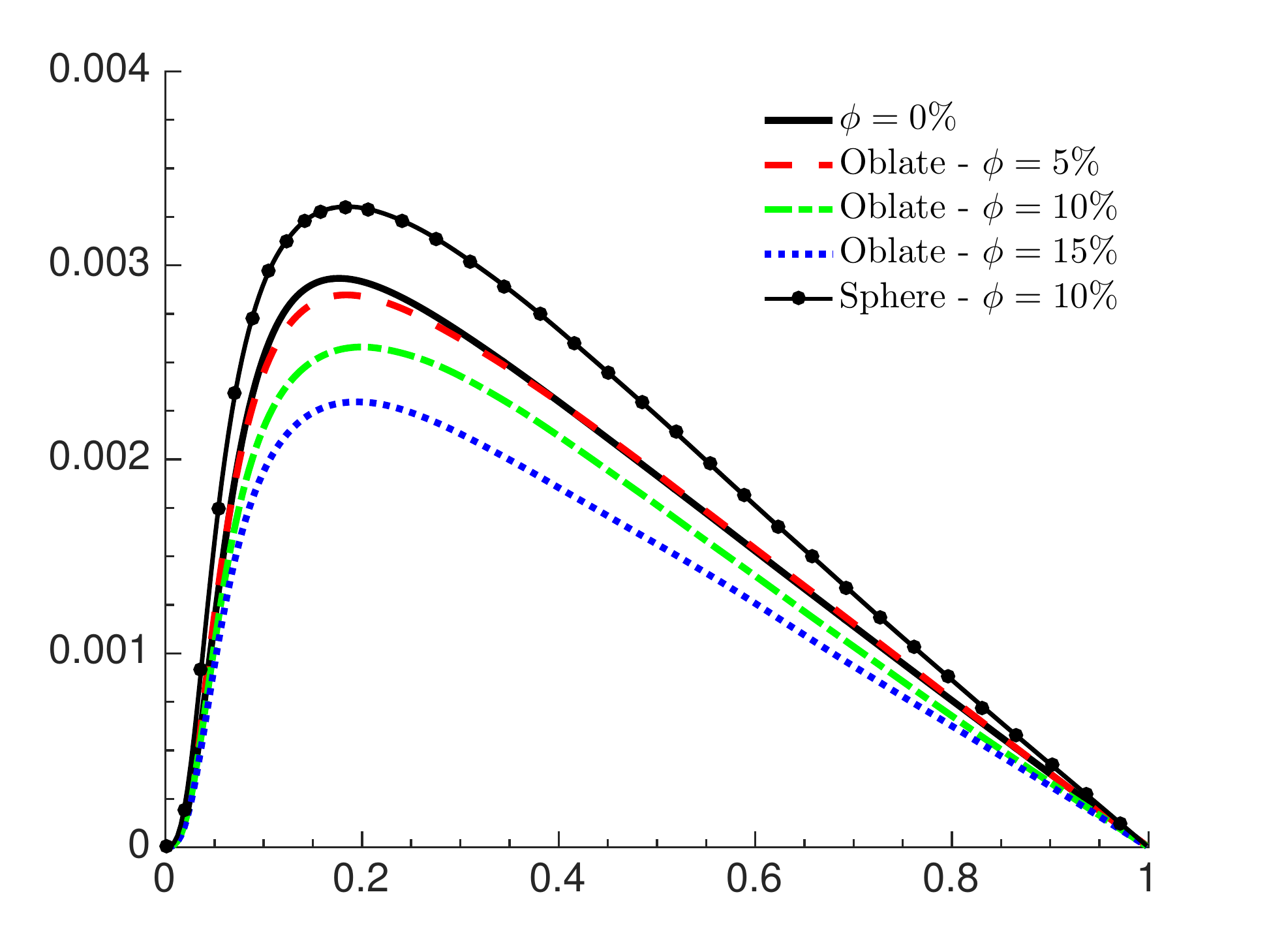} 
      \put(-395,120){\footnotesize $(c)$}
   \put(-200,120){\footnotesize $(d)$}
   \put(-292,-5){{$y / h$}}   
   \put(-98,-5){{$y / h$}}  
   \put(-388,60){\rotatebox{90}{${w^\prime_f}_{rms}$}}
   \put(-195,58){\rotatebox{90}{$-\langle u^\prime_f v^\prime_f \rangle$}}       
  \caption{Root-mean-square velocity fluctuations and Reynolds shear stress for the fluid phase, scaled in outer units: $(a)$ streamwise ${u^\prime_f}_{rms}$; $(b)$ wall-normal ${v^\prime_f}_{rms}$; $(c)$ spanwise ${w^\prime_f}_{rms}$; $(d)$ Reynolds shear stress $\langle u^\prime_f v^\prime_f \rangle$.}
\label{fig:RMS}
\end{figure}

\begin{figure}
   \centering
   \includegraphics[width=0.495\textwidth]{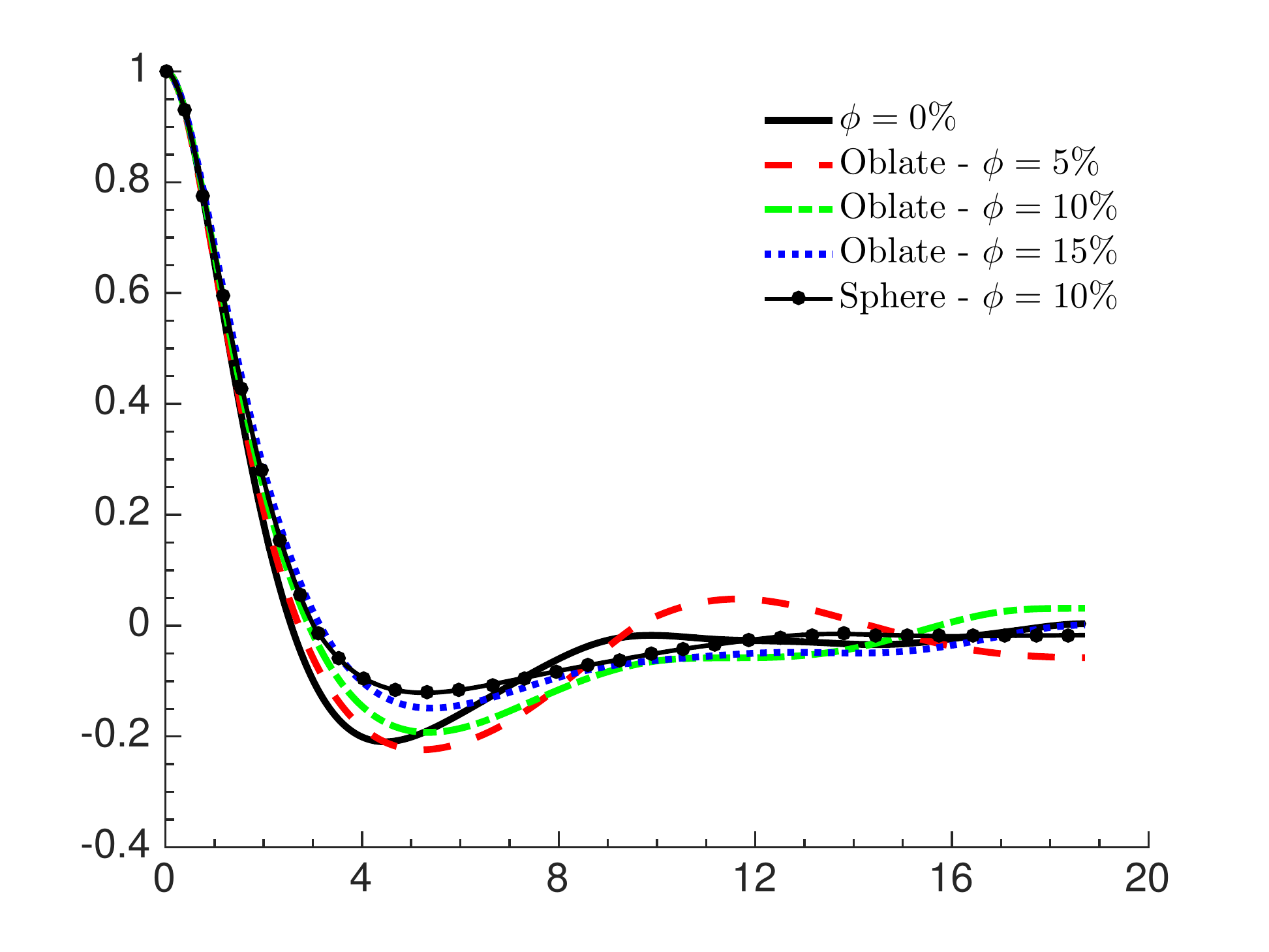}
   \includegraphics[width=0.495\textwidth]{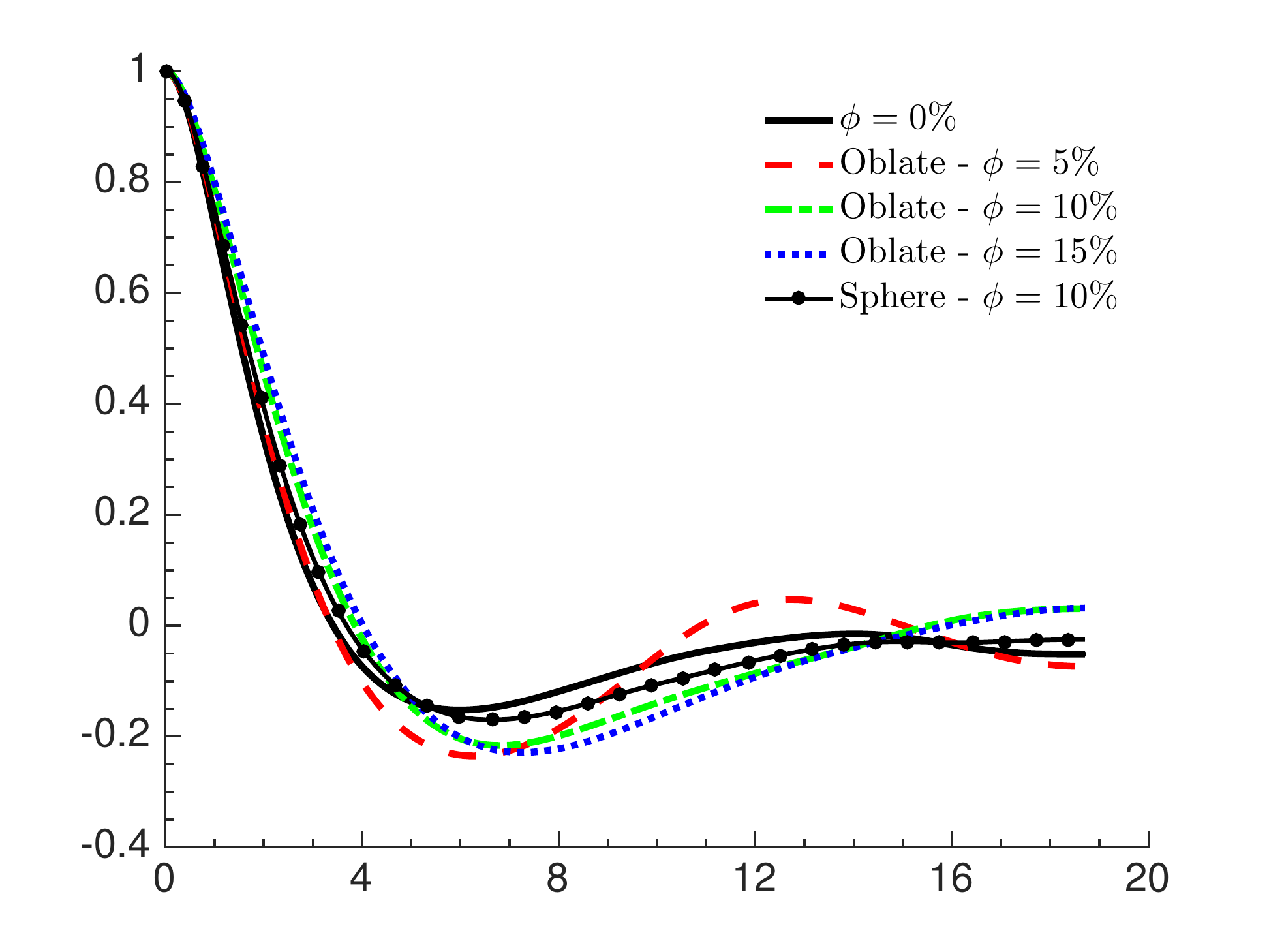} 
   \put(-301,-5){{$z / R_{max}$}}   
   \put(-107,-5){{$z / R_{max}$}} 
   \put(-388,66){\rotatebox{90}{$R_{uu}$}}
   \put(-195,66){\rotatebox{90}{$R_{uu}$}}           
     \put(-395,120){\footnotesize $(a)$}
   \put(-200,120){\footnotesize $(b)$} \\     
   \includegraphics[width=0.495\textwidth]{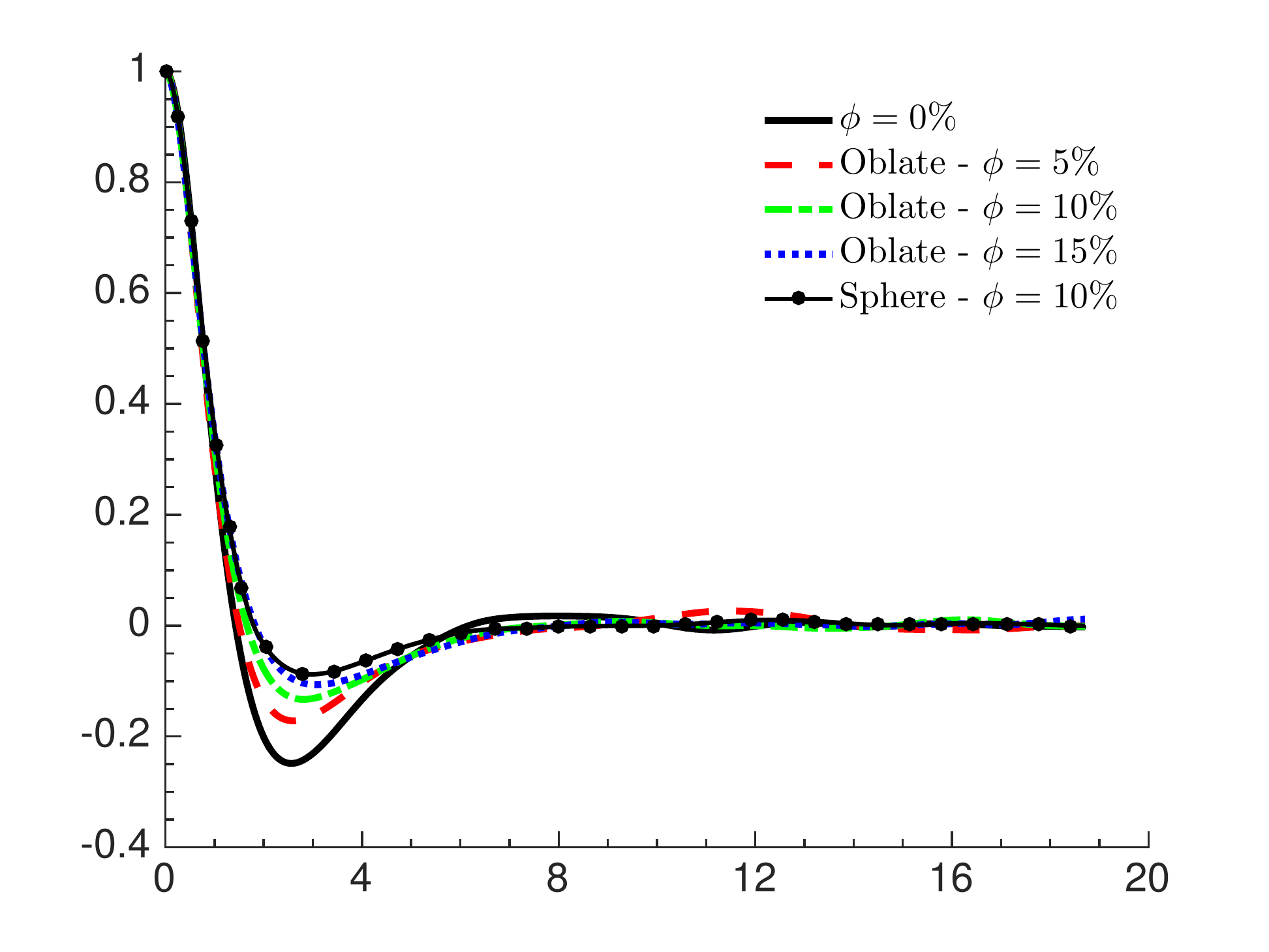}
   \includegraphics[width=0.495\textwidth]{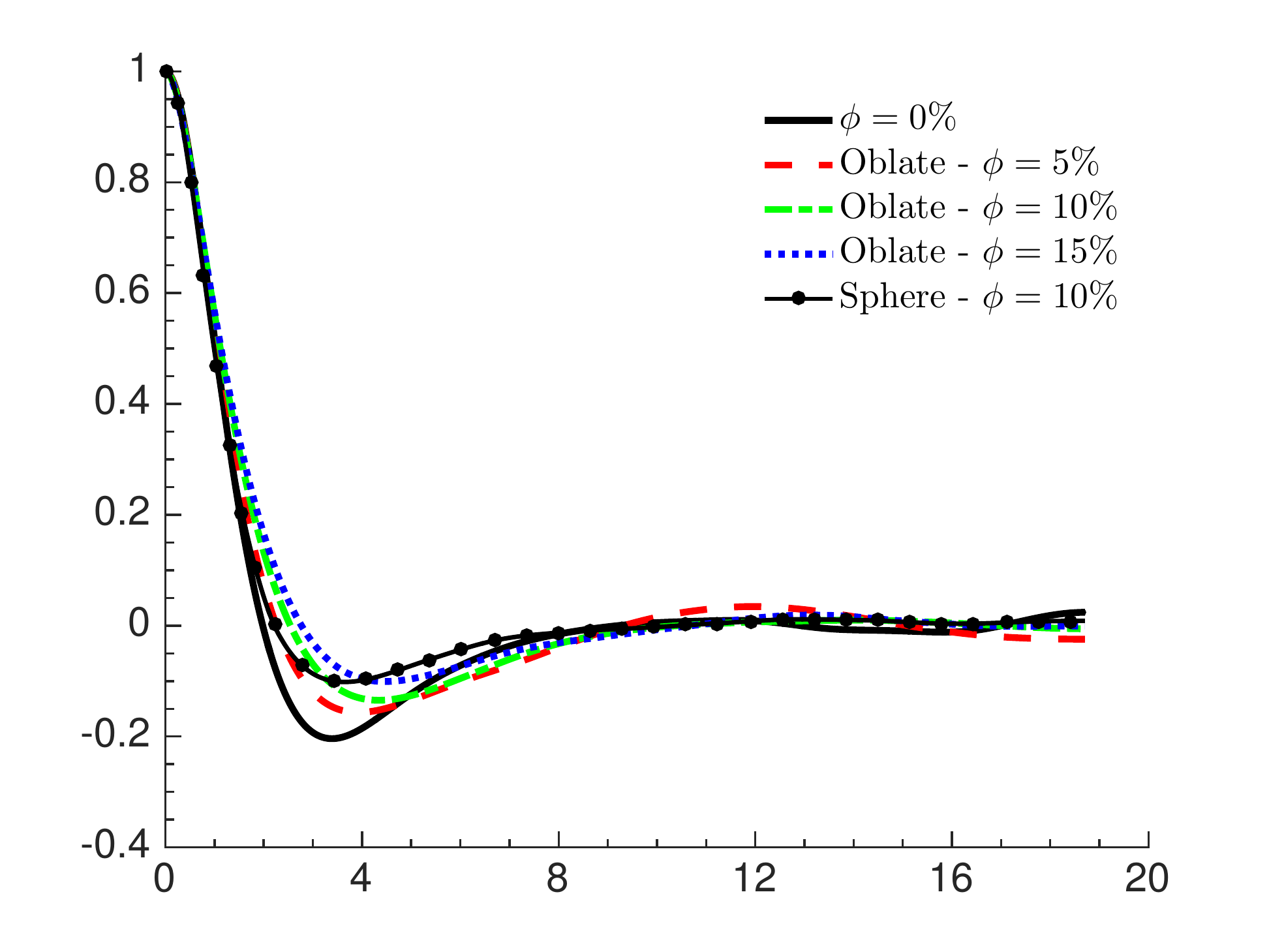}
      \put(-395,120){\footnotesize $(c)$}
   \put(-200,120){\footnotesize $(d)$}
   \put(-388,66){\rotatebox{90}{$R_{vv}$}}
   \put(-195,66){\rotatebox{90}{$R_{vv}$}}     
   \put(-301,-5){{$z / R_{max}$}}   
   \put(-107,-5){{$z / R_{max}$}} \\ [5pt]        
  \caption{ Correlations of the velocity fluctuations versus the spanwise separation, normalized by the major radius of the oblate particles $z / R_{max}$, for the different cases under consideration: $(a)$ streamwise--streamwise component $R_{uu}$ at $y^+ = 20$; $(b)$ $R_{uu}$ at $y^+ = 40$; $(c)$ wall-normal--wall-normal component $R_{vv}$ at $y^+ = 20$ and $(d)$ $R_{vv}$ at $y^+ = 40$.}
\label{fig:Correlations}
\end{figure}
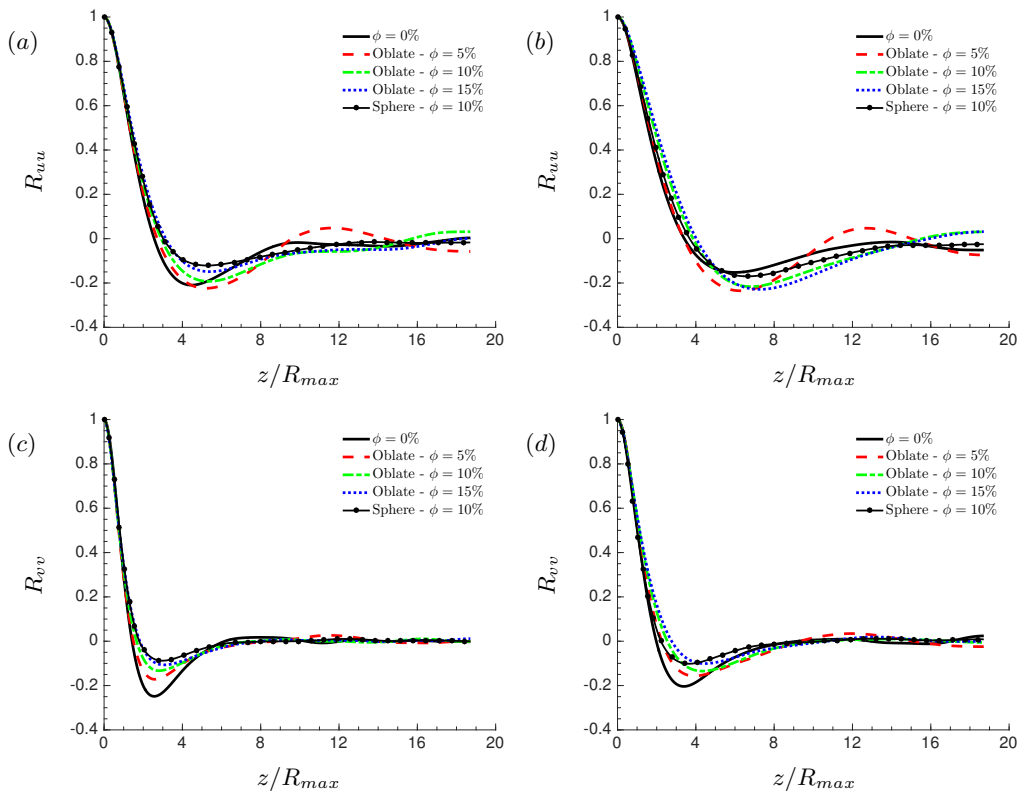

Drag reduction with small (smaller than the Kolmogorov length scale) rigid fiber additives has been reported in the literature \citep[e.g.][]{Paschkewitz2004,Gillissen2008}. These authors associate the drag reduction to the attenuation of the turbulence and the increase of its anisotropy, which results in higher streamwise velocity fluctuations and lower spanwise and the wall-normal velocity fluctuations (when scaled in inner units) with respect to the single phase flow, similarly to what observed for polymer additives \citep{De2002,Nowbahar13}. 
\cite{De2002} explained this by revealing that in most of the field, polymers are extracting energy from the turbulence, resulting in a reduction of spanwise and wall-normal velocity fluctuations. In the streamwise direction, however, an increase in the velocity fluctuations is caused by larger streamwise vortices located further from the wall.
The root-mean-square (rms) of the the fluid velocity fluctuations and the Reynolds shear stresses for suspensions of finite-size oblates are depicted in figure~\ref{fig:RMS} for 
the different cases studied here. 
The data reveal that the peak of the turbulent velocity fluctuations is reduced in the presence of oblate particles with respect to the single phase flow for all three components. In the cross-stream directions, the fluctuations decrease with the volume fraction of oblate particles, while an increase is observed for spheres at $\phi = 10\%$. The peak in the streamwise velocity fluctuations displays a reduction for both spherical and oblate particles with respect to the single phase flow, while a slight increase of the fluctuations is observable in the regions far from the wall. This slight increase, present for both spherical and oblate particles, can be related to the movement of particles in the wall-normal direction, where they accelerate or decelerate the flow based on their streamwise velocity.
A clear turbulence attenuation is shown by the Reynolds shear stress profiles of oblate particles. The turbulence attenuation is weak at $\phi = 5\%$ and becomes more pronounced at higher volume fractions;  a considerable reduction in the Reynolds shear stress is observed for the case at $\phi = 15\%$. On the contrary, a significant increase in the turbulence activity is observed for spheres at $\phi = 10\%$.  Interestingly,  $v^\prime_f$ and $w^\prime_f$ change significantly with respect to the single-phase flow in the close vicinity of the wall, which is peculiar of spherical particles and missing in the case of oblates. 
This indicates that the effect of particles on the turbulence very close to the wall is considerably higher for spherical particles than for oblates due to the absence of a particle layer close to the wall, as documented later. As discussed in \cite{Costa2016}, accounting for the particle dynamics in this region is critical for an accurate prediction of the overall drag.

Finally, we examine how the presence of the  oblate particles affects the turbulent structures near the wall by computing the two-point spatial correlation of the velocity field. The autocorrelations of the streamwise and wall-normal velocity, $R_{uu}$ and $R_{vv}$, are depicted in figure \ref{fig:Correlations} versus the spanwise separation, normalized by the major radius of the oblate particles. Panels $(a)$ and $(b)$ of this figure show $R_{uu}$ at two different distances from the wall $y^+ = 20$ and $y^+ = 40$, while \ref{fig:Correlations}$(c,d)$ report $R_{vv}$ at the same wall distances. The data reveal the characteristics of typical turbulence structures near the wall, i.e. quasi-streamwise
vortices and low-speed streaks \citep{Kim1987,Pope2001,Brandt2014}. 
The results at $y^+ = 20$ show an increase in the minimum location of $R_{uu}$, i.e.\ widening of the near wall streaks, whose size becomes independent of the volume fraction, and appears to be more correlated to the particle size. This is in contrast with the results for spherical particles reported in \cite{Picano2015}, where the spacing increases monotonously with the volume fraction. 
We also observe a decrease of the minimum of the correlation, i.e.\ less pronounced streaks,
 as the volume fraction increases, while further away from the wall ($y^+ = 40$), this effect disappears, the peak is still distinct, but the streamwise velocity streaks are considerably wider in the particle laden flows. 
 The results for $R_{uu}$ are consistent with the turbulence attenuation reported above for oblate particles as the width and the spacing of the streamwise velocity streaks increases in the drag-reducing turbulent flows \citep{Stone2002}, corresponding to an increase in the extent of the buffer layer. 
 The wall-normal autocorrelations $R_{vv}$ indicate a progressive smoothening of the local minima with the volume fraction in the cases of oblate particles at both wall-normal locations under consideration, indicating a more random flow in terms of coherent turbulence structures.

\subsection{Effective viscosity and stress budget}

To better understand the effect of particles on the fluid turbulence one may consider the idealized case of a suspension with effective viscosity $\nu_e$. The effective viscosity of a particle suspension is always higher than the viscosity for the single phase flow: The ratio between the effective suspension viscosity and the fluid viscosity $\nu_r=\nu_e/\nu$ is typically estimated via empirical relations at volume fractions larger than few percents, an example being the Eilers fit \citep{Stickel2005}, $\nu_r = {[1 + 1.25 \phi / (1-\phi/0.6)]}^2$. 
The effective viscosity estimated from simulations of spherical and oblate particles in laminar flow at $Re_b = 1000$ and $\phi=10\%$ (see appendix~\ref{app:Laminar_Simulations})
is in good agreement with the mentioned fit ($2\%$ under-predicted for oblates and $4\%$ over-predicted for spheres), and this relation is therefore used to estimate the effective viscosity used for the analysis of suspensions in turbulent flow.

The results in \cite{Picano2015} show that spherical particles increase the turbulence activity in the flow up to $\phi = 10\%$, an effect which overcomes the increase of the effective viscosity of the suspension. In fact, the increase in the effective viscosity can result in turbulence attenuation, however, the cases in \cite{Picano2015} show an enhancement in the presence of spherical particles. 
For oblate particles, instead, we observe a clear turbulence attenuation and an associated overall drag reduction. This raises the question of whether the increased viscosity of the suspension is the only reason for the turbulence attenuation or the specific particle shape also contribute to dampen  the turbulence. To answer this question, we perform  simulations of single-phase flow at bulk Reynolds number equal to the effective bulk Reynolds number $Re^e_b=2hU_b/\nu_e$, calculated for the corresponding particle volume fraction. 
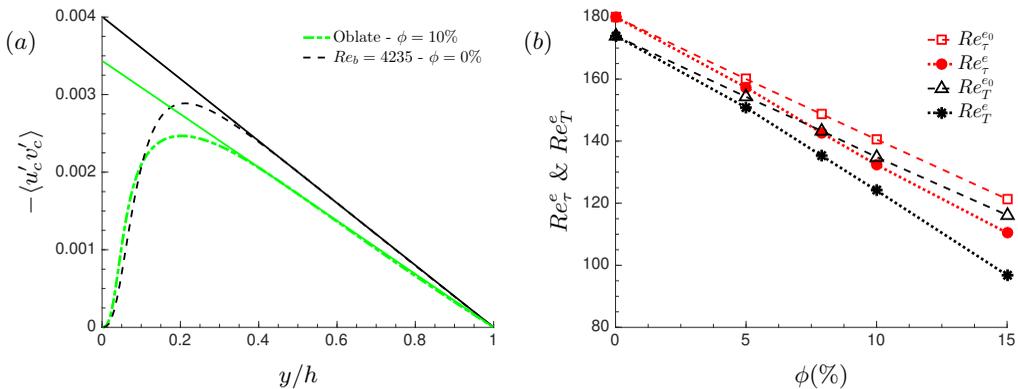
\begin{figure}
   \centering
   \includegraphics[width=0.495\textwidth]{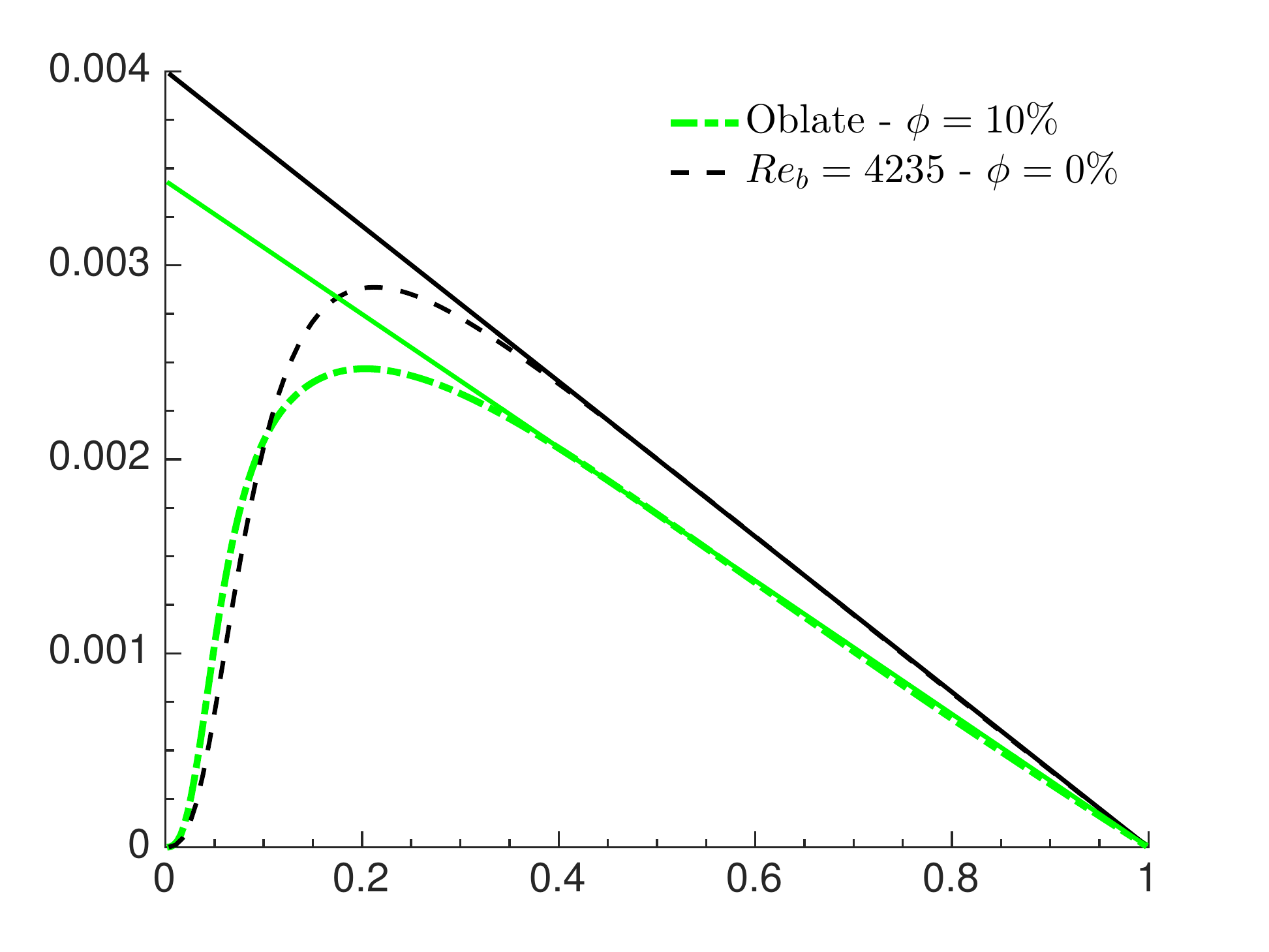}
   \includegraphics[width=0.495\textwidth]{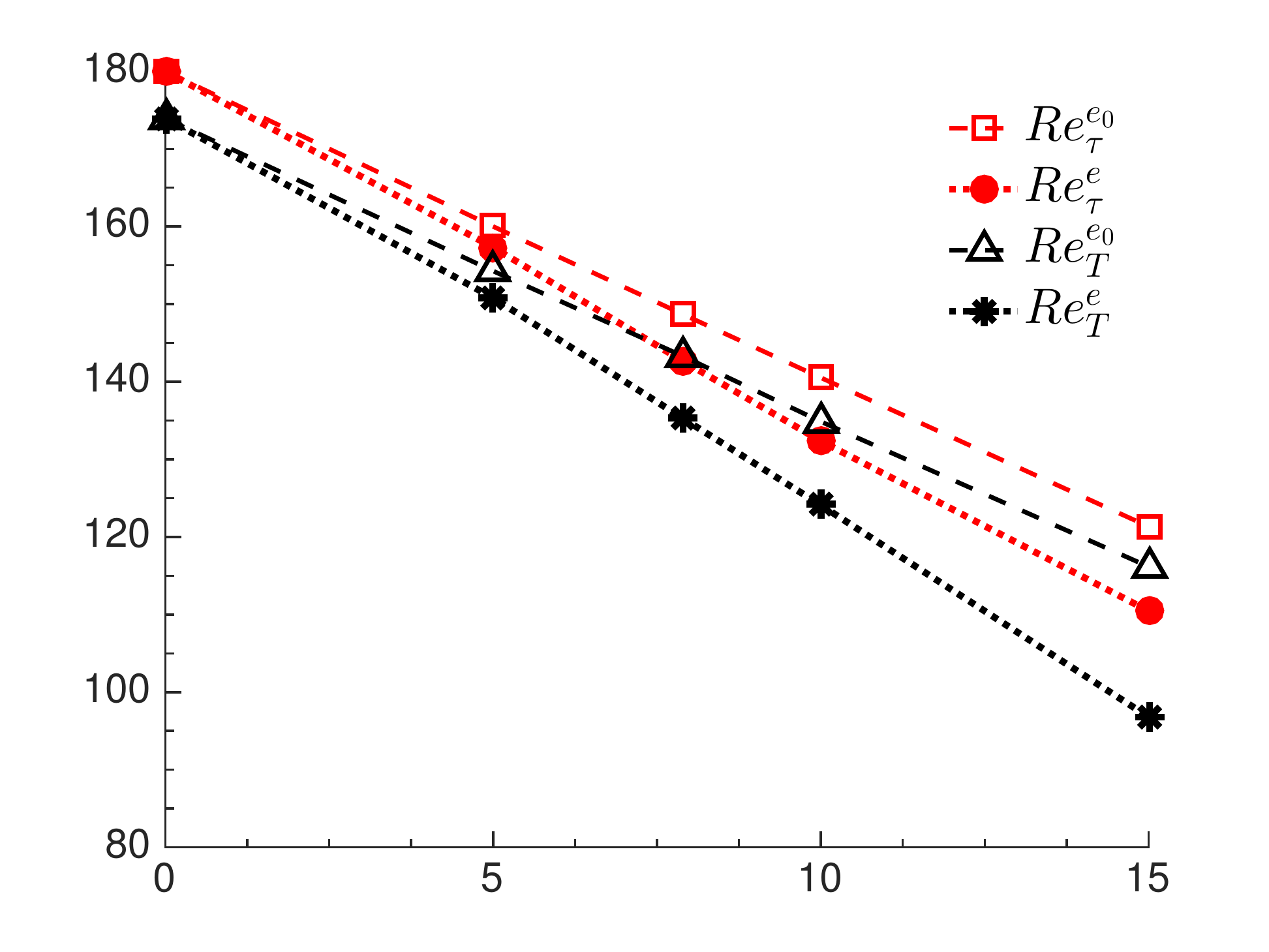}   
 \put(-390,58){\rotatebox{90}{$-\langle u^\prime_c v^\prime_c \rangle$}}
 \put(-190,51){\rotatebox{90}{$Re^e_\tau$ \& $Re^e_T$}}
 \put(-98,-5){{$\phi (\%)$}}
 \put(-292,-5){{$y / h$}}
   \put(-395,120){\footnotesize $(a)$}
   \put(-200,120){\footnotesize $(b)$}
  \vspace{3pt}        
  \caption{$(a)$ Reynolds stress of the combined phase, scaled in outer units for oblate particles at $\phi = 10\%$, compared to a single-phase flow simulation with $\nu = \nu_e |{{}_{\phi = 10\%}} $. The solid lines show the linear fitting of the slope of the profiles at the centreline ($y/h =1$). $(b)$ The effective turbulent friction Reynolds number $Re^e_T$ and the effective friction Reynolds number $Re^e_\tau$ versus the volume fraction $\phi$. These value are compared to the effective turbulent friction Reynolds number $Re^{e_0}_T$ and the effective friction Reynolds number  $Re^{e_0}_\tau$, obtained from single-phase simulations of a fluid with  viscosity equal to the effective viscosity of the suspension.}
 \label{fig:ReStEffective}
\end{figure}

\begin{table}
  \begin{center}
\def~{\hphantom{0}}
  \begin{tabular}{lccccc}
   & &  \multicolumn{4}{c}{Oblate - $\mathcal{AR} = 1/3$}  \\
   \cline{3-6} \\
       \,\,\,\,\,\,\,\,\,  \,Case & $\,\,\,\, \phi=0\% \,\,\,\,\,\,\,\,\,\,$  & $\phi=5\% \,\,$ & $\phi=7.9\% \,\,$  & $\phi=10\% \,\,$  & $\phi=15\%$   \\[3pt]
$ \,\,\,\,\,\,\,\,\,\,\,\,\, \,\,\, \nu_r$  & $\,\,\,\, 1 \,\,\,\,\,\,\,\,\,\,$ & $1.14 \,\,$ & $1.24 \,\,$ & $1.32 \,\,$ & $1.56$  \\[3pt]
$ \,\,\,\,\,\,\,\,\,\,\,\, \,  Re^e_\tau$  & $ \,\,\,\, 180 \,\,\,\,\,\,\,\,\,\,$ & $157.3 \,\,$ & $142.7 \,\,$ & $132.4 \,\,$ & $110.5$  \\[3pt]
$ \,\,\,\,\,\,\,\,\,\,\,\, \,  Re^e_T$  & $\,\,\,\, 174 \,\,\,\,\,\,\,\,\,\,$ & $150.8 \,\,$ & $135.3 \,\,$ & $124.1 \,\,$ & $96.8$ \\[3pt]
$ \,\,\,\,\,\,\,\,\,\,\,\, \, Re^e_b$  & $\,\,\,\, 5600 \,\,\,\,\,\,\,\,\,\,$ & $4908 \,\,$ & $4515 \,\,$ & $4235 \,\,$ & $3584$ \\[3pt]
$ \,\,\,\,\,\,\,\,\,\,\,\, \,  Re^{e_0}_\tau$  & $\,\,\,\, 180 \,\,\,\,\,\,\,\,\,\,$ & $160 \,\,$ & $148.7 \,\,$ & $140.5 \,\,$ & $121.4$ \\[3pt]
$ \,\,\,\,\,\,\,\,\,\,\,\, \,  Re^{e_0}_T$  & $\,\,\,\, 174 \,\,\,\,\,\,\,\,\,\,$ & $154.1 \,\,$ & $143.6 \,\,$ & $134.6 \,\,$ & $115.7$ \\[3pt]
  \end{tabular}
  \caption{The ratio between effective suspension viscosity and the fluid viscosity $\nu_r$, the effective friction Reynolds number $Re^e_\tau \equiv hU_* / \nu_e$, the effective turbulent friction Reynolds number $Re^e_T \equiv h U^T_* / \nu_e$, the effective suspension bulk Reynolds number $Re^e_b \equiv 2h U_b / \nu_e$ and the effective turbulent friction Reynolds number and effective friction Reynolds number $Re^{e_0}_T$ and $Re^{e_0}_\tau$, obtained from single phase simulations by only accounting for the effective suspension viscosity, for all cases with oblate particles.}
 \label{tab:TurbulenceReynolds}
 \end{center}
\end{table}

Figure~\ref{fig:ReStEffective}$(a)$ shows the Reynolds stress of the combined phase $\langle u^\prime_c v^\prime_c \rangle = \Phi \langle u^\prime_p v^\prime_p \rangle + (1-\Phi) \langle u^\prime_f v^\prime_f \rangle$ for the case with oblate particles at $\phi = 10\%$, compared to a single phase flow simulation with $\nu = \nu_e |{{}_{\phi = 10\%}} $. The data reveal that in the presence of oblate particles, the turbulence activity is even lower than that predicted by the simulation at $Re^e_b$, indicating that the effective viscosity is not the only responsible for the turbulence attenuation shown above.

As in  \cite{Picano2015}, to quantify the level of turbulence activity, we define the turbulent friction velocity $U^T_* = \sqrt{\mathrm{d} \langle u^\prime_c v^\prime_c \rangle \,/\, \mathrm{d}  \left(y/h\right)} |_{y/h = 1} $, calculated as the square root of the wall-normal derivative of the Reynolds stress profile at the centreline of the channel. The slope of the Reynolds stress profile at the centreline is shown in figure~\ref{fig:ReStEffective}$(a)$. Note that the turbulent friction velocity approximates the wall friction velocity for single phase flows at high bulk Reynolds number, the error scaling with $1/Re_b$ \citep{Pope2001}. This estimate accounts only for the effect of turbulence and not for the viscous or the particle-induced stress and therefore helps to quantify the role of the turbulence activity on the overall drag.

The effective turbulent friction Reynolds number, defined as $Re^e_T = U^T_* h / \nu_e$, is reported in table~\ref{tab:TurbulenceReynolds} for all cases under consideration together with the turbulent friction Reynolds number obtained from the single phase simulations $Re^{e_0}_T$ (with $\nu = \nu_e$). 
In addition to these two, we also depict in figure ~\ref{fig:ReStEffective}$(b)$ the friction Reynolds number based on the effective viscosity, $Re^e_\tau= Re_\tau / \nu_r$ and $Re^{e_0}_\tau$, the friction Reynolds number obtained from the single-phase simulations. These are shown for increasing volume fraction $\phi$ and oblate particles. 
We conclude from the figure that the oblate particles reduce the turbulence activity to lower values than those obtained by only accounting for the effective suspension viscosity, resulting in an overall drag reduction. The observed drag reduction is therefore related to the specific dynamics of the oblate particles and their interactions with the turbulent velocity field. 


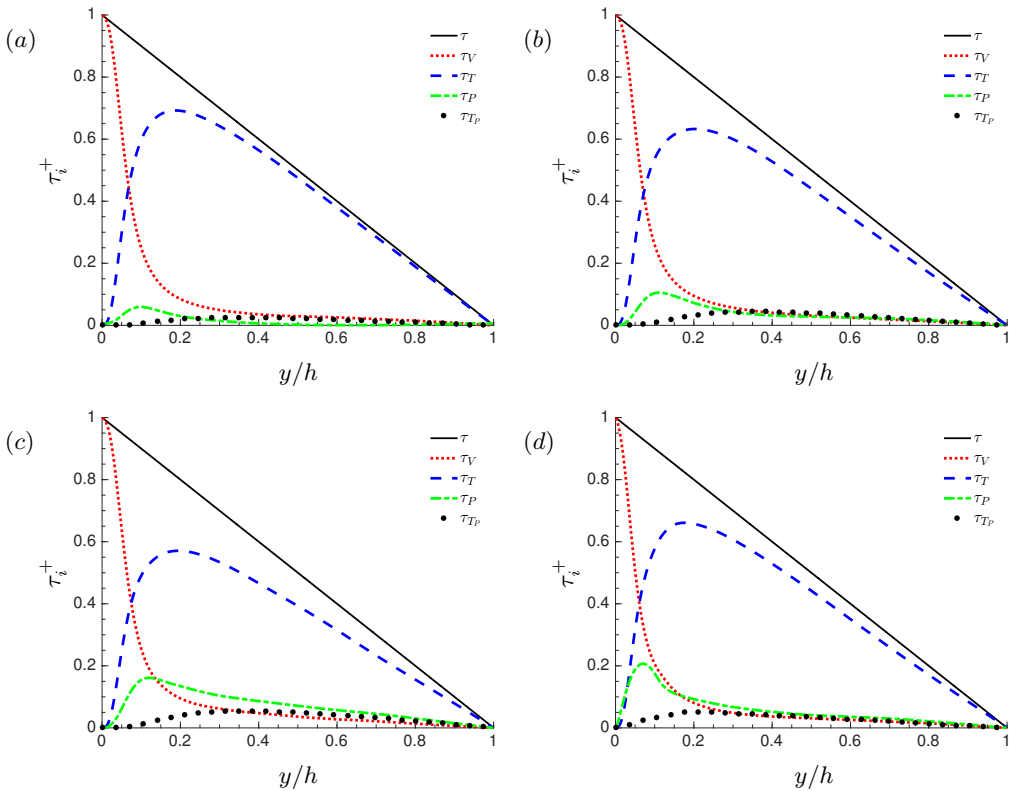
\begin{figure}
   \centering
   \includegraphics[width=0.495\textwidth]{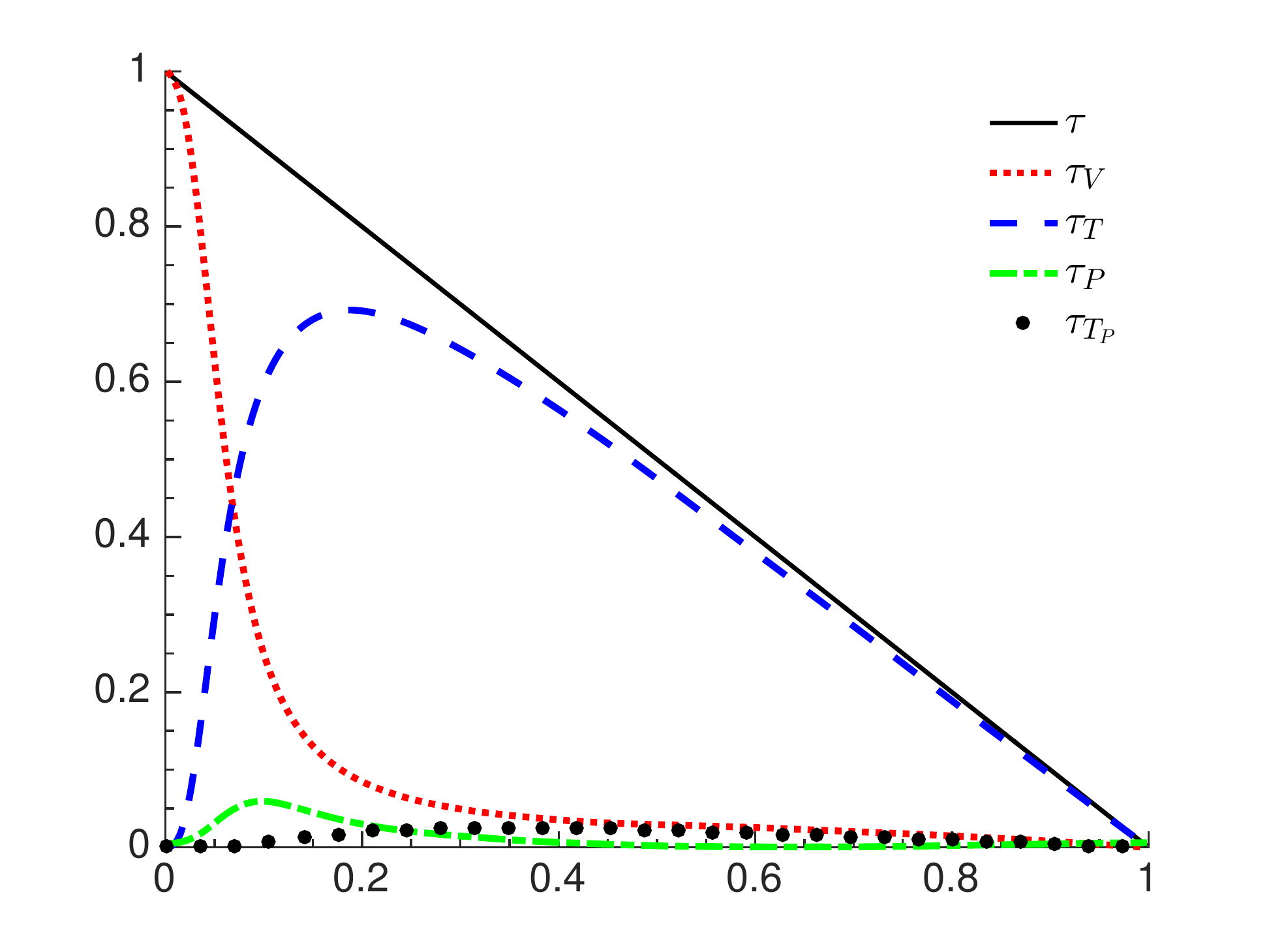}
   \includegraphics[width=0.495\textwidth]{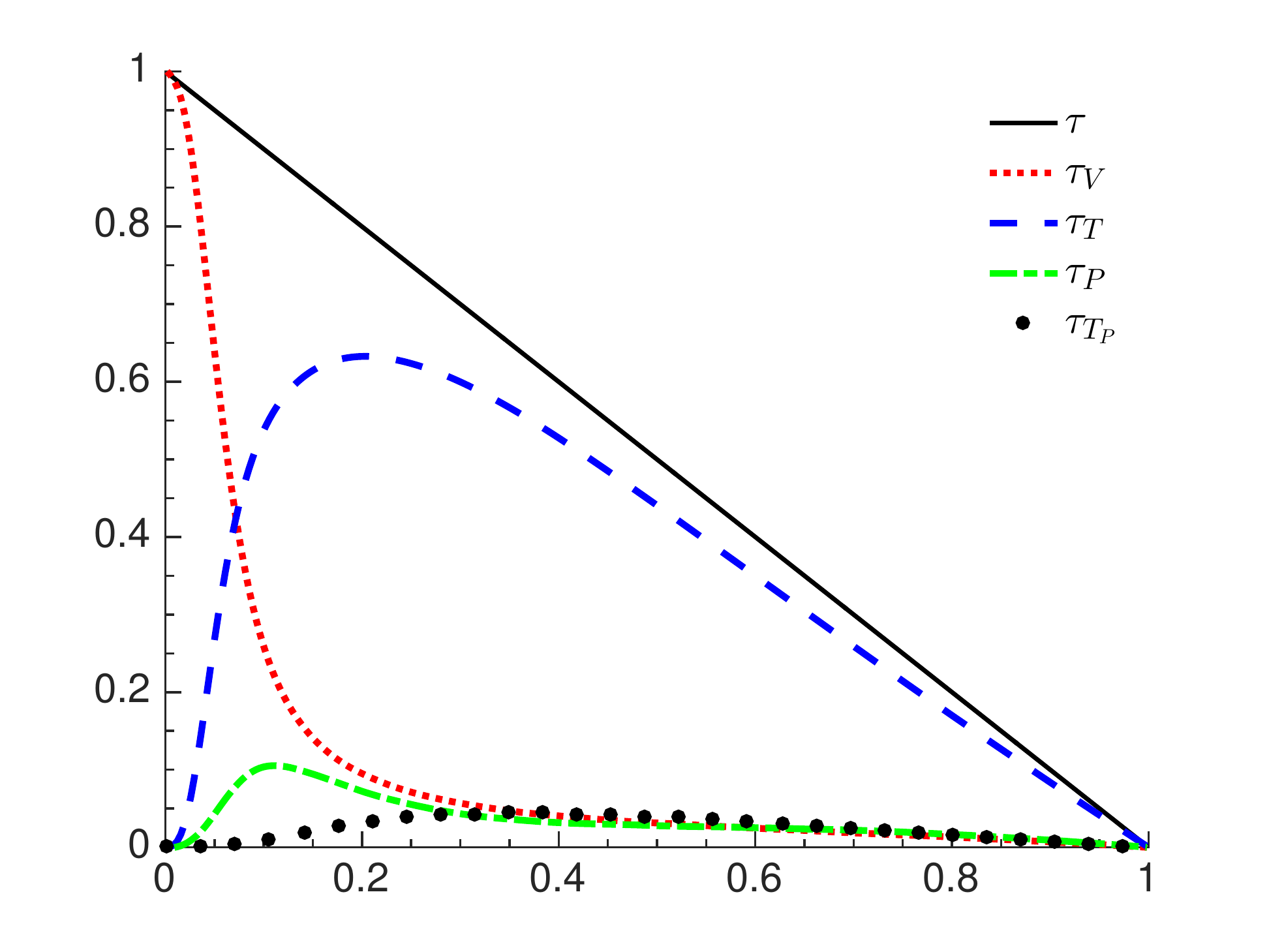} 
   \put(-292,-5){{$y / h$}}   
   \put(-98,-5){{$y / h$}} 
   \put(-382,68){\rotatebox{90}{$\tau^+_i$}}
   \put(-189,68){\rotatebox{90}{$\tau^+_i$}}           
   \put(-395,120){\footnotesize $(a)$}
   \put(-200,120){\footnotesize $(b)$} \\  
   \includegraphics[width=0.495\textwidth]{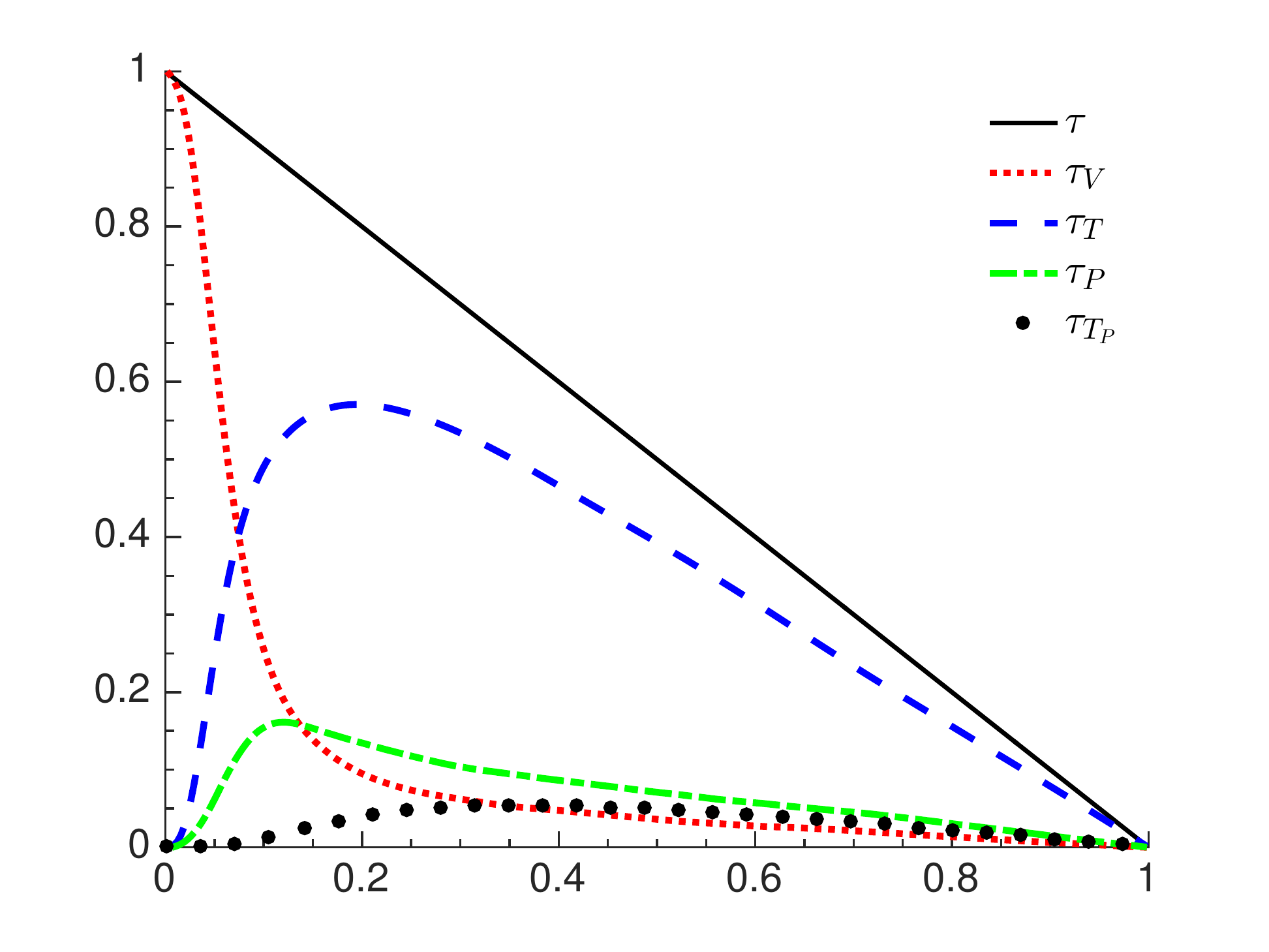}
   \includegraphics[width=0.495\textwidth]{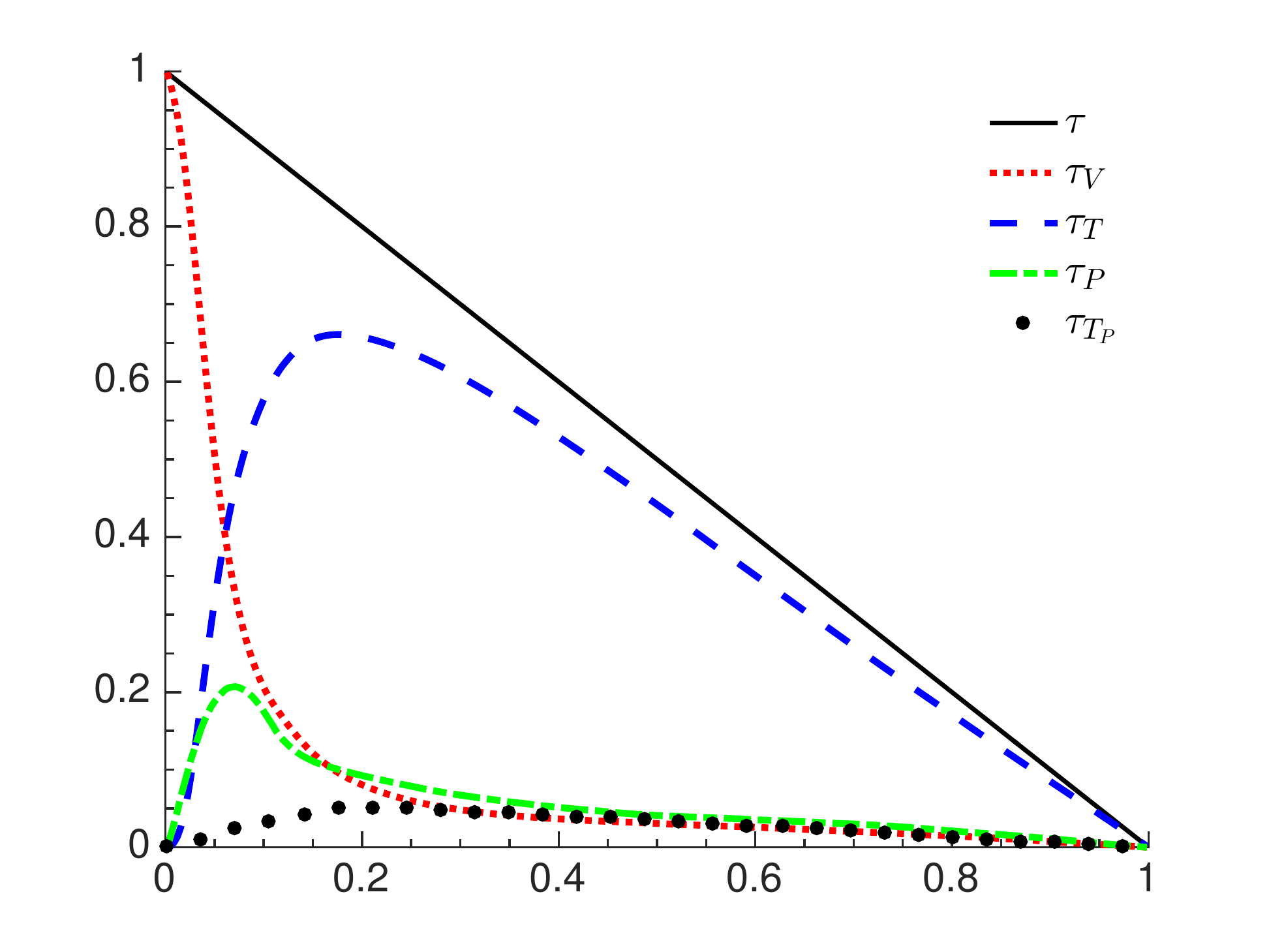}
   \put(-395,120){\footnotesize $(c)$}
   \put(-200,120){\footnotesize $(d)$}
   \put(-292,-5){{$y / h$}}   
   \put(-98,-5){{$y / h$}}
   \put(-382,68){\rotatebox{90}{$\tau^+_i$}}
   \put(-189,68){\rotatebox{90}{$\tau^+_i$}}  \\ [5pt]           
  \caption{Momentum budget, normalized with $\rho_f U^2_*$, for oblate particles at $(a)$  $\phi = 5\%$, $(b)$ $\phi = 10\%$, $(c)$ $\phi = 15\%$ and $(d)$ spheres at $\phi = 10\%$. Here, $\tau$ is the total stress, $\tau_V$ denotes viscous stress, $\tau_T = - \langle u^\prime_c v^\prime_c \rangle$ the turbulent Reynolds shear stress of the combined phase and $\tau_P$ the particles induced stress. The particles Reynolds shear stress $\tau_{T_p} = - \langle u^\prime_p v^\prime_p \rangle$ is also depicted with dots. }
\label{fig:StressShare}
\end{figure}

The results presented so far indicate that the turbulence attenuation and the absence of a particle layer close to the wall (documented in the next section) are responsible for the drag reduction in the flow laden with oblate particles. \cite{Picano2015} showed that for spherical particles at $\phi = 20\%$, the turbulence activity reduces while the total drag still increases. This is attributed to the increase of the particle-induced stress at high volume fractions.  To better understand the effect of particle-induced stress for the cases with oblate particles, we perform here a stress budget analysis similar to that in \cite{Lashgari2014} and \cite{Picano2015}.  

Based on the formulation proposed in \cite{Zhang2010}, we can write the mean momentum balance in the channel as
 \begin{equation}
\label{eq:Fl}  
\tau/\rho_f \, = \, U^2_* \left( 1 - \frac{y}{h} \right ) \, =  \, \nu (1-\Phi) \frac{\mathrm{d} U_f}{\mathrm{d} y} \, - \, \left[ \, \Phi \langle u^\prime_p v^\prime_p \rangle \, + \, (1-\Phi) \langle u^\prime_f v^\prime_f \rangle \right] \, +  \, \frac{\Phi}{\rho}  \langle \sigma^p_{xy} \rangle \, \, \, , \\ [10pt]
\end{equation}
where $\tau$ is the total stress, see Appendix in \cite{Picano2015} for a derivation. The first term on the right hand side of the budget above is the viscous shear stress $\tau_V$, the second and the third term are together the turbulent Reynolds shear stress of the combined phase $\tau_T = - \langle u^\prime_c v^\prime_c \rangle$ and the fourth term indicates the particle-induced stress $\tau_P$. The momentum transfer pertaining each term, normalized by $\rho_f U^2_*$, is depicted versus $y/h$ in figure~\ref{fig:StressShare} for oblate particles at $\phi = 5\%$, $10\%$ and $15\%$ and for spheres at $\phi = 10\%$. 
The contribution due to the particle-induced stress is always lower than that from the Reynolds stress for oblate particles. 
As the volume fraction increases, the relative momentum transfer due to the Reynolds stress decreases, yet the particle-induced stress does not increase enough to compensate for the reduction in turbulence activity.  The particle-induced stress is considerably higher near the wall for spheres than for oblate particles, as expected by the absence of the particle layer close to the wall in the cases of oblates. 
The momentum budget analysis performed here shows that, unlike the case of spheres, the effect of the particle-induced stress on the total drag is small for oblate particles and the Reynolds shear stress is the main factor determining the overall drag.  This confirms that the specific dynamics of the oblate particles is closely related to the turbulence attenuation and therefor, drag reduction.

\subsection{Particle dynamics}

\begin{figure}
 \centering
 \includegraphics[width=0.495\textwidth]{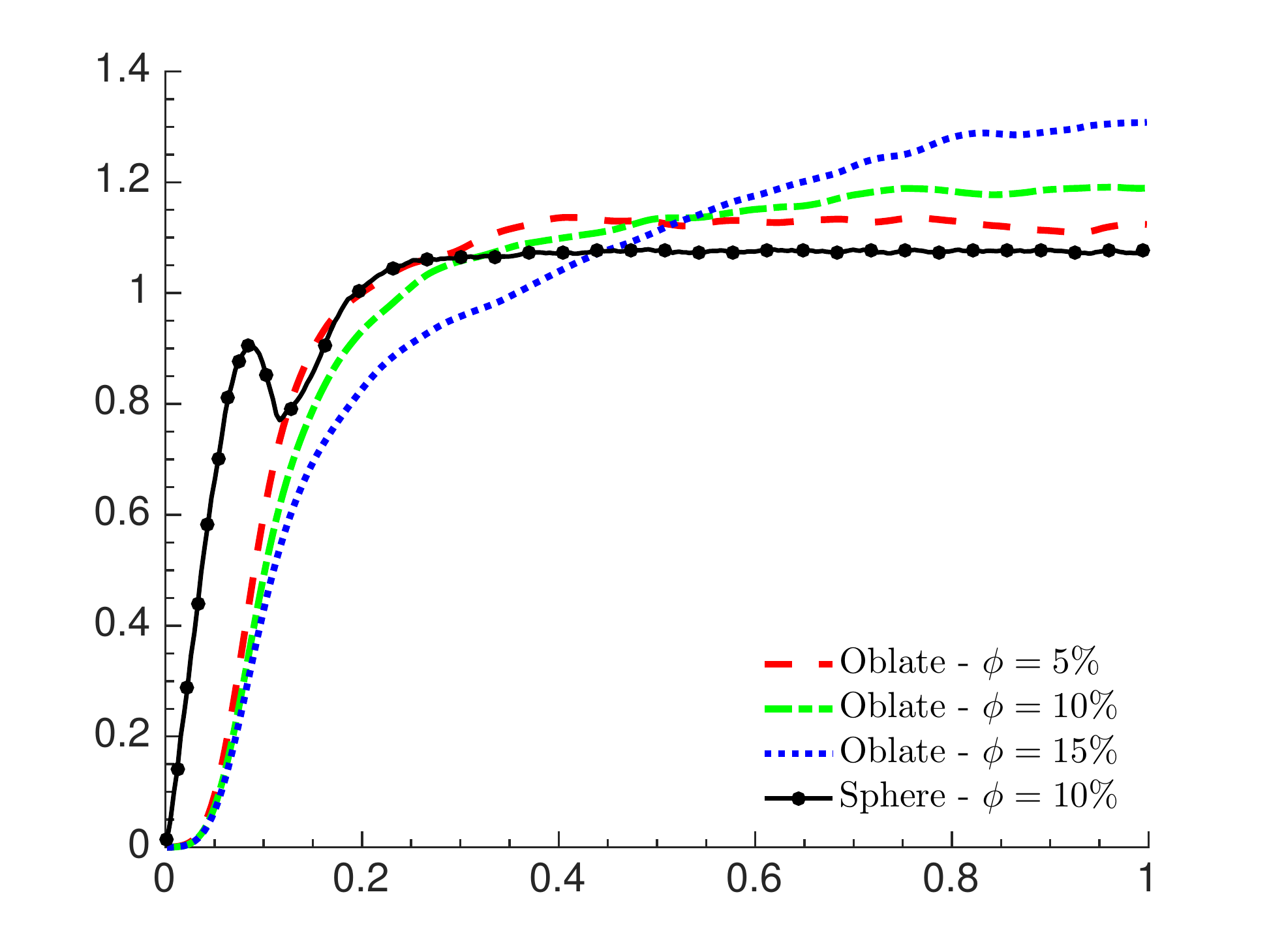}
 \includegraphics[width=0.495\textwidth]{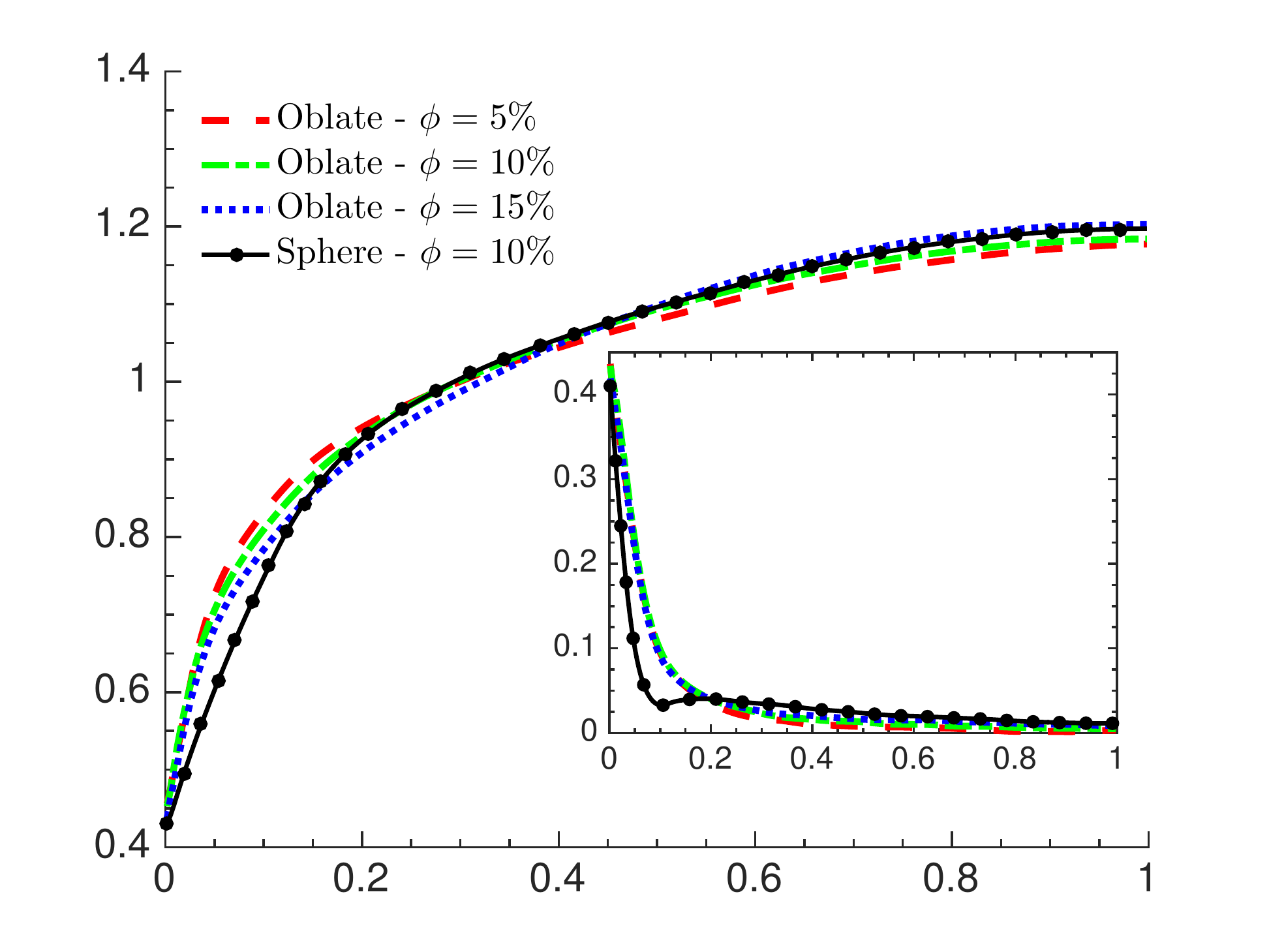} 
 \put(-382,60){\rotatebox{90}{$ \Phi (y) / \phi$}}
 \put(-190,60){\rotatebox{90}{$U_p / U_b$}}
 \put(-98,-5){{$y / h$}}
 \put(-292,-5){{$y / h$}}
 \put(-114.5,35){\rotatebox{90}{ \tiny $(U_p - U_f) \, / \, U_b$}}
 \put(-67,22){\tiny $y / h$} 
   \put(-395,120){\footnotesize $(a)$}
   \put(-200,120){\footnotesize $(b)$}
  \vspace{3pt}     
 \caption{Profiles of solid-phase averaged data versus $y / h$: $(a)$ mean local volume fraction $\Phi (y)$, normalized by total volume fraction $\phi$ and $(b)$ mean particle velocity profiles, normalized by the fluid bulk velocity $U_b$. The inset in $(b)$ shows the difference between the particle and the fluid mean velocity. }
 \label{fig:concentration}
\end{figure}

 The mean local volume fraction $\Phi (y)$, normalized by the bulk volume fraction $\phi$, is depicted in figure~\ref{fig:concentration}$(a)$.  Spherical particles display a local maximum at a distance slightly larger than 1 particle radius from the wall. \cite{Picano2015} attribute this local maximum to the formation of a particle layer at the wall, due to the wall-particle interactions that stabilise the particle position. \cite{Costa2016} explain that the presence of this particle layer always results in drag increase, which is therefore higher than what can be predicted by only accounting for the effective suspension viscosity.  
Interestingly, the particle layer at the wall is not present in the flow with oblate particles. The local volume fraction, $\Phi$, is considerably lower in the region close to the wall ($y/h < 0.1$) for oblates than for spherical particles.  A minor migration towards the channel center is observed for oblate particles in turbulent flow and this effect is more pronounced when increasing the volume fraction.  Interestingly, the oblate distribution is uniform in laminar flow, while spherical particles tend to migrate towards the channel centre, see figure~\ref{fig:phi_Lam} in appendix~\ref{app:Laminar_Simulations}. 

Figure~\ref{fig:concentration}$(b)$ shows the mean particle velocity profiles. The difference between the particle and the fluid mean velocity, the local slip, is depicted in the inset of the figure. 
As the velocity of the particles is not zero at the wall, significant slip velocity is found close to the wall. Interestingly, a local minimum of the velocity difference is observed for spheres at the location of the particle layer close to the wall. Spherical particles, trapped in the particle layer, experience a smaller slip velocity due to the higher local volume fraction at this location, thus creating the mentioned local minimum in the mean velocity difference. This particle layer disappears for the case with oblates and the difference in the mean velocity decays monotonically with decreasing wall-normal distance.

\begin{figure}
   \centering
   \includegraphics[width=0.495\textwidth]{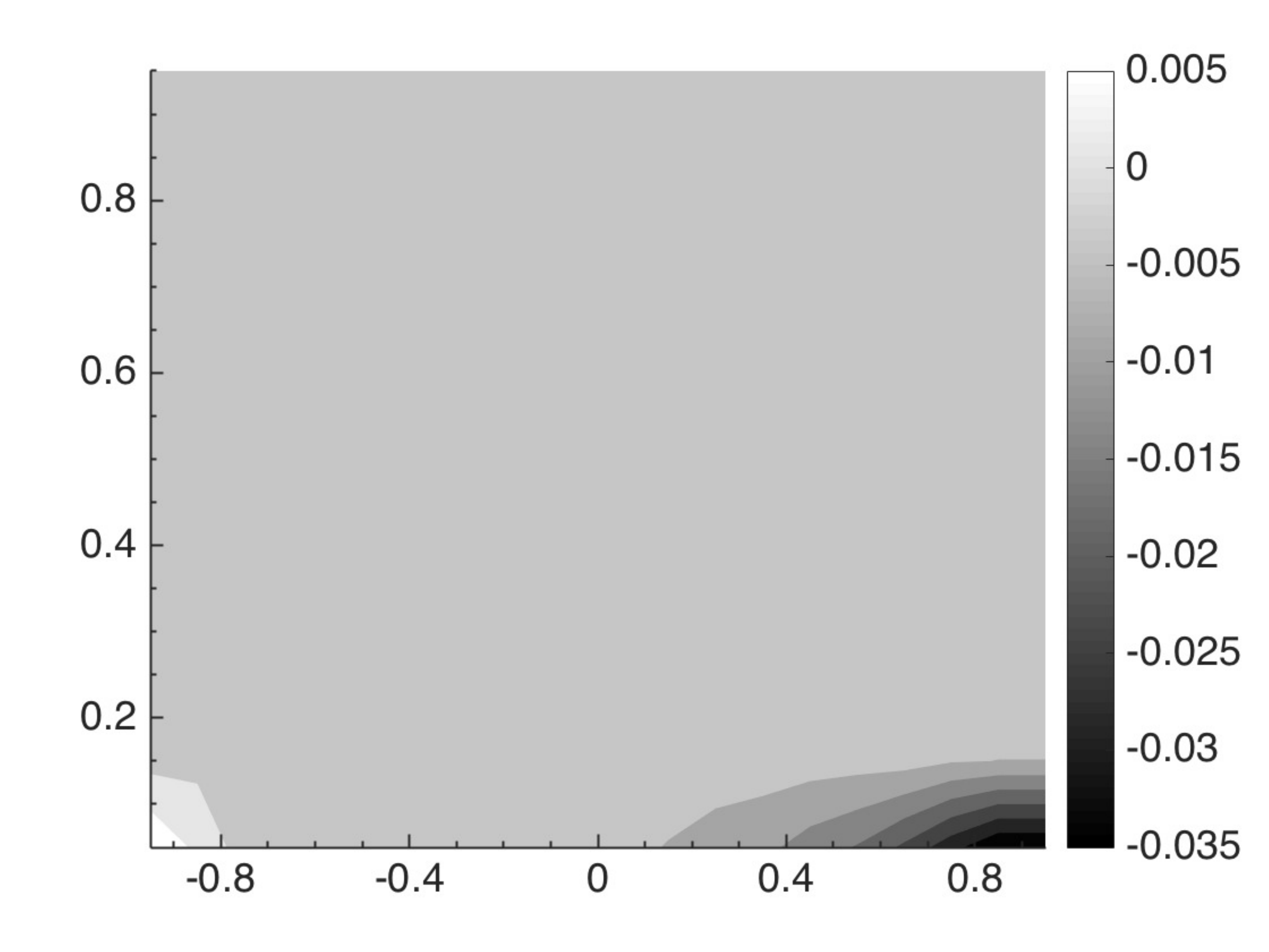}
   \includegraphics[width=0.495\textwidth]{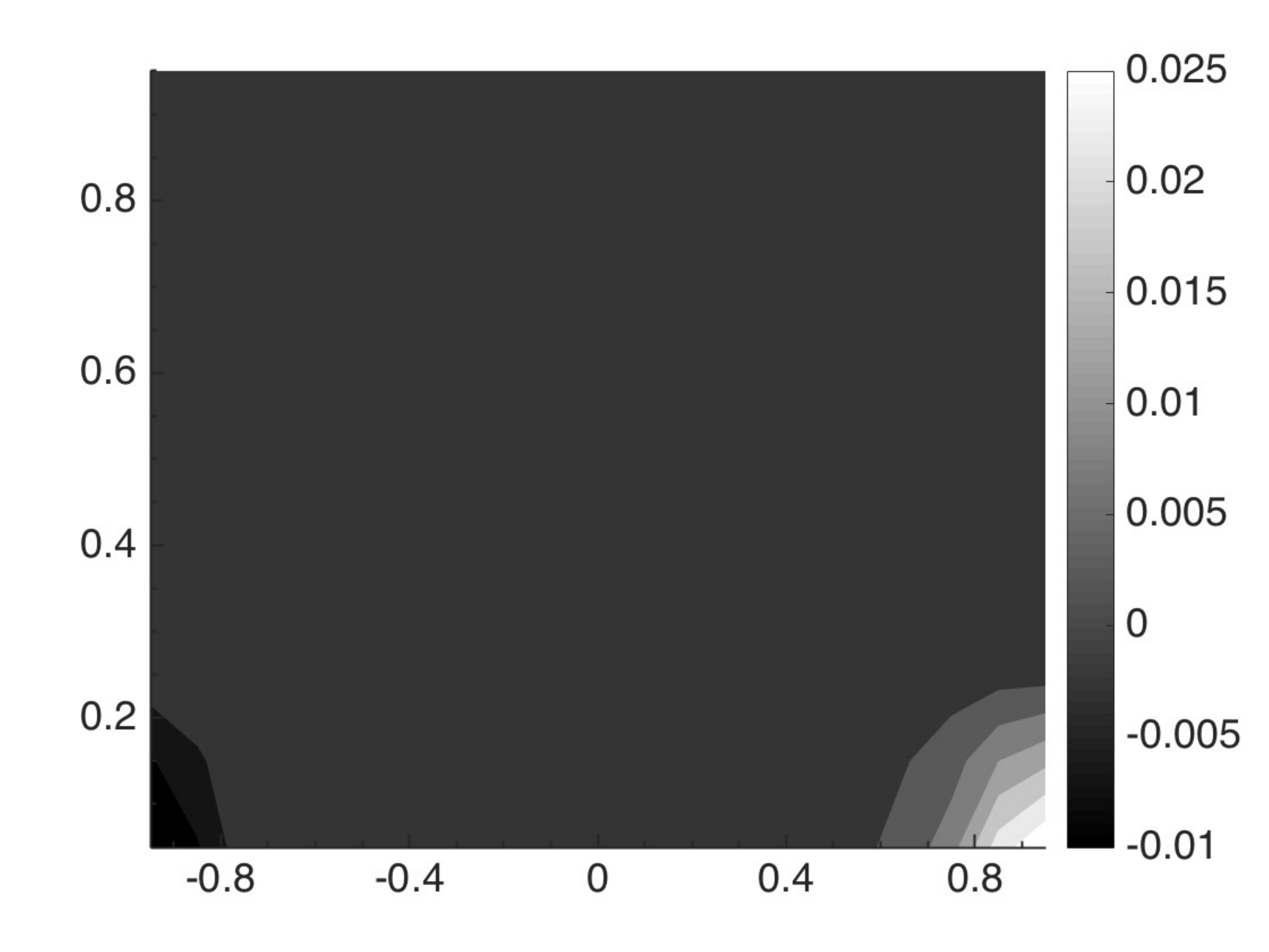} 
   \put(-298,-2){{$o^x_p$}}   
   \put(-104,-2){{$o^x_p$}} 
   \put(-382,66){\rotatebox{90}{$y / h$}}
   \put(-189,66){\rotatebox{90}{$y / h$}}           
   \put(-390,120){\footnotesize $(a)$}
   \put(-193,120){\footnotesize $(b)$} \\  
   \includegraphics[width=0.495\textwidth]{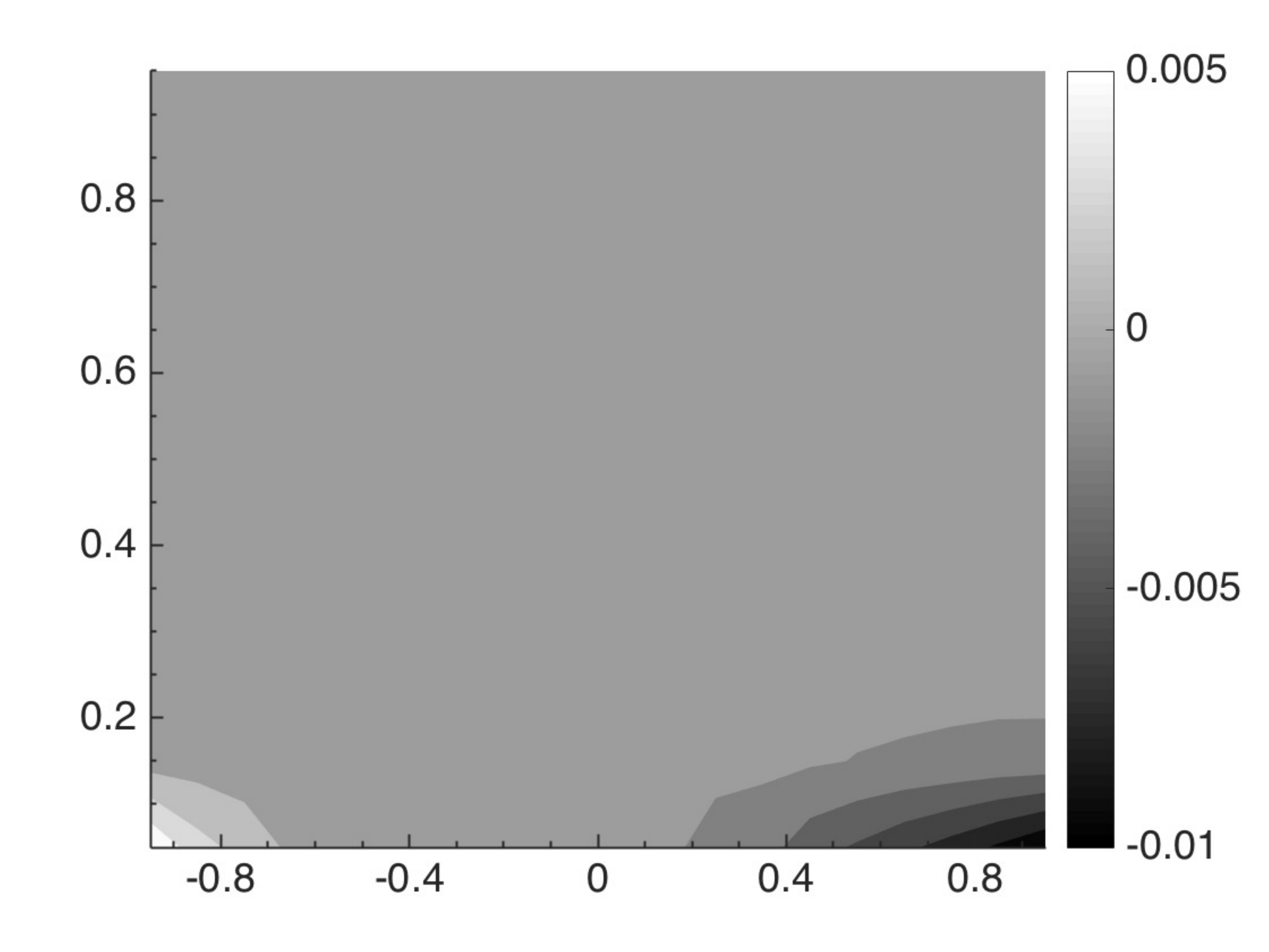}
   \includegraphics[width=0.495\textwidth]{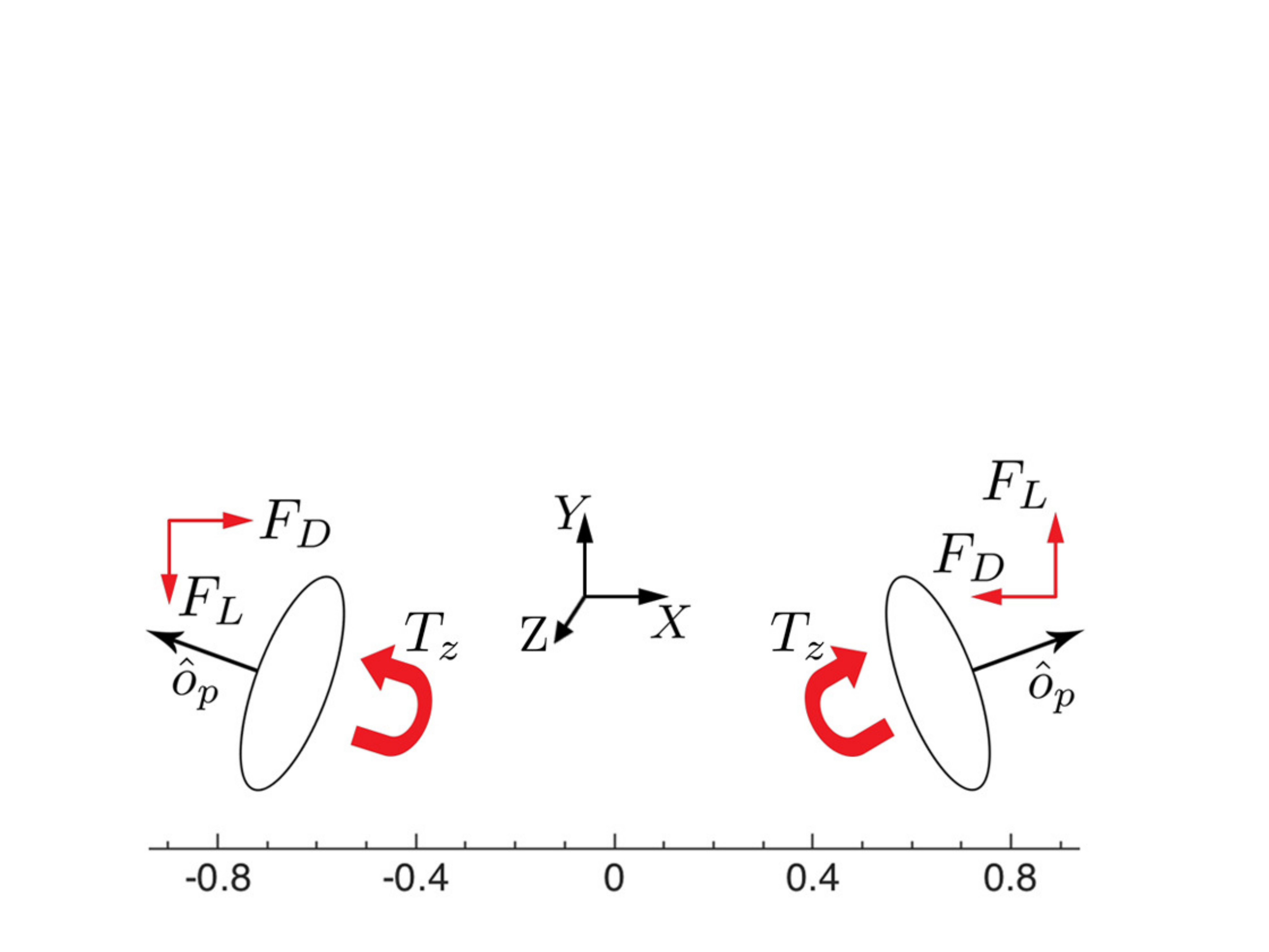}
   \put(-390,120){\footnotesize $(c)$}
   \put(-193,120){\footnotesize $(d)$}
   \put(-298,-2){{$o^x_p$}}   
   \put(-104,-2){{$o^x_p$}} 
   \put(-382,66){\rotatebox{90}{$y / h$}}  \\ [5pt]           
  \caption{Contours of the mean forces and torques acting on the particles as a function of distance to the wall, $y/h$, and the streamwise component of the particle orientation vector $\hat{o}^x_p$ : $(a)$ Drag force $F_D$  and $(b)$ Lift force $F_L$  in units of $\rho_f U^2_b D^2_{eq}$; $(c)$ Spanwise torque $T_z$ in units of $\rho_f U^2_b D^3_{eq}$. A schematic of the action of the forces and torques is given in in panel $(d)$ for the cases of positive and negative $\hat{o}^x_p$.}
\label{fig:ForceTorque}
\end{figure}

\begin{figure}
   \centering
   \includegraphics[width=0.495\textwidth]{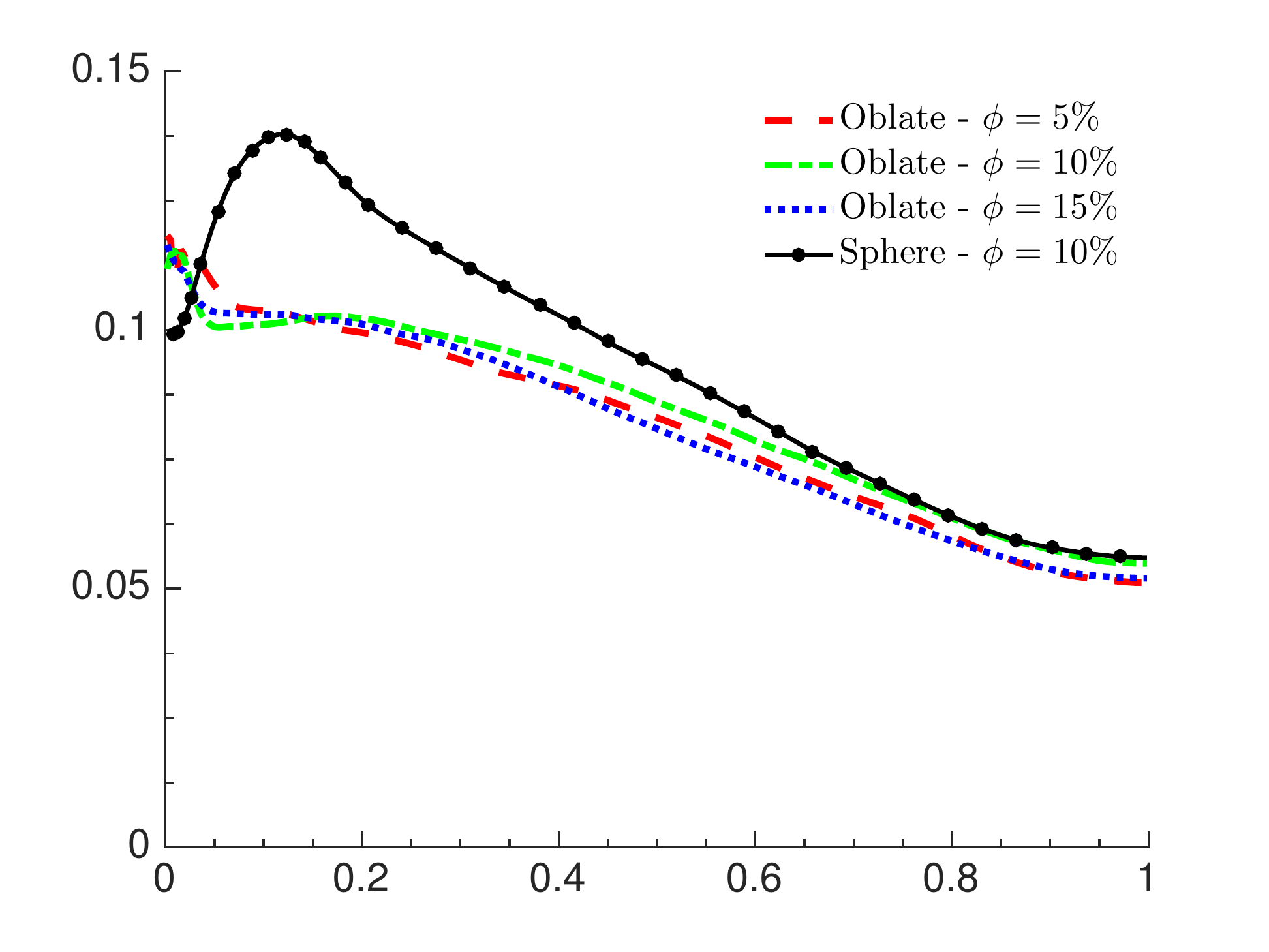}
   \includegraphics[width=0.495\textwidth]{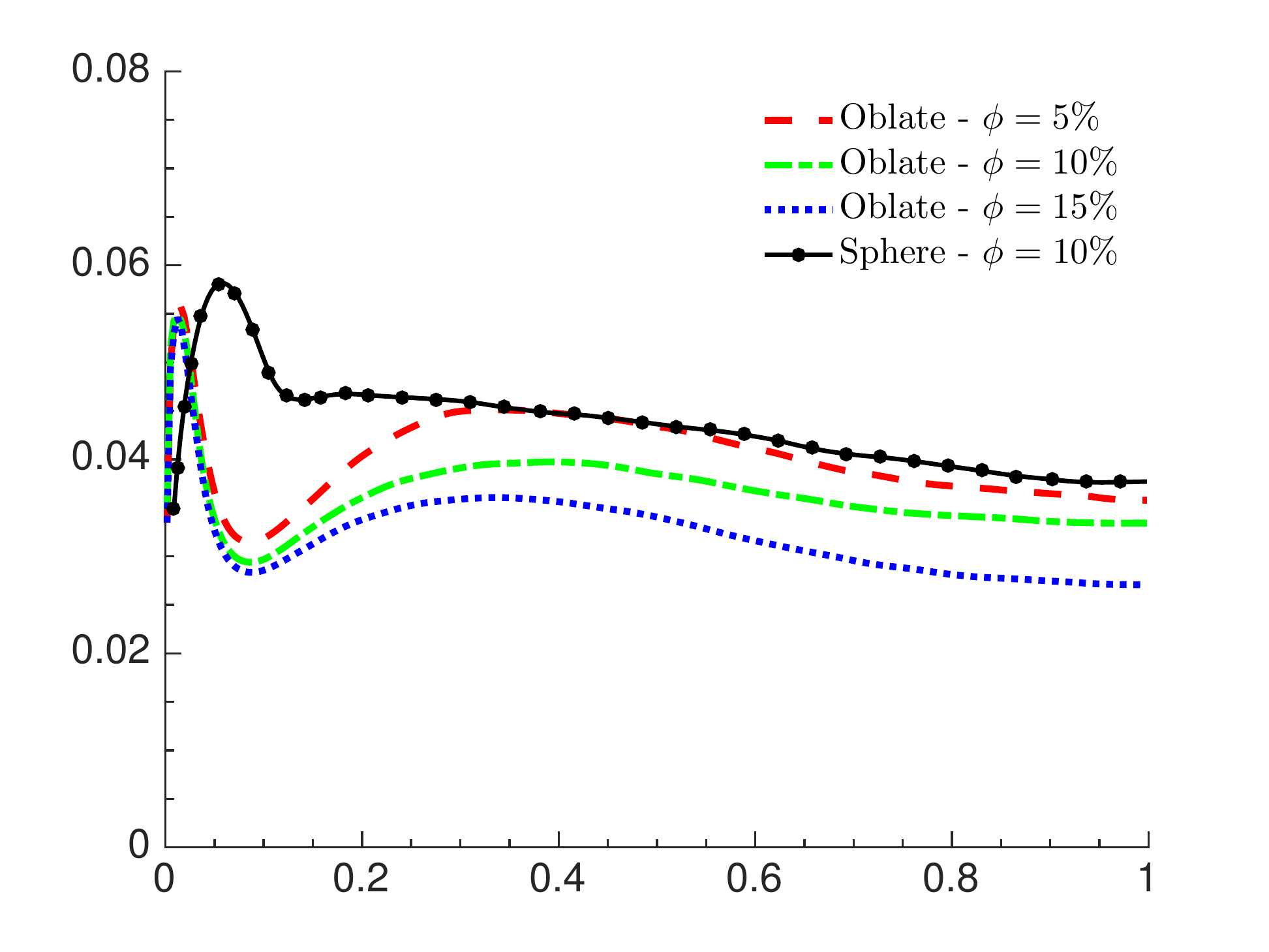} 
   \put(-292,-5){{$y / h$}}   
   \put(-98,-5){{$y / h$}} 
   \put(-388,62){\rotatebox{90}{${u^\prime_p}_{rms}$}}
   \put(-195,62){\rotatebox{90}{${v^\prime_p}_{rms}$}}           
     \put(-395,120){\footnotesize $(a)$}
   \put(-200,120){\footnotesize $(b)$} \\     
   \includegraphics[width=0.495\textwidth]{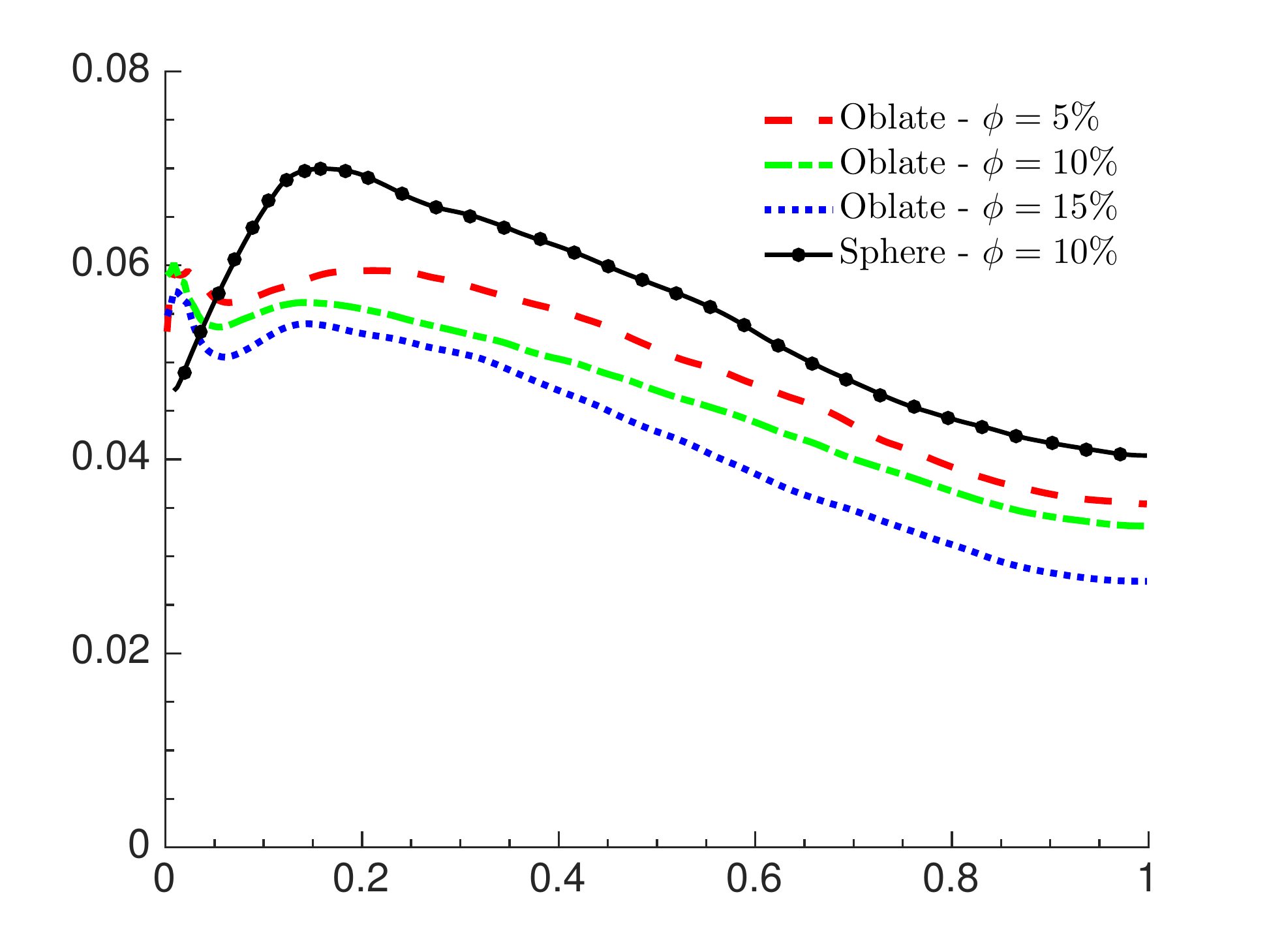}
   \includegraphics[width=0.495\textwidth]{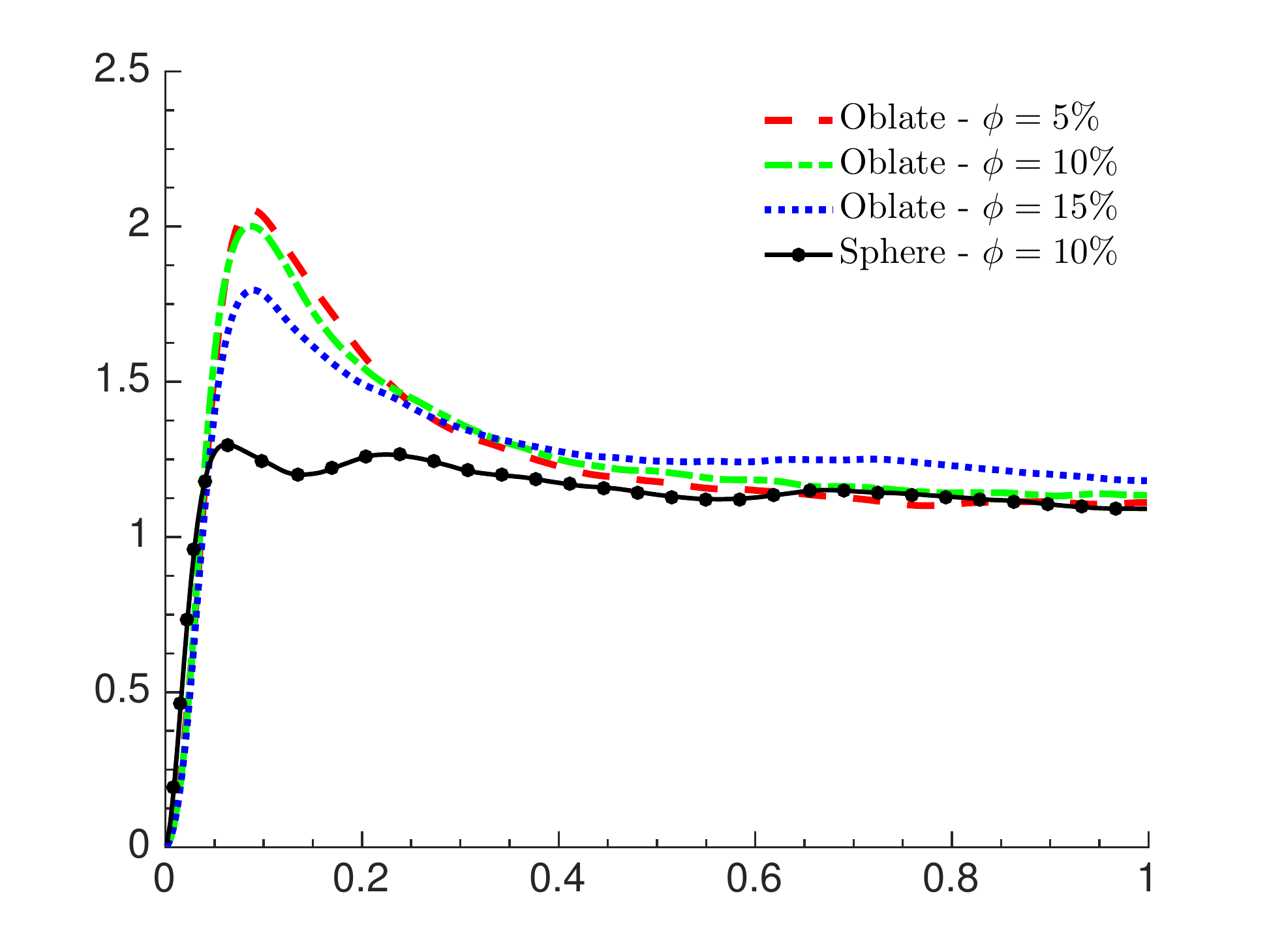}
      \put(-395,120){\footnotesize $(c)$}
   \put(-200,120){\footnotesize $(d)$}
   \put(-388,62){\rotatebox{90}{${w^\prime_p}_{rms}$}}
   \put(-195,59){\rotatebox{90}{$K_f / K_p$}}       
   \put(-292,-5){{$y / h$}}   
   \put(-98,-5){{$y / h$}} \\ [5pt]        
  \caption{Intensity of the solid-phase velocity fluctuation components, scaled in outer units (bulk velcoity $U_b$), for the different cases under consideration: $(a)$ streamwise ${u^\prime_f}_{rms}$; $(b)$ wall-normal ${v^\prime_f}_{rms}$; $(c)$ spanwise ${w^\prime_f}_{rms}$ component. Panel $(d)$  displays the ratio between the turbulent kinetic energy of the fluid and that of the solid phase $K_f / K_p$.}
\label{fig:ParticleRMS}
\end{figure}
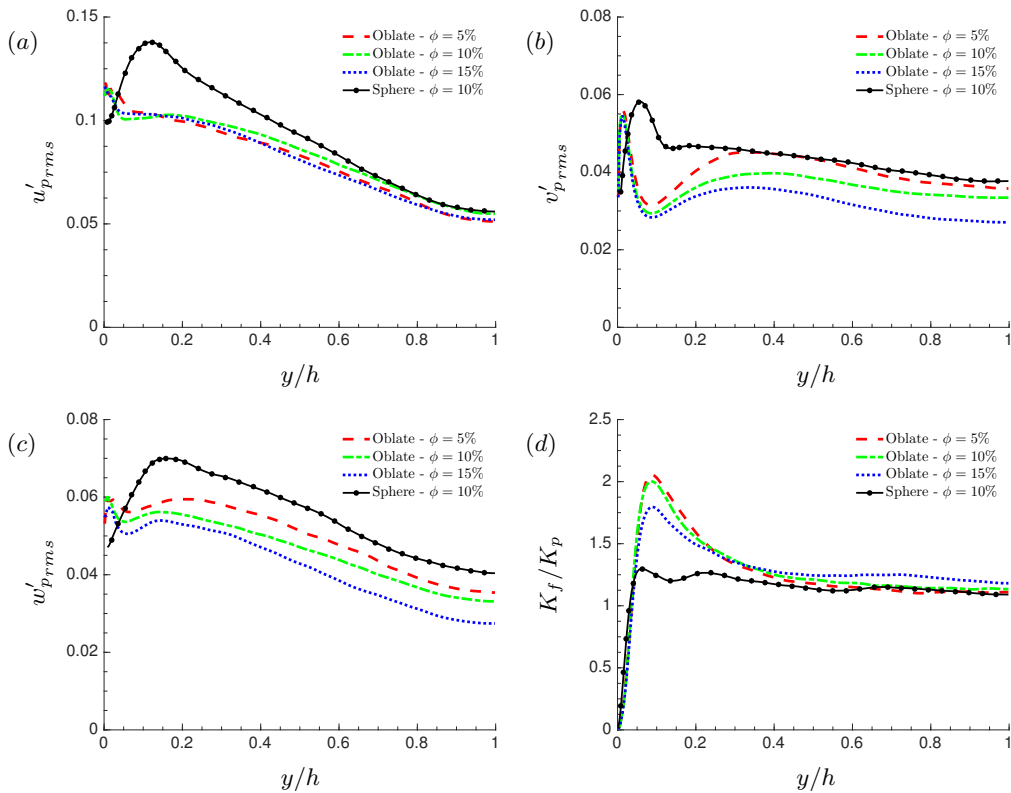

The reasons behind the absence of the particle layer for oblates are investigated here using a force and torque analysis on the case at $\phi=15\%$. We compute the mean drag force $F_D$, lift force $F_L$ and spanwise torque $T_z$ acting on the particles as a function of distance to the wall, $y/h$, and of the streamwise component of the particle orientation vector $\hat{o}^x_p$. The orientation vector $\hat{o}_p$ is defined as unit vector parallel to the particle symmetric axis and pointing towards the channel center.  Contours of the normalized  $F_D$, $F_L$ and $T_z$ are depicted in figure~\ref{fig:ForceTorque}$(a-c)$. 
The results show that when particles are sufficiently close to the wall (high mean velocity gradients), the forces and the torques acting on the particle change according to  their orientation as sketched in~\ref{fig:ForceTorque}$(d)$. Particles, on average, have a negative spanwise angular velocity due to the mean flow gradient and the analysis presented here, interestingly, indicates that when these have a positive $\hat{o}^x_p$, they are lifted by the flow towards the channel center whereas the opposite is true when they have a negative  $\hat{o}^x_p$. Note, however, that the magnitude of the upward lift is significantly higher than the downward force. This difference in the magnitude can explain the absence of particle layer for oblates. 
 It can also be concluded from figure~\ref{fig:ForceTorque}$(d)$ that particles near the wall  with a negative $\hat{o}^x_p$ tend to align parallel to the wall and accelerate towards it; particles with positive $\hat{o}^x_p$, conversely,
  accelerate towards the channel center. This indicates that oblate particles are most likely parallel to the wall in its vicinity. This stable configuration and its importance to drag reduction are addressed later in this section.

To gain further insight on 
the turbulence attenuation in the presence of oblates we investigate the particle collective behaviour. 
Figure~\ref{fig:ParticleRMS} shows the rms of the solid phase velocity fluctuations. The data clearly reveal that all the three components of the particle 
velocity fluctuations are considerably smaller than in the case of spherical particles, except for a very small region close to the wall, where the rare collisions with 
the wall have a large impact on the statistics. It should also be noted that, contrary to the case of spheres, oblate particles cannot roll on the wall and any collision with the wall can create a large particle velocity 
fluctuations. A maximum of the wall-normal particle velocity fluctuations $v^\prime_p$ very close to the wall and a local minimum at $y/h \approx 0.1$ are observed for oblate particles, whereas a local maximum and a weak local minimum at $y/h \approx 0.15$ can be seen for spheres. The ratio between the turbulent kinetic energy of the fluid and of the solid phase $K_f / K_p = ({u^\prime_f}^2 + {v^\prime_f}^2 + {w^\prime_f}^2) / ({u^\prime_p}^2 + {v^\prime_p}^2 + {w^\prime_p}^2)$, depicted in figure~\ref{fig:ParticleRMS}$(d)$,  shows that particle velocities tend to fluctuate 
less than the fluid at the same position except for the region close to the wall where velocities fluctuate more due to the absence of a no-slip condition. This tendency 
is considerably higher for oblate particles, as shown by the  peak of the velocity fluctuations close to the wall, $y \approx 0.1 $.

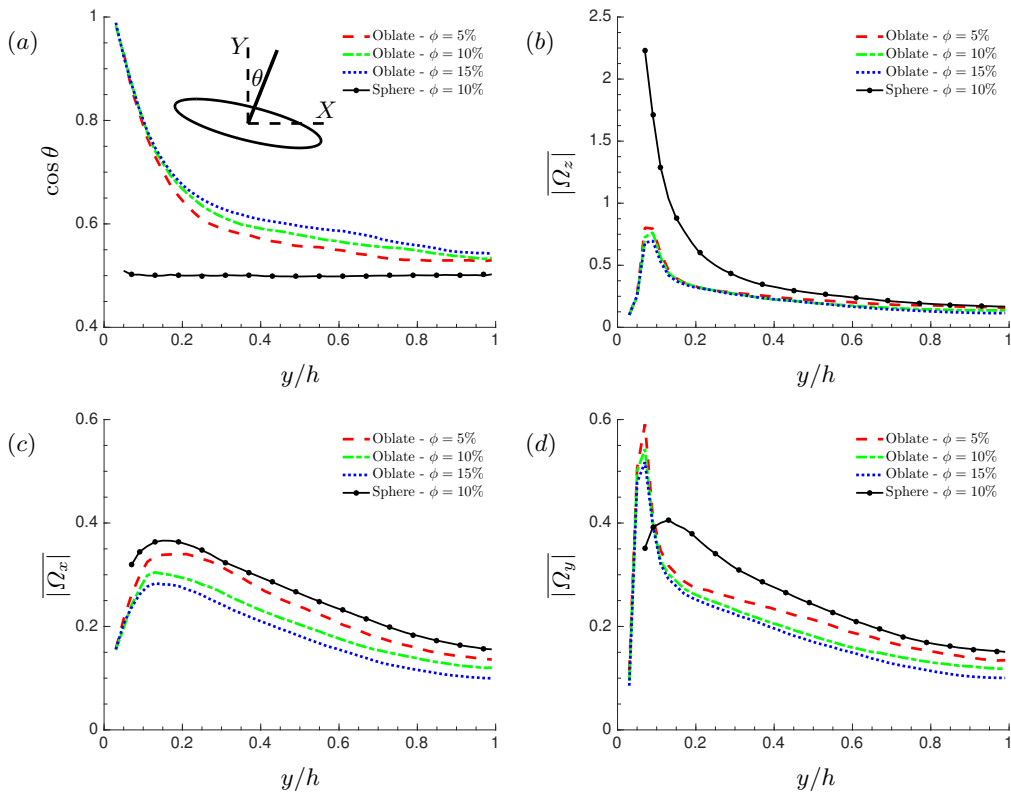
\begin{figure}
   \centering
   \includegraphics[width=0.495\textwidth]{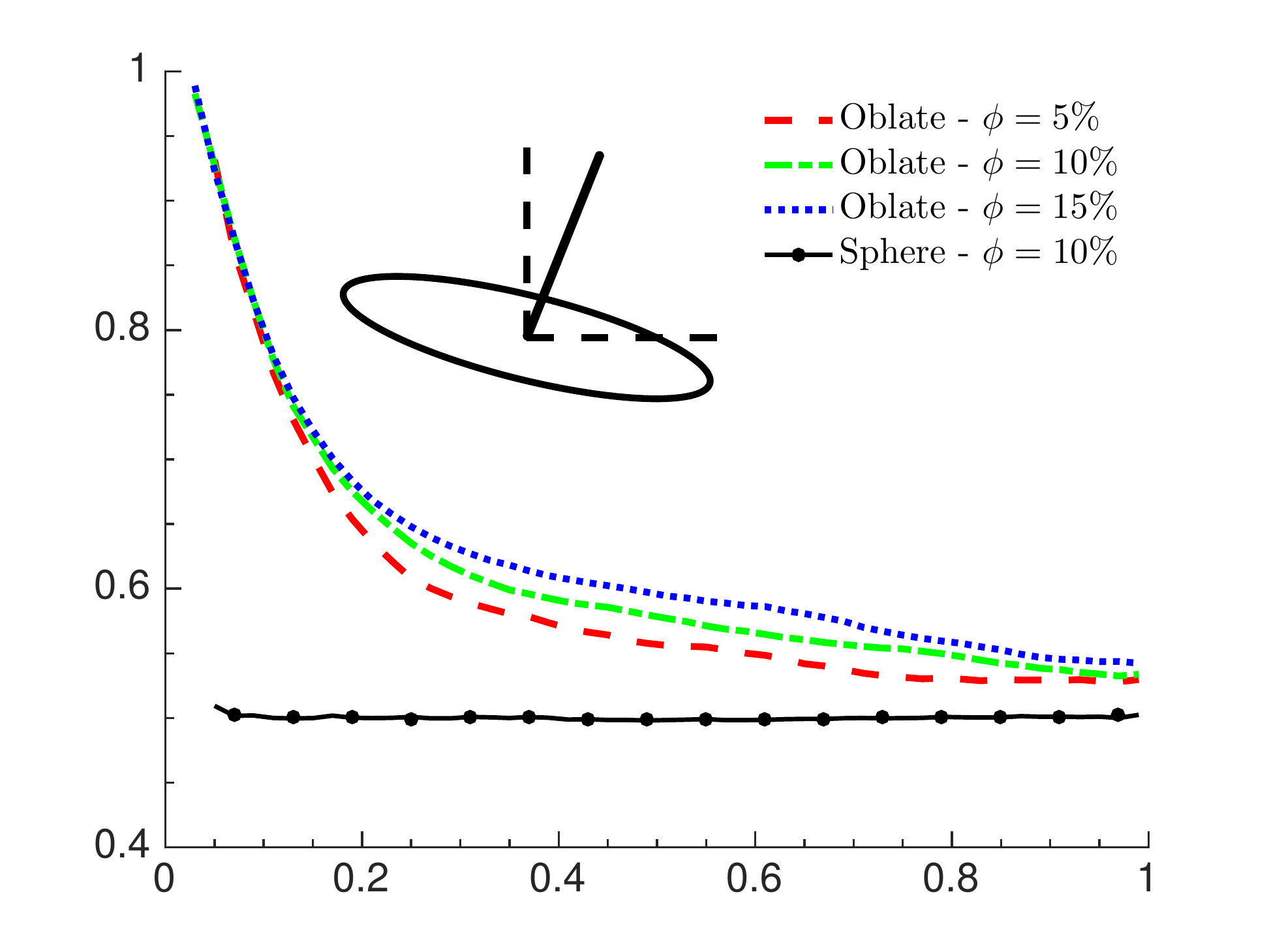}
   \includegraphics[width=0.495\textwidth]{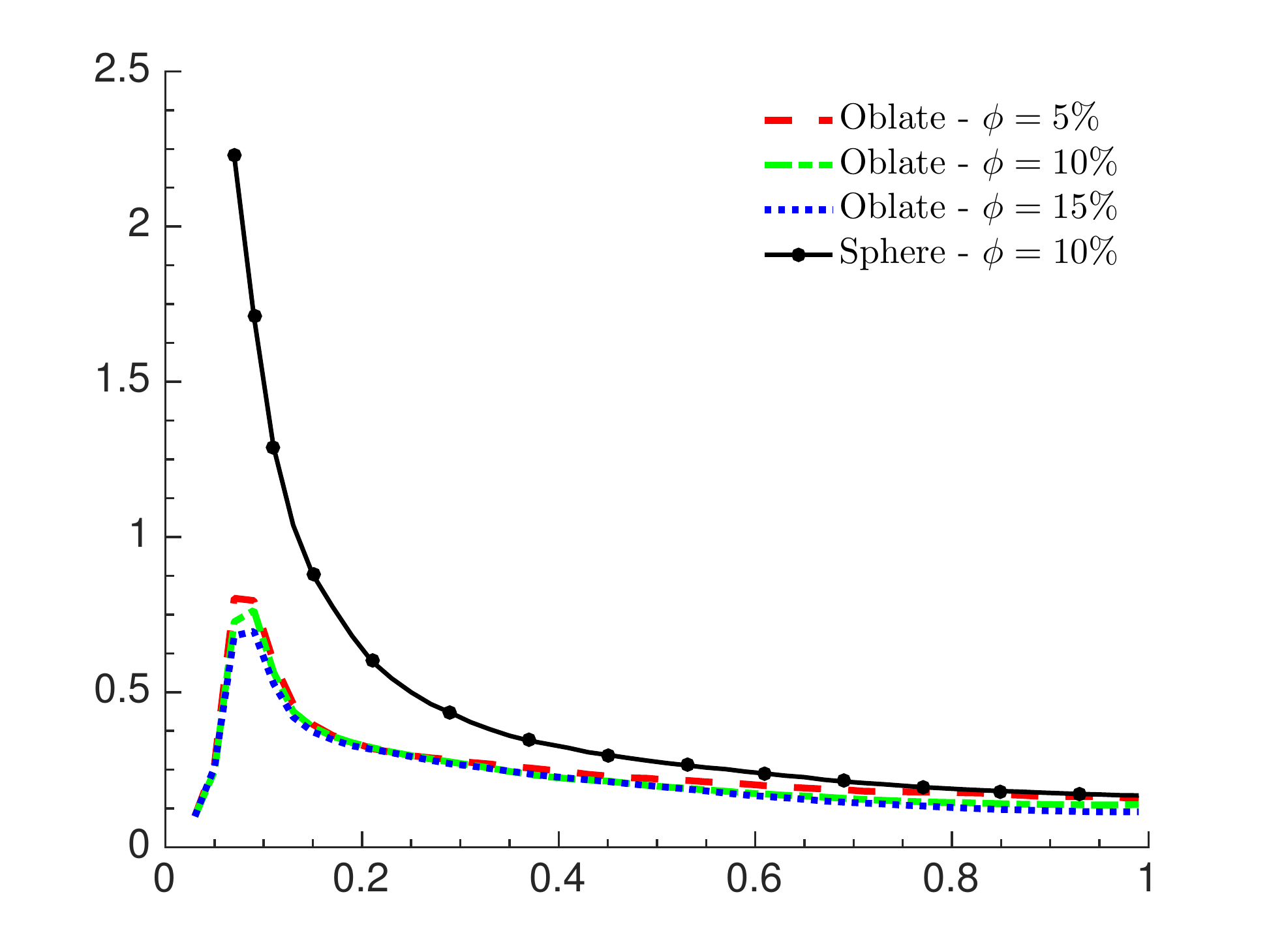}
   \put(-292,-5){{$y / h$}}   
   \put(-98,-5){{$y / h$}}  
   \put(-303,106){{$\theta$}}  
   \put(-311,118){{$Y$}} 
   \put(-279,94){{$X$}}          
   \put(-382,65){\rotatebox{90}{$\cos \theta$}}
   \put(-192,65){\rotatebox{90}{$\overline{|\Omega_z|}$}}        
   \put(-395,120){\footnotesize $(a)$}
   \put(-200,120){\footnotesize $(b)$} \\       
   \includegraphics[width=0.495\textwidth]{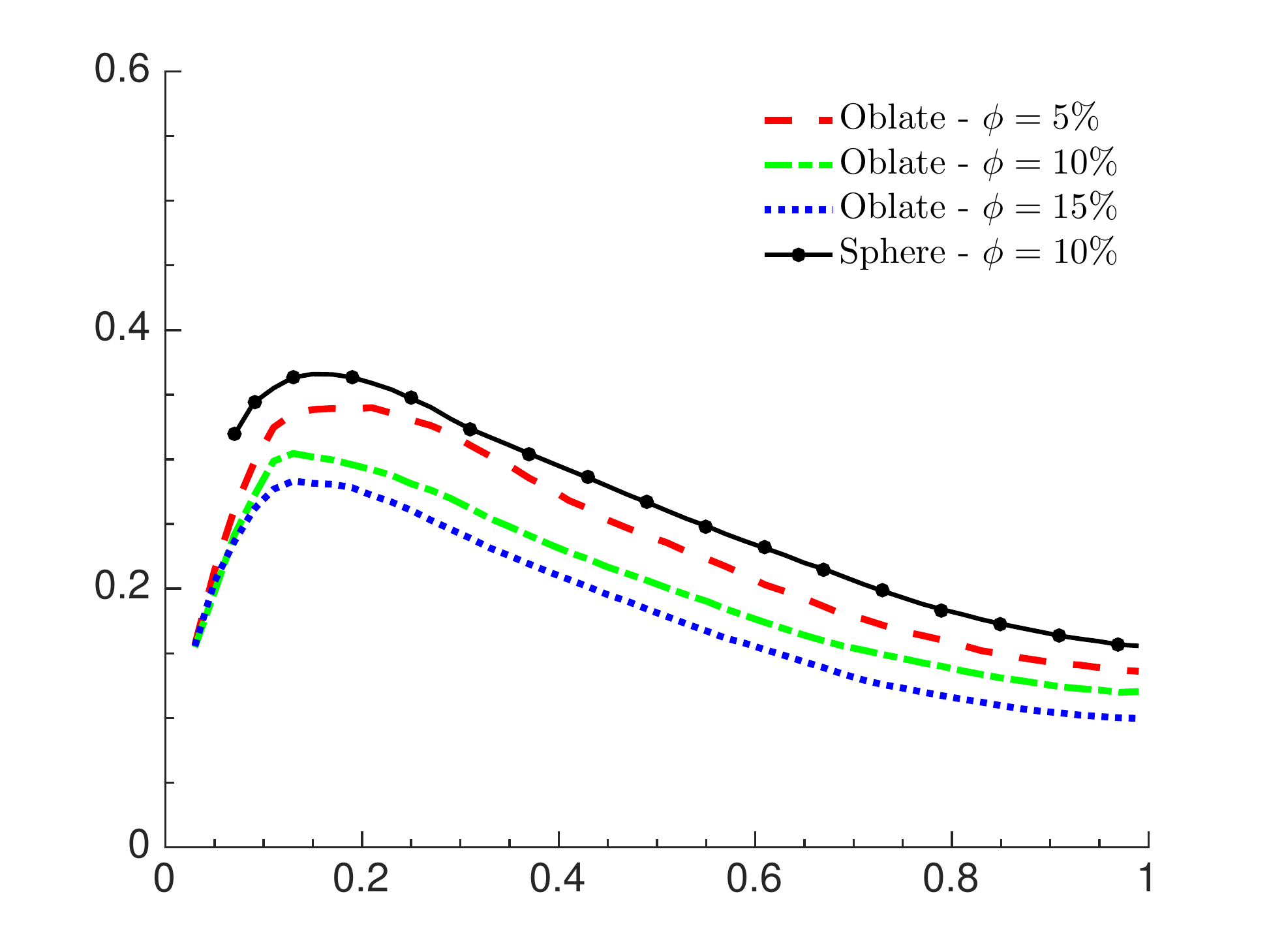}
   \includegraphics[width=0.495\textwidth]{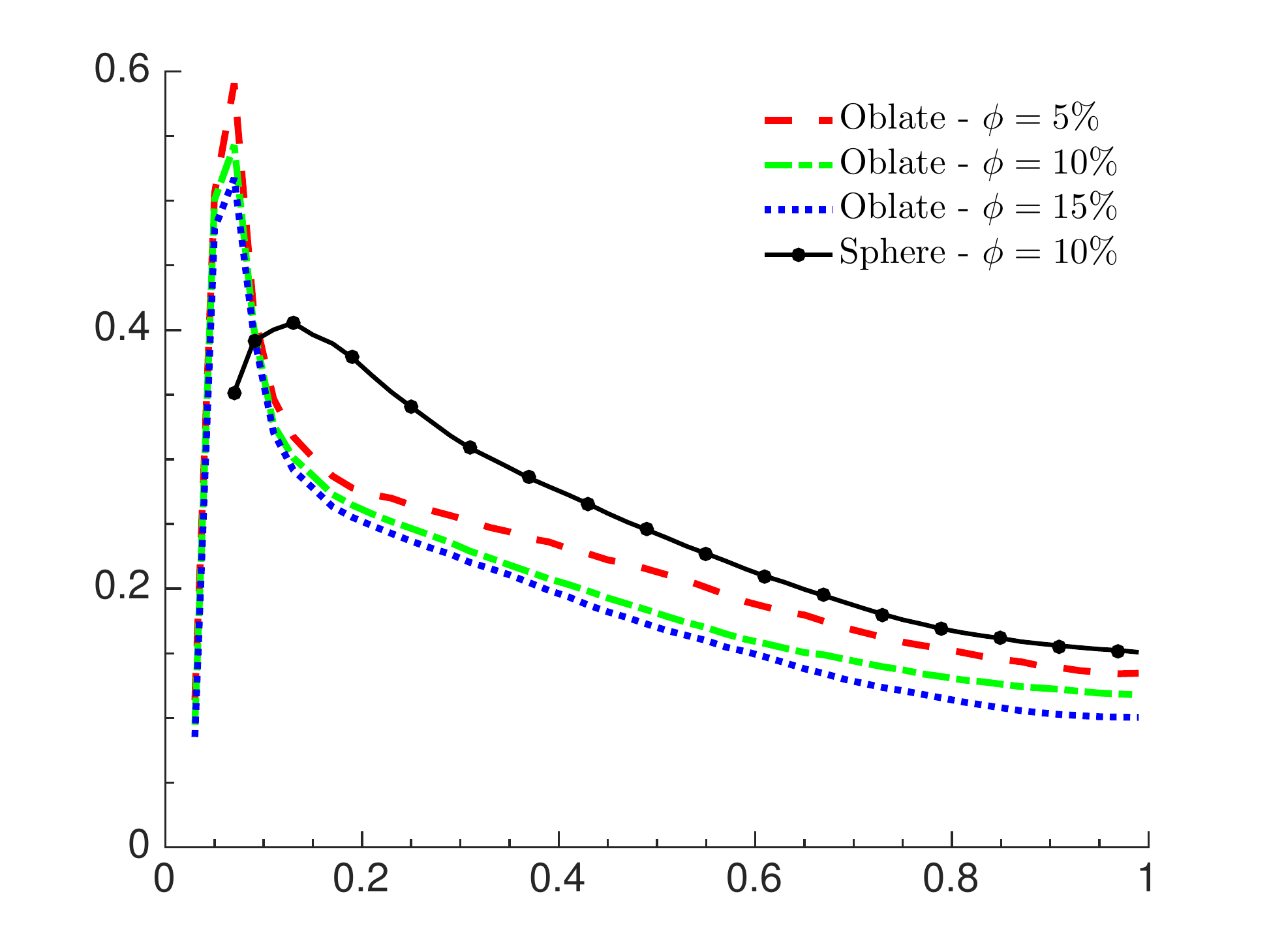} 
   \put(-395,120){\footnotesize $(c)$}
   \put(-200,120){\footnotesize $(d)$}
   \put(-382,65){\rotatebox{90}{$\overline{|\Omega_x|}$}}        
   \put(-192,65){\rotatebox{90}{$\overline{|\Omega_y|}$}}              
   \put(-292,-5){{$y / h$}}   
   \put(-98,-5){{$y / h$}} \\ [5pt]                
  \caption{$(a)$ Cosine of the mean particle inclination angle, measured with respect to the wall, $\theta$ versus $y/h$. Mean particle absolute value of angular velocity, given in outer units for: $(b)$ spanwise direction $\overline{|\Omega_z|}$, $(c)$ streamwise direction $\overline{|\Omega_x|}$ and $(d)$ wall-normal direction $\overline{|\Omega_y|}$}
\label{fig:ParticleOrient}
\end{figure}

The cosine of the mean particle inclination angle, $\theta$, measured with respect to the wall, is reported in figure~\ref{fig:ParticleOrient}$(a)$ versus the distance from the wall. Note that values of $\cos \theta$ close to 1 indicate that the particles tend to be aligned with their semi-minor (symmetry) axis normal to the wall, whereas the oblates sit with the major axis normal to the wall if $\theta \approx 0$.
From the figure, we note the clear tendency of oblate particles to be, on average, parallel to the wall; far from the wall they tend to be more isotropic, even though they still show some preferential orientation. This preferential orientation far from the wall is more pronounced in the laminar regime when the fluid velocity fluctuations are weak (see appendix~\ref{app:Laminar_Simulations}).  For spheres, the poles are fixed arbitrary to show their uniform orientation. 

The mean absolute value of the particle angular velocities are reported in figure~\ref{fig:ParticleOrient}.
As seen in~\ref{fig:ParticleOrient}$(b)$, displaying the rotation rate in the spanwise direction, $\overline{|\Omega_z|}$,
oblate particles have significantly lower rotation rates than spheres close to the wall. 
The mean particle angular velocity in the streamwise and wall-normal direction, see figure~\ref{fig:ParticleOrient}$(c)$ and $(d)$, 
are lower in the case of oblate particles,
except for the near wall-region where the wall-normal angular velocity is larger than that of spheres.

\begin{figure}
   \centering
   \includegraphics[width=0.495\textwidth]{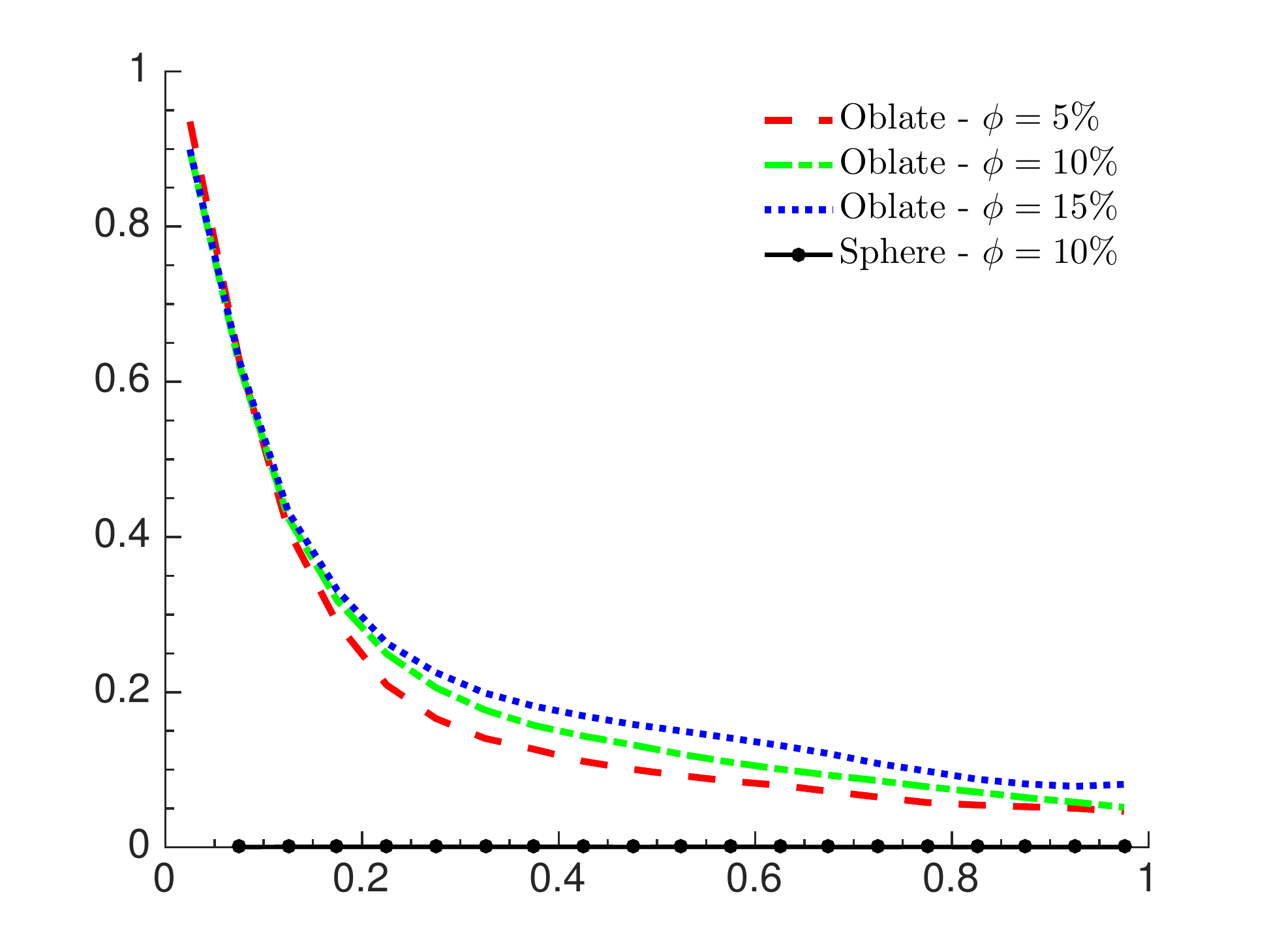}
   \includegraphics[width=0.495\textwidth]{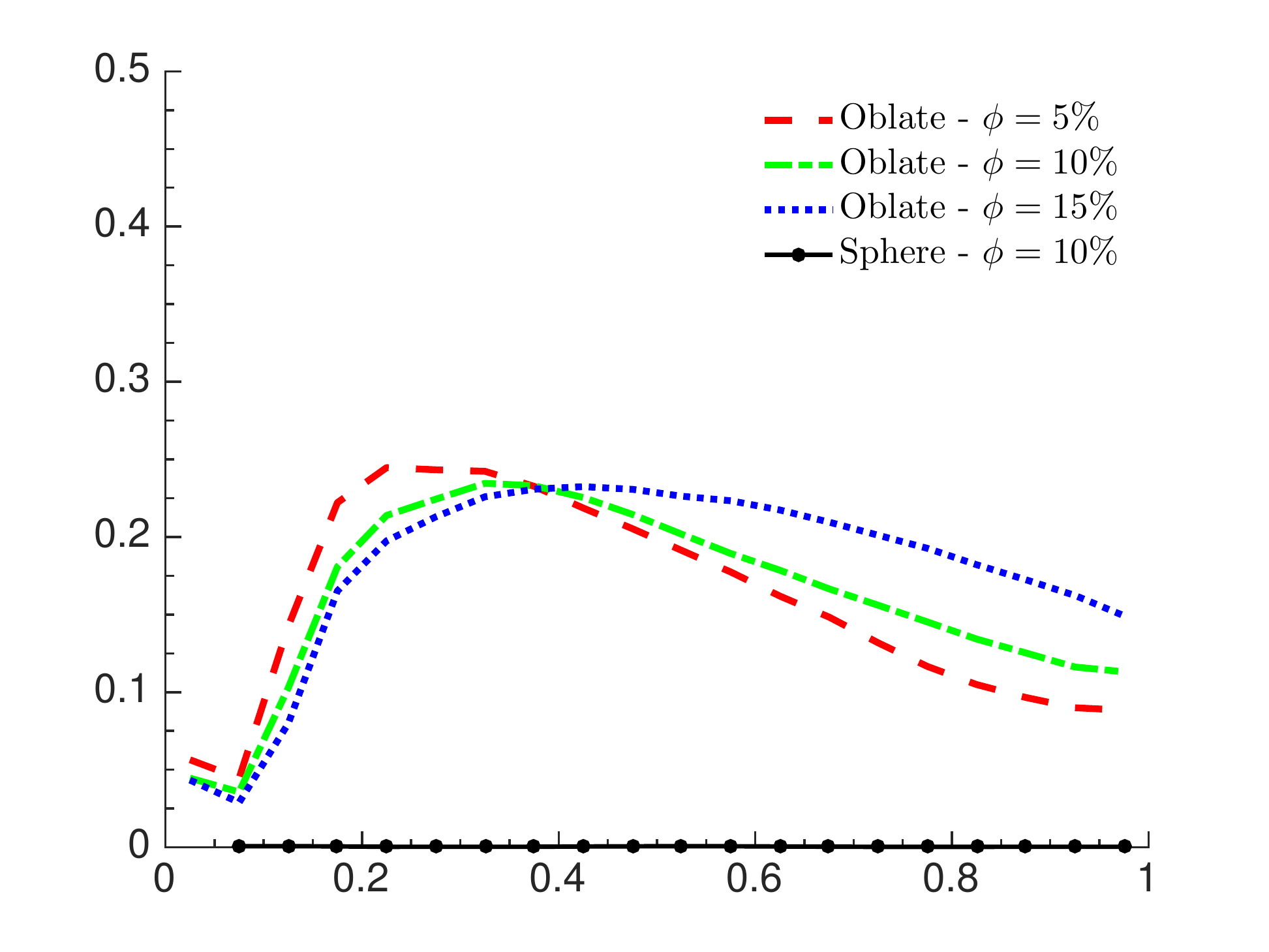}   
 \put(-385,70){\rotatebox{90}{$\lambda$}}
 \put(-190,71){\rotatebox{90}{$\zeta$}}
 \put(-98,-5){{$y / h$}}
 \put(-292,-5){{$y / h$}}
   \put(-395,120){\footnotesize $(a)$}
   \put(-200,120){\footnotesize $(b)$}
  \vspace{3pt}        
  \caption{$(a)$ The nematic order parameter $\lambda$ and $(b)$ the biaxial parameter $\zeta$ versus $y/h$ for the simulated cases. Results for spheres is also depicted in the figure to indicate a suspension with fully isotropic particle orientation.}
\label{fig:NemTur}
\end{figure}

 As depicted in figure~\ref{fig:ParticleOrient}$(a)$, oblate particles are, on average, parallel to the wall; to fully characterise the particle relative orientation 
 we therefore consider the properties of their order parameter tensor \citep{Prost1995},
  \begin{equation}
\label{eq:Q}  
\mathbf{Q} = \langle \hat{o} \otimes \hat{o}  -  \textbf{I} /3  \rangle \, , \\ [5pt] 
\end{equation} 
where $\hat{o}$ is the unit vector associated with the particle symmetry axis and $\textbf{I}$ the identity tensor.  
The eigenvalues ($\lambda_i$) of the tensor $\mathbf{Q}$ are equal to zero in the case of a fully isotropic particle orientation, while the tendency to align in a certain direction (nematic order) reflects in non-vanishing real eigenvalues. The eigenvalues ($\lambda_1 > \lambda_2 > \lambda_3$) of the symmetric tensor $\mathbf{Q}$ can be expressed  in the following form:
\begin{equation}
\label{eq:Lambda}  
\lambda_1 = \frac{2}{3} \,  \lambda \, \,  , \,\,\,\,\, \lambda_2 = -  \frac{1}{3} \,  (\lambda + \zeta) \, \, , \,\,\,\,\, \lambda_3 = -  \frac{1}{3} \,  (\lambda - \zeta) \, , \\ [5pt]
\end{equation} 
with $\lambda$ the scalar nematic order parameter and $\zeta$ the biaxial index. It can be shown \citep{Prost1995} that $\lambda$ varies between $- 0.5 < \lambda < 1$, where a positive $\lambda$ indicates that particles tend to be oriented in one direction, while a negative $\lambda$  that the orientation vector preferentially lies in one plane. 
The parameter $\zeta$ shows the bi-axiality of the order parameter tensor. $\zeta \approx 0$ indicates that there is only one preferential direction associated with the eigenvector of largest eigenvalue ($\lambda_1$), and a non-zero value of $\zeta$ creates a nematic phase which is considered bi-axial, meaning that there exist two preferential directions. 
Here we divide the half-channel height $h$ into $20$ regions and compute the order parameter tensor for each of these regions. The values of $\lambda$ and $\zeta$ obtained in each slab are displayed in figure~\ref{fig:NemTur}$(a)$ and $(b)$ versus the channel height for all the investigated cases. The results reveal, as expected from figure~\ref{fig:ParticleOrient}$(a)$, a large value of $\lambda$ close to the wall ($\lambda \approx 1$), which decays at the center of the channel. Thus, particles are preferentialy aligned parallel to the wall in its vicinity.
From the values of $\zeta$ throughout the channel, we find a region  with larger $\zeta$ ($0.2 < y/h < 0.4$), where the tensor $\mathbf{Q}$ can be considered biaxial. This  means that particles tend to be preferentially oriented also in a second direction, which is found to be in the spanwise direction (particles are rolling with their symmetry axis, oriented in the spanwise direction). From a phenomenological point of view it appears that particles close to the wall leave the area, rolling in the streamwise direction, therefore particles preferentially align with their symmetric axis in the spanwise direction. This information can be obtained from the eigenvector associated with $\lambda_2$.

\begin{figure}
   \centering
   \includegraphics[width=0.60\textwidth]{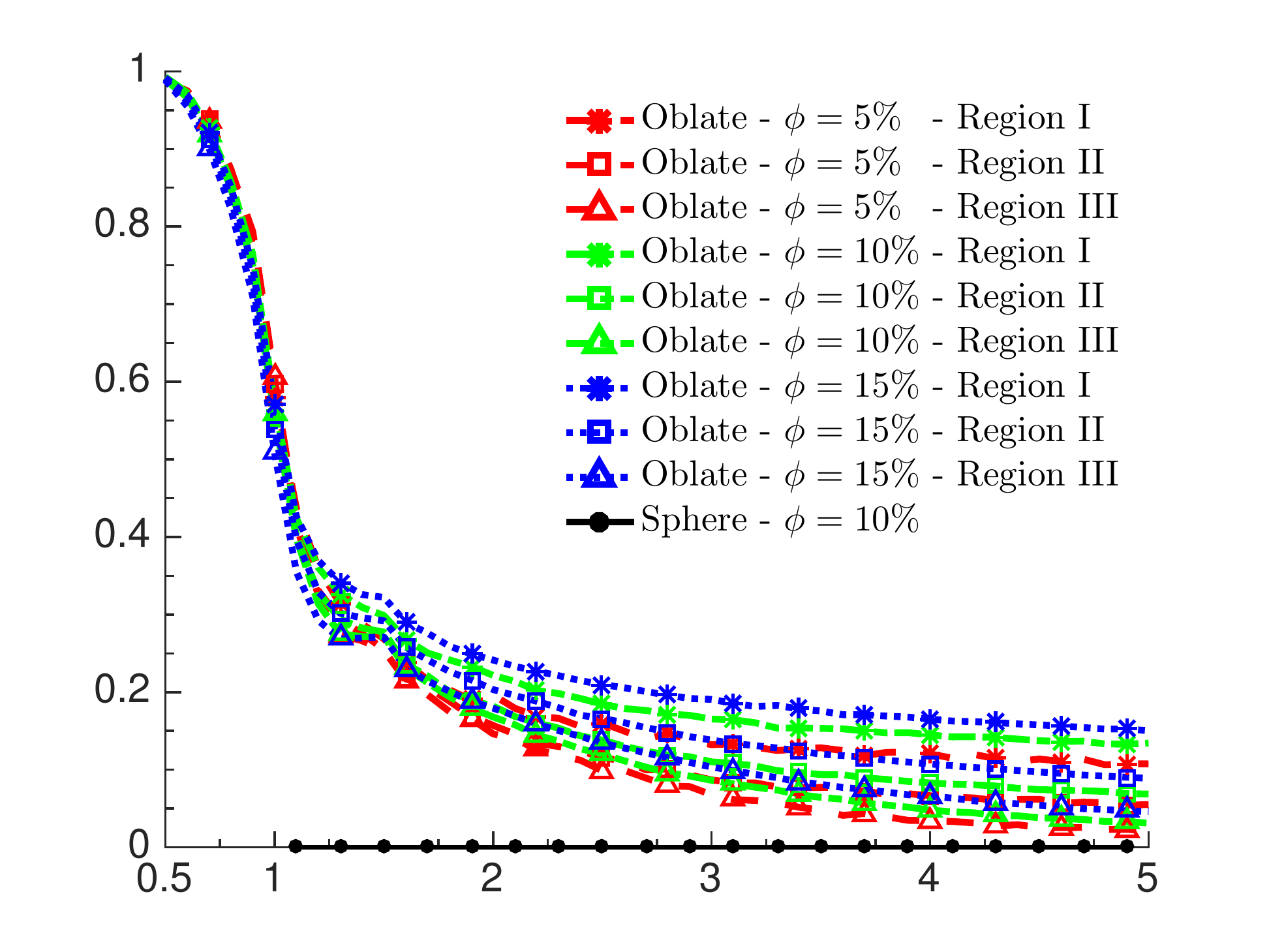}
   \put(-233,63){\rotatebox{90}{ \large $O.C.F\,(r)$}}
   \put(-128,-3){{ \large $r/D_{eq}$}}
  \caption{Orientational correlation function $O.C.F\,(r)$, versus center separation $r/D_{eq}$ for the simulated cases in $3$ regions, where regions I to III refer to $y/h < 1/3$, $1/3 < y/h < 2/3$ and $2/3 < y/h < 1$ respectively.}
\label{fig:OCF}
\end{figure}

Next, we analyze the particle relative orientation by calculating the orientation correlation function ($O.C.F.$),
\begin{equation}
\label{eq:OCF}  
O.C.F. \,(r)= \langle 2 \, | \hat{o}_p \cdot \hat{o}_q |  -  1  \rangle \, , \\ [5pt] 
\end{equation} 
where $\hat{o}_p$ and $\hat{o}_q$ denote the orientations of particles p and q at distance $r$ between their centers. 
$O.C.F. \,(r) = 1$ indicates particle pairs at distance $r$ perfectly aligned while $O.C.F. \,(r) = -1$ particles with the symmetry axis orthogonal to each other. This observable is zero for a suspension with random particle orientation.  Figure~\ref{fig:OCF} depicts $O.C.F.$ versus the separation $r/D_{eq}$ for the simulated cases in $3$ regions, where regions I to III refer to the near-wall region, $y/h < 1/3$, a region between the wall and the core, $1/3 < y/h < 2/3$ and the region around the channel centreline $2/3 < y/h < 1$. 
The results show that particles become more aligned with respect to each other as the volume fraction $\phi$ increases;  the alignment increases as the distance from the wall reduces. A small peak is observed in the figure  around $1.2 < r/D_{eq} < 1.5 \approx 2 r_{max}$, which indicates particles sitting on top of each other, 
parallel to the wall and with their centers shifted by less than the major oblate diameter.
  The particles therefore seem to form structures that are more reluctant to rotate. 


The results so far indicate that oblate particles experience considerably smaller angular velocities close to the wall (except for $\overline{|\Omega_y|}$) and stay prevalently with their major axes parallel to the wall. These two facts may explain the turbulence attenuation observed in the presence of oblate particles. As the volume fraction $\phi$ increases, oblate particles create a sort of strong \emph{shield} that dampens the turbulent activity close to the wall, preventing the outer layer turbulence from a direct interaction with the wall. 
Indeed, in the simulations reported in \cite{Ardekani20162} we examined the effect of the particle rotation on the turbulence dynamics. To this end, we have performed a simulation at $\phi=10\%$ in which the oblate particles can translate but not rotate and are kept always parallel to the wall (semi-minor axis normal to the wall). 

\begin{figure}
   \centering
   \includegraphics[width=0.60\textwidth]{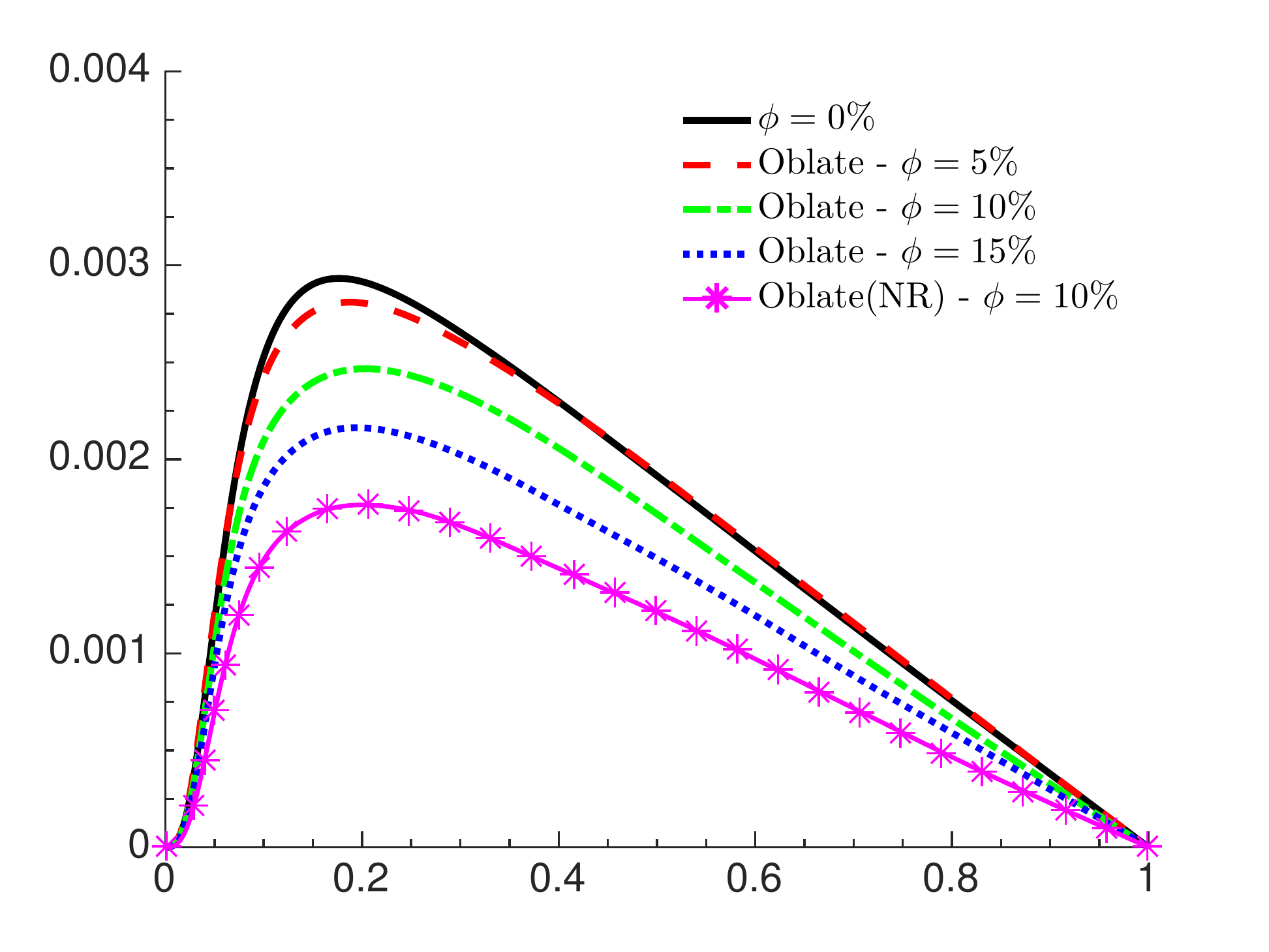}
   \put(-235,65){\rotatebox{90}{ \large $-\langle u^\prime_c v^\prime_c \rangle$}}
   \put(-125,-3){{ \large $y / h$}}
  \caption{Reynolds stress profile of the combined phase for the non-rotative (NR) oblate case, compared to the previous cases with free oblates.}
\label{fig:NR}
\end{figure}
The results of this simulation display an even larger turbulence attenuation, see figure~\ref{fig:NR} where the Reynolds stresses of the combined phase of the non-rotating (NR) and freely rotating oblate particles are compared. As a consequence of the reduced Reynolds stresses,  the friction Reynolds number considerably decreases, $Re_\tau=163$, corresponding to $18\%$ drag reduction, as also reported  in figure~\ref{fig:Drag}. It should be noted that it might be feasible to operate with a suspension of particles kept at a preferential orientation
using an external magnetic field \citep{Rosensweig2013}.



\vspace{-10pt}
\section{Final remarks}\label{sec:Final_remarks}

We have reported results from simulations of turbulent channel flow of suspensions of finite size oblate spheroidal particles with $\mathcal{AR}=1/3$ at different volume fractions, $5$, $7.9$, $10$ and $15\%$.  
The numerical approach proposed by \cite{Ardekani2016} is used for the simulations of oblate spheroidal particles presented here. This is based on the IBM method for the fluid-solid interactions with lubrication and contact models for the short-range particle-particle (particle-wall) interactions. 

Drag reduction with respect to single-phase turbulence is found in the presence of oblate particles, contrary to the results for spherical particles shown in \cite{Picano2015}, for which the drag increases with the particle volume fraction. 
The two important factors determining the overall drag are the
 turbulence activity in the suspension and the particle-induced stresses, which are not only due to the increase of the suspension effective viscosity in the presence of particles but also to their distribution across the channel. 
In fact, \cite{Picano2015} have shown that for suspensions of spheres a particle layer forms close to the wall, causing high particle-induced stresses in this region. The particle-induced stresses compensate for the reduced Reynolds shear stress observed at high volume fractions ($\phi > 10\%$), resulting in an overall drag increase. \cite{Costa2016}, indeed showed that a combination of higher effective viscosity and the presence of the mentioned particle layer is at the origin of the overall drag increase in suspensions of spheres. Interestingly, such particle layer is not present in the flow laden with oblate particles and this makes the turbulence activity the most important factor determining the overall drag. 

Attenuation of the turbulence activity is observed in the presence of oblate particles, an effect more pronounced as the volume fraction increases. It is well known that the effective viscosity of a particle suspension is always higher than that of the single phase flow, which can cause turbulence attenuation.
 In this study, however, we show that the presence of oblate particles reduces the turbulence activity to lower values than those obtained  by only accounting for the effective suspension viscosity.  
To show this, we perform simulations of single-phase flow at bulk Reynolds numbers calculated with the effective viscosity for the corresponding particle volume fraction, concluding that the specific dynamics of the oblate particles and their interactions with the turbulent velocity field considerably reduce the turbulence activity of the suspension. In fact, turbulence attenuation is sufficiently high that, despite the increase in effective viscosity, overall drag reduction is achieved.

 We explain the turbulence attenuation observed in the presence of oblate particles by noting that oblate particles experience considerably lower angular velocities close to the wall and stay prevalently aligned parallel to the wall. At high volume fractions, the oblate particles create a kind of strong \emph{shield} that dampens the turbulent fluctuations close to the wall, separating the outer layer turbulence from direct interaction with the wall. To show the isolated effect of the particle rotation and the orientation on the turbulent dynamics, we perform a simulation at $\phi=10\%$ in the same flow geometry, in which the oblate particles are free to translate but cannot rotate and are kept always parallel to the wall (semi-minor axis normal to the wall). This simulation reveals an even larger turbulence attenuation and therefore increased drag reduction when compared to the cases where oblate particles are free to rotate.

Examining the particle relative orientation, e.g. by calculating the nematic order and the orientation correlation function of the center separation $r$ between the particles,
we show that the particles are preferentially parallel to the wall in its vicinity and are also increasingly parallel to each other as the volume fraction $\phi$ increases, with the highest relative alignment in the region close to the wall. The relative alignment in this region is of utmost importance in reducing the local turbulence fluctuations.  The analysis of the forces acting on the particles further shows that particles aligning with the wall are attracted towards it, while they are lifted towards the centerline when inclined.

We have shown here that finite-size oblates reduce drag in turbulent channel flow as already observed for small fibers and polymers. The physical mechanisms are, however, different as the latter two seem to act on the small scales in the flow. In the future, it will be therefore interesting to study the interactions among finite-size fibers or prolate particles and turbulence as done in \cite{Do2014} at low volume fractions.

\vspace{-10pt}
\section*{Acknowledgements}
This work was supported by the European Research Council Grant No. ERC-2013-CoG-616186, TRITOS. The authors acknowledge computer time provided by SNIC (Swedish National Infrastructure for Computing) and the support from the COST Action MP1305: Flowing matter.

\begin{figure}
   \centering
   \includegraphics[width=0.60\textwidth]{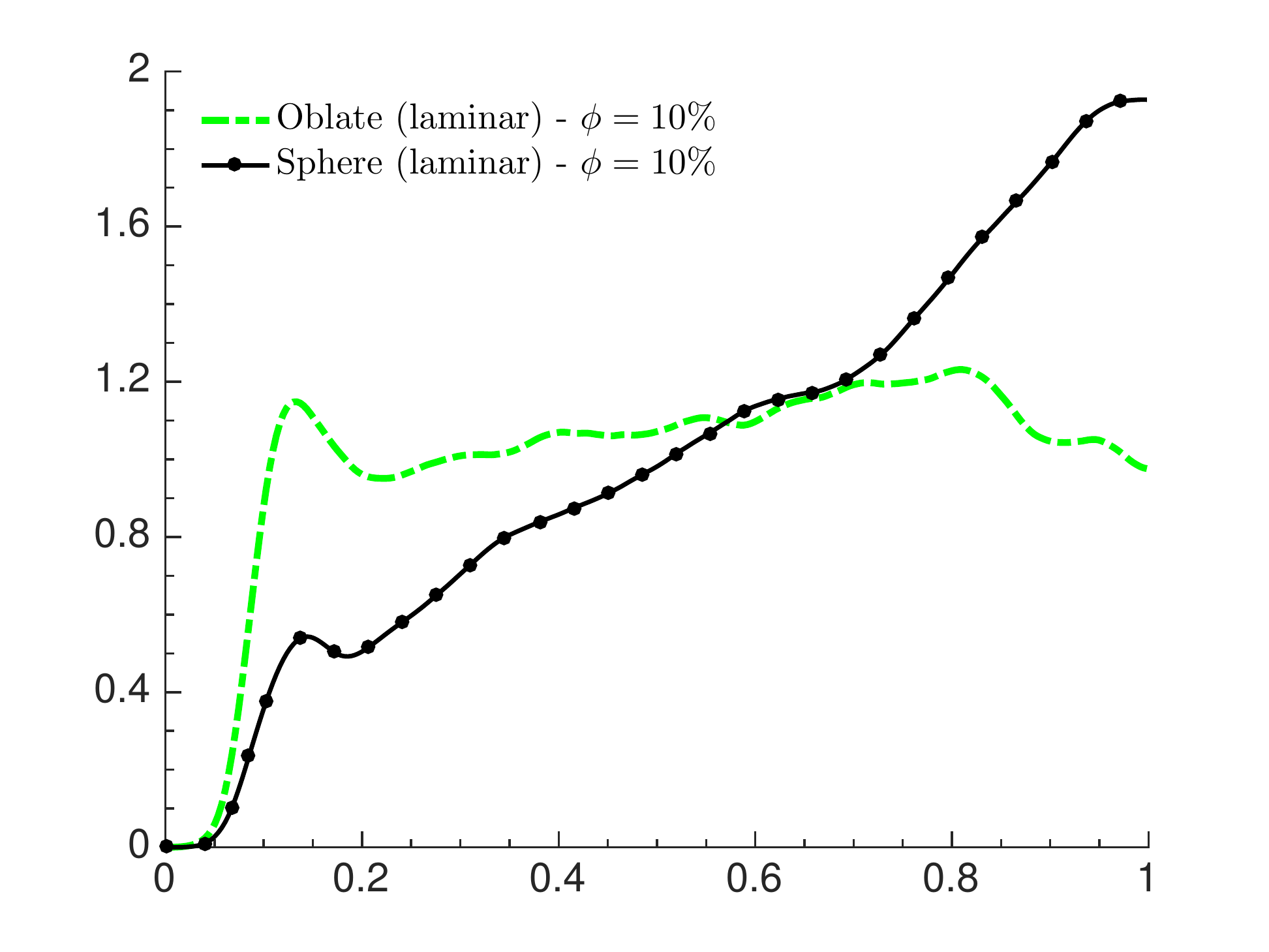}
   \put(-230,70){\rotatebox{90}{ \large $ \Phi (y) / \phi$}}
   \put(-123,-3){{ \large $y/h$}}
  \caption{Mean local volume fraction $\Phi (y)$, normalized by total volume fraction $\phi$ versus $y/h$ for the laminar flow of spherical and oblate particles in channel.}
\label{fig:phi_Lam}
\end{figure}

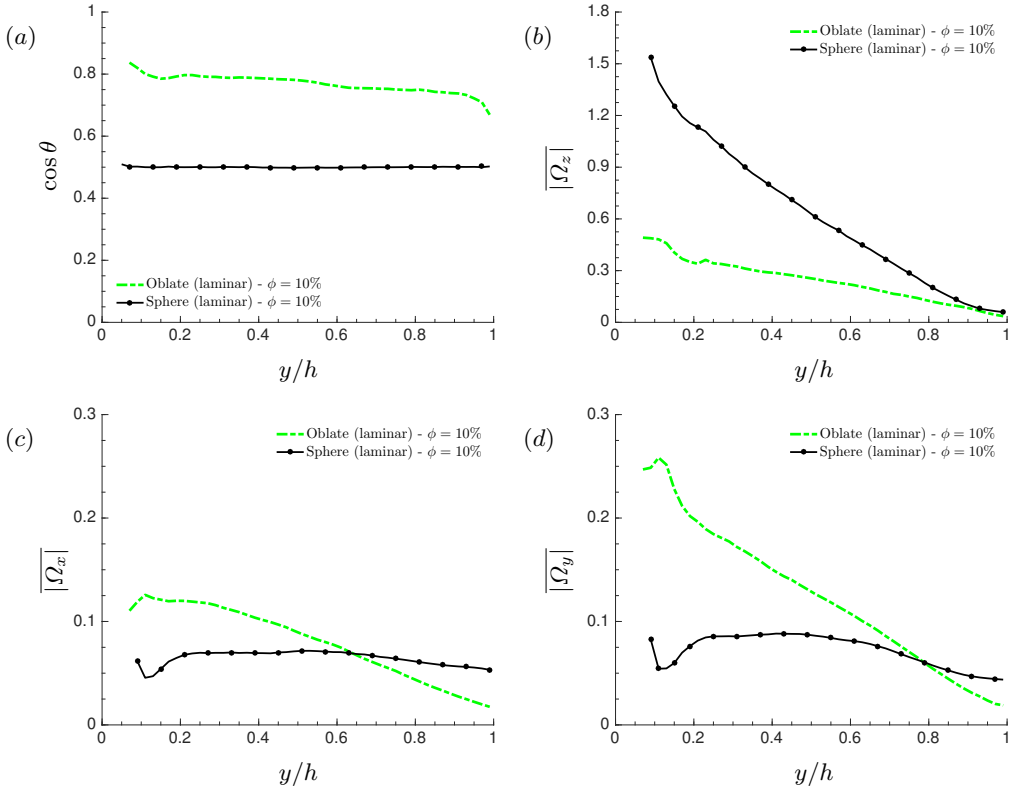
\begin{figure}
   \centering
   \includegraphics[width=0.495\textwidth]{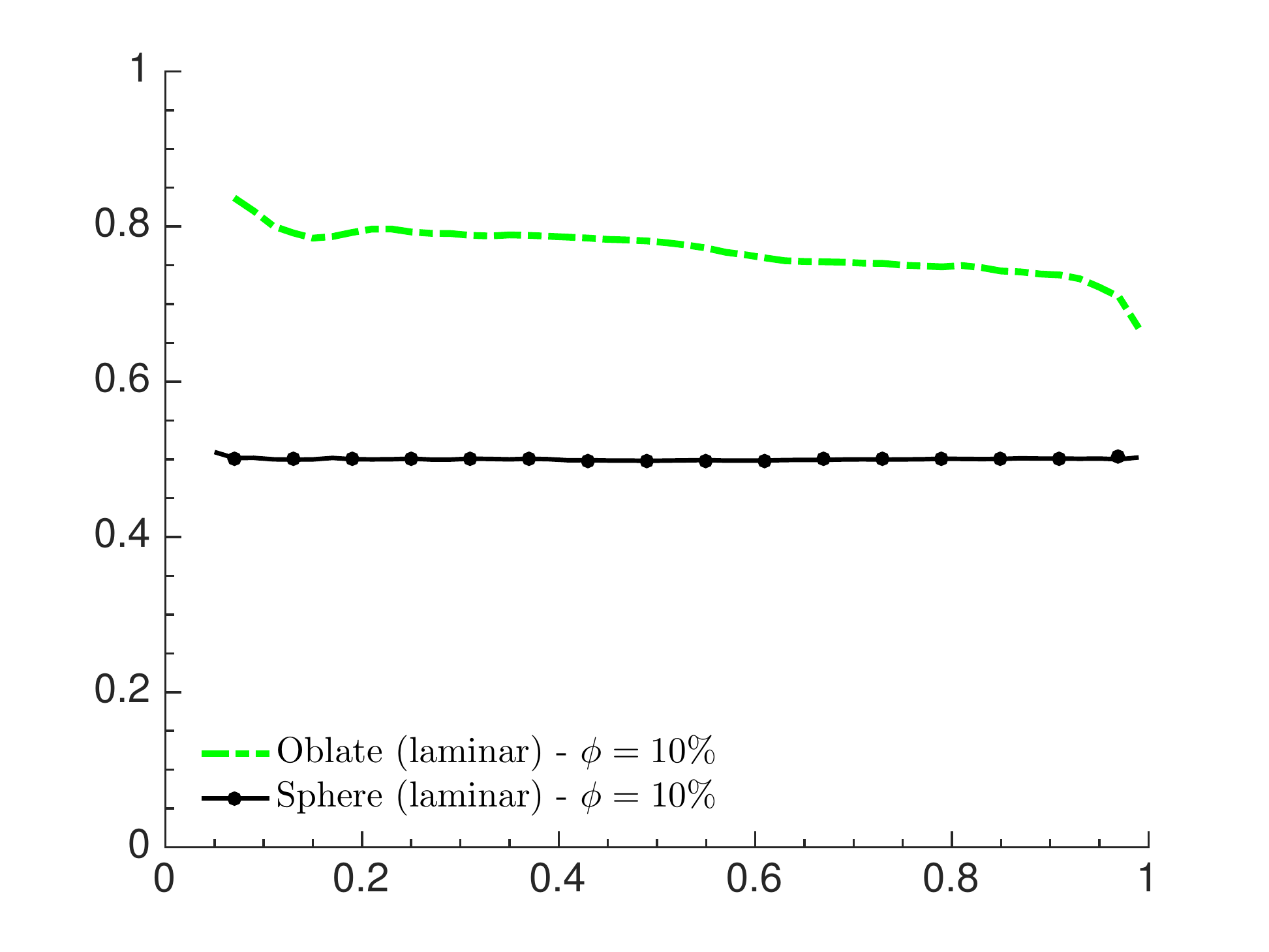}
   \includegraphics[width=0.495\textwidth]{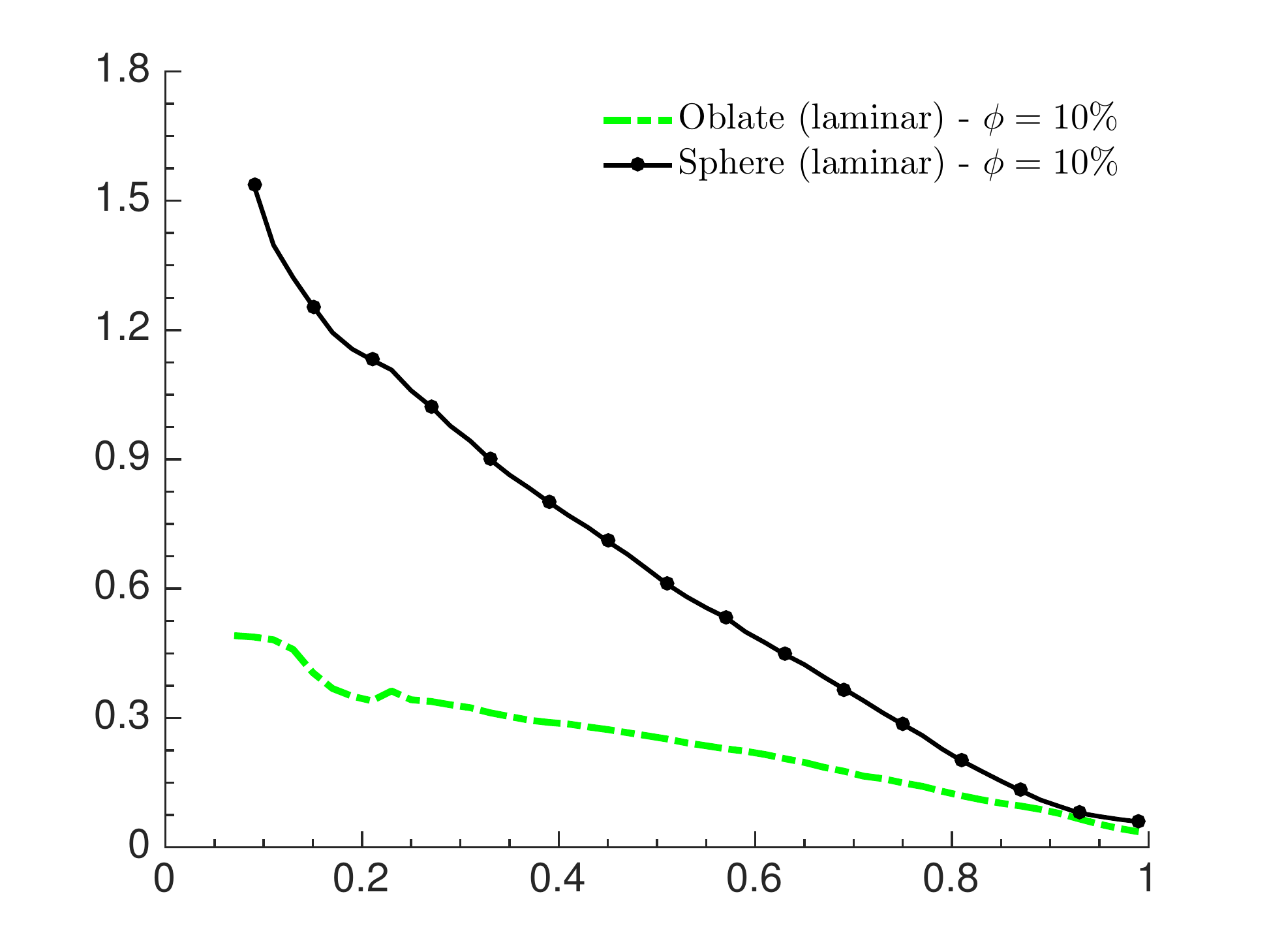}
   \put(-292,-5){{$y / h$}}   
   \put(-98,-5){{$y / h$}}  
   \put(-382,65){\rotatebox{90}{$\cos \theta$}}
   \put(-192,65){\rotatebox{90}{$\overline{|\Omega_z|}$}}        
   \put(-395,120){\footnotesize $(a)$}
   \put(-200,120){\footnotesize $(b)$} \\       
   \includegraphics[width=0.495\textwidth]{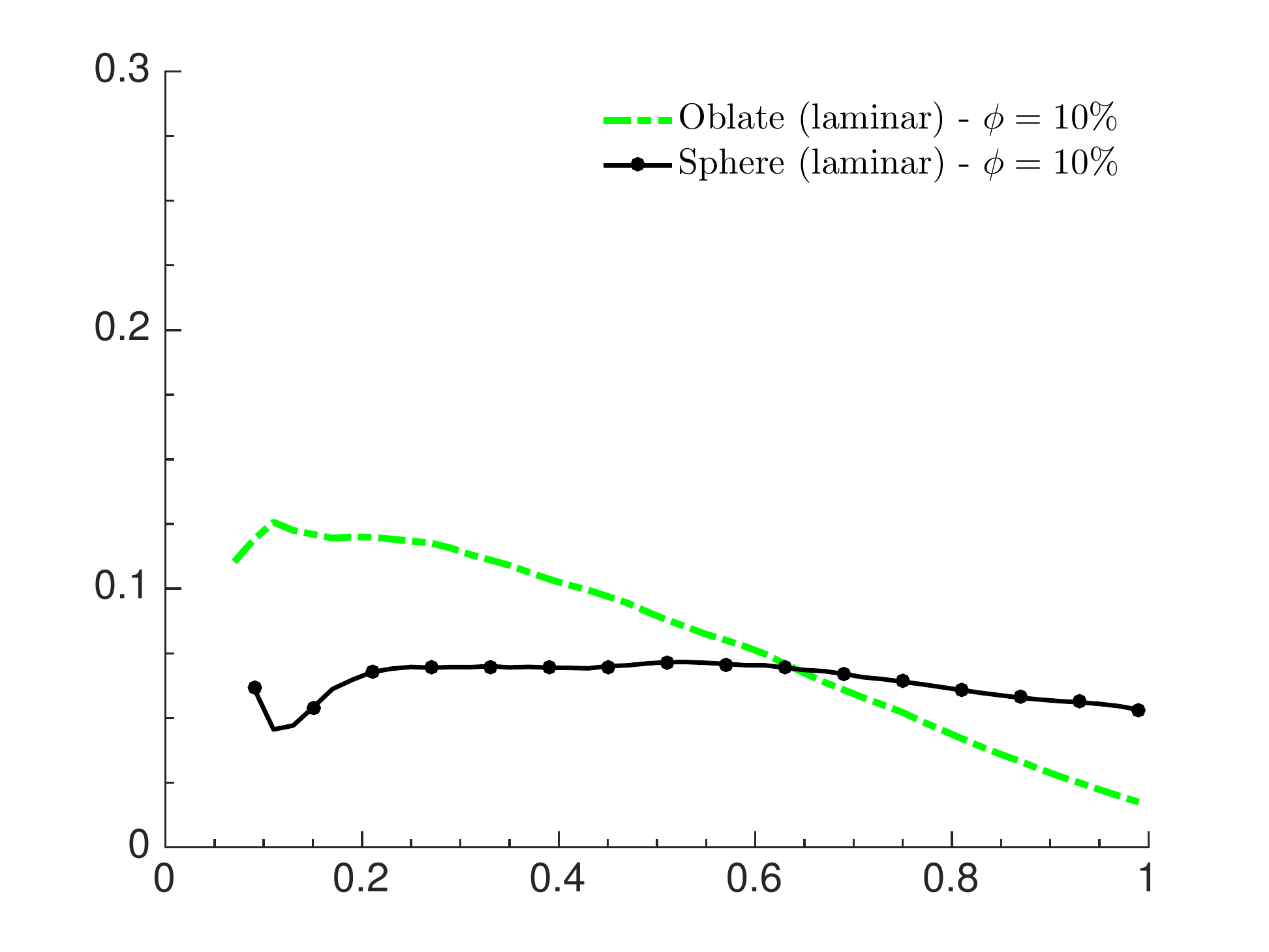}
   \includegraphics[width=0.495\textwidth]{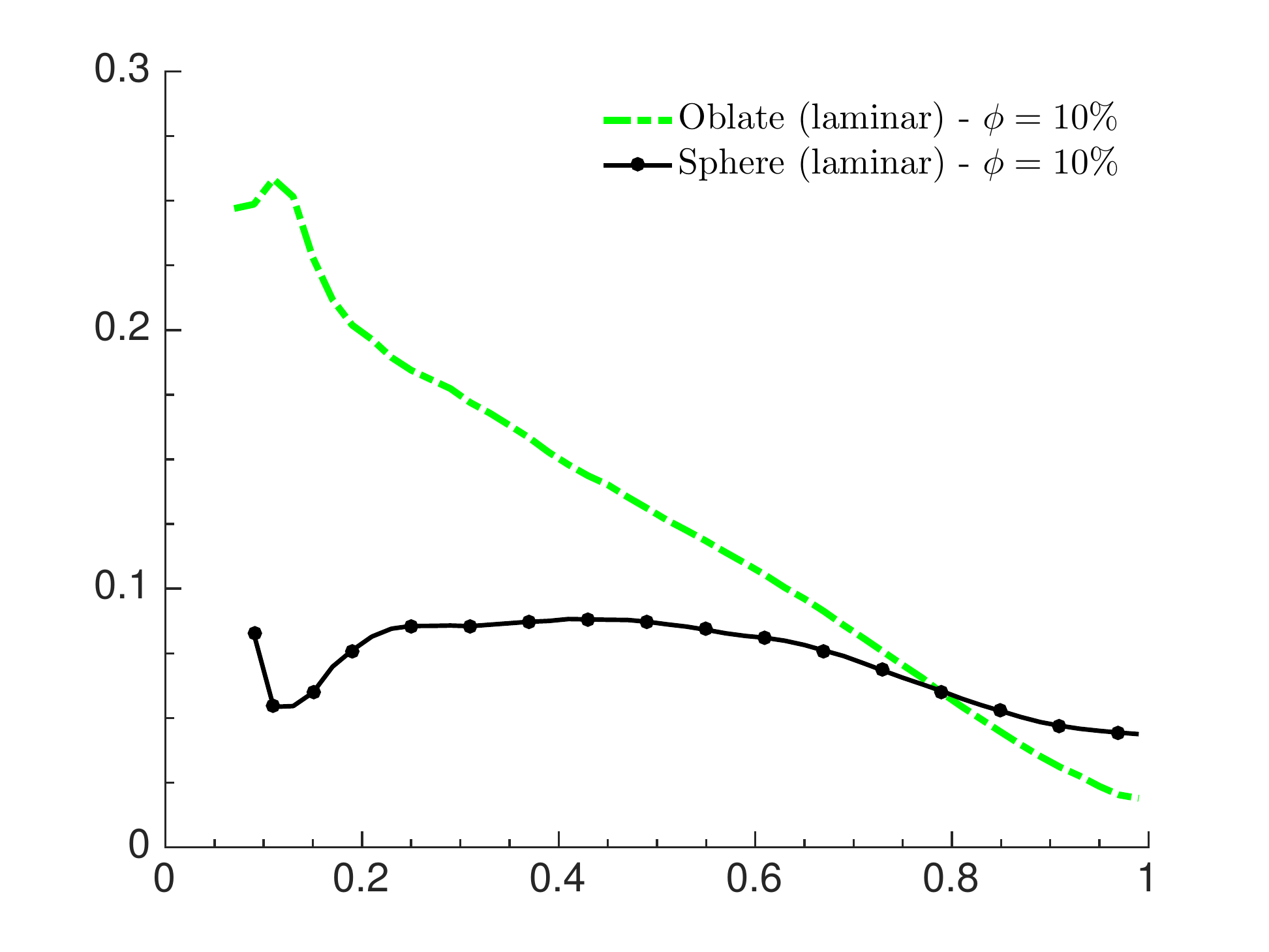} 
   \put(-395,120){\footnotesize $(c)$}
   \put(-200,120){\footnotesize $(d)$}
   \put(-382,65){\rotatebox{90}{$\overline{|\Omega_x|}$}}        
   \put(-192,65){\rotatebox{90}{$\overline{|\Omega_y|}$}}              
   \put(-292,-5){{$y / h$}}   
   \put(-98,-5){{$y / h$}} \\ [5pt]                
  \caption{$(a)$ The cosine of the mean particle inclination angle, measured with respect to the wall, $\theta$ versus $y/h$. Mean particle absolute value of angular velocity, normalized with $U_b/h$ in: $(b)$ spanwise direction $\overline{|\Omega_z|}$, $(c)$ streamwise direction $\overline{|\Omega_x|}$ and $(d)$ wall-normal direction $\overline{|\Omega_y|}$}
\label{fig:ParticleOrient_Lam}
\end{figure}

\appendix
\section{Laminar flow of suspensions} \label{app:Laminar_Simulations} 

In this appendix, we report results for the laminar flow of suspensions of spheres and oblate particles. The data are compared here and used as reference when discussing the turbulent cases above.
 The simulations are performed at $Re_b = 1000$ and $\phi = 10\%$ in the same pressure-driven plane channel.

The suspension effective viscosity, quantified here by the friction Reynolds number, $Re_\tau$, in analogy to the turbulent cases, 
is slightly larger for the oblate particles, $Re_\tau= 44.6$ against  $Re_\tau=43.3$ for spheres (i.e.\ $15\%$ increase for oblates and $12\%$ for spheres with respect to the laminar single phase flow). 
Figure~\ref{fig:phi_Lam} displays the wall-normal profiles of the local volume fraction, $\Phi$ for the two types of particles under considerations.
The data clearly show that spheres migrate towards the channel center, displaying a local maximum close to the wall, as also reported by \cite{Lashgari2016};
interestingly, this migration disappears  for disc-like particles whose distribution is more uniform throughout the channel.
Two local maxima are observed for oblates: one close to the wall ($y/h=0.1$) and the other one close to the channel center ($y/h=0.8$). 
The higher local volume fraction close to the wall can explain the higher drag for the laminar flow of oblate particles.                 
 The analysis of the forces and torques presented in the text has been repeated for laminar flow to observe 
that the difference between the positive and negative lift force
is much less than in the turbulent regime, probably due to the lower velocity gradients close to the wall, which may explain why the particles have a more or less uniform distribution.

\begin{figure}
   \centering
   \includegraphics[width=0.495\textwidth]{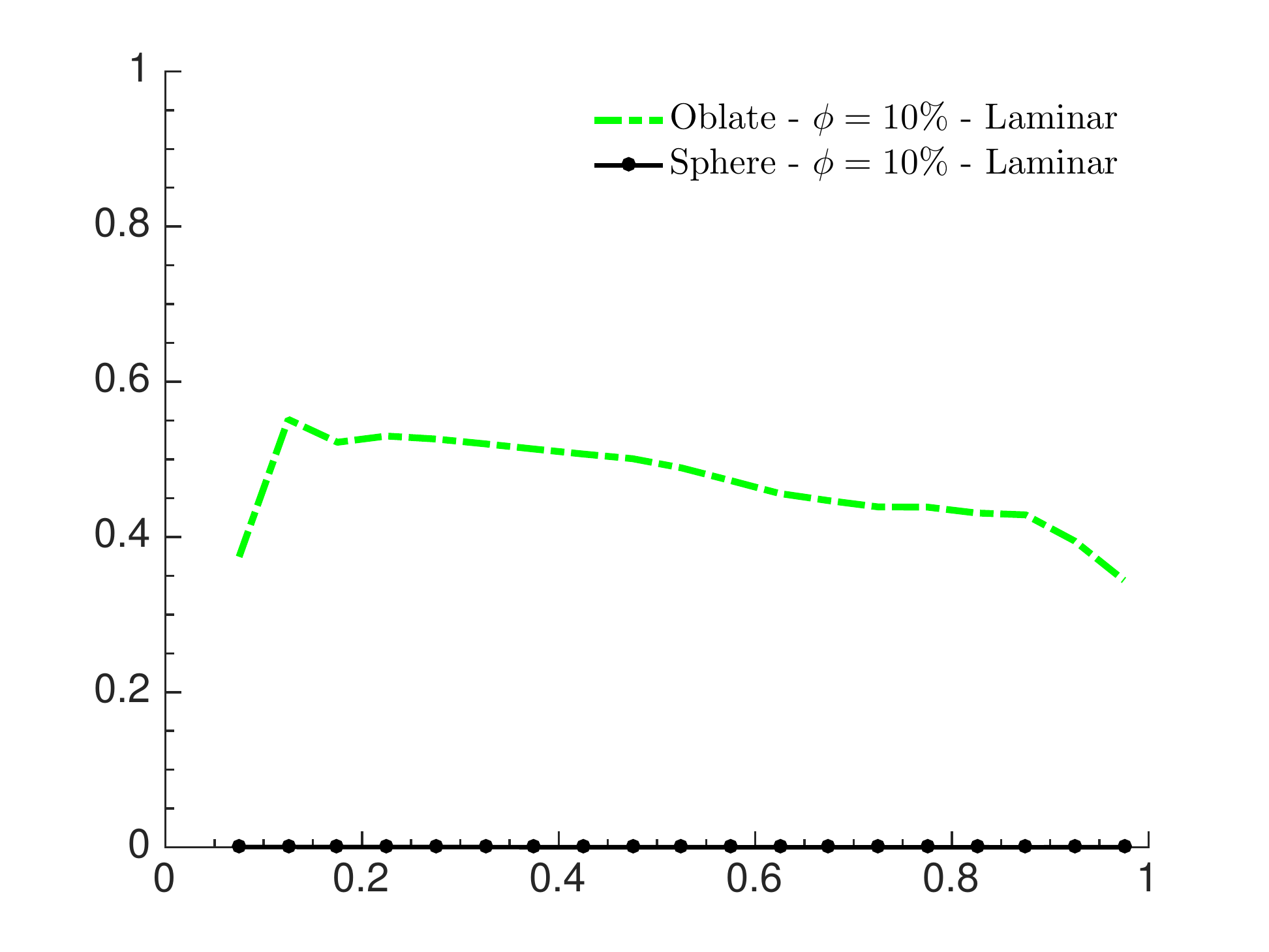}
   \includegraphics[width=0.495\textwidth]{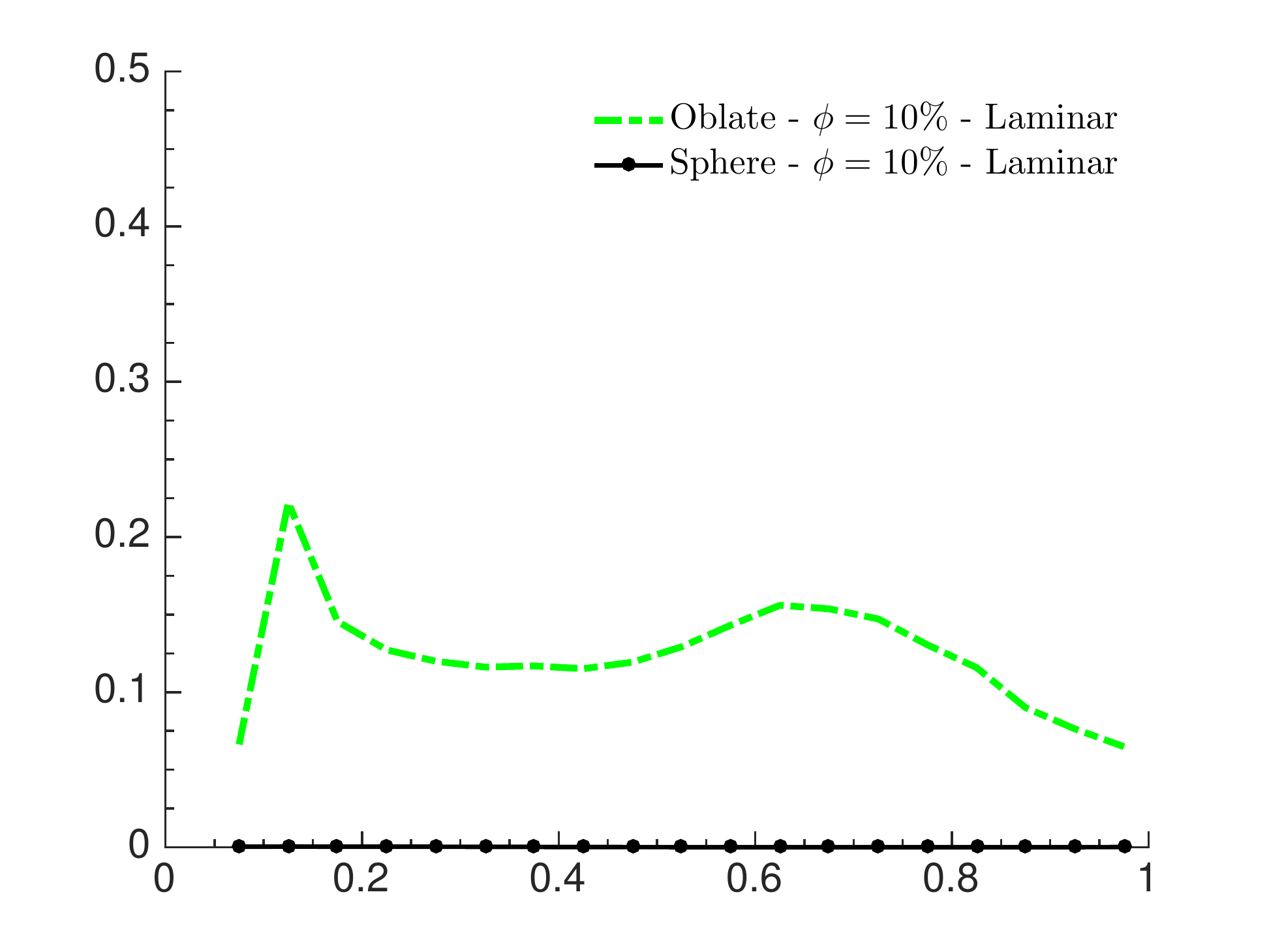}   
 \put(-385,70){\rotatebox{90}{$\lambda$}}
 \put(-190,71){\rotatebox{90}{$\zeta$}}
 \put(-98,-5){{$y / h$}}
 \put(-292,-5){{$y / h$}}
   \put(-395,120){\footnotesize $(a)$}
   \put(-200,120){\footnotesize $(b)$}
  \vspace{3pt}        
  \caption{$(a)$ The nematic order parameter $\lambda$ and $(b)$ the biaxial parameter $\zeta$ versus $y/h$ for the laminar cases, studied here. Results for spheres are also depicted to confirm their fully isotropic particle orientation.}
\label{fig:NemLam}
\end{figure}

\begin{figure}
   \centering
   \includegraphics[width=0.60\textwidth]{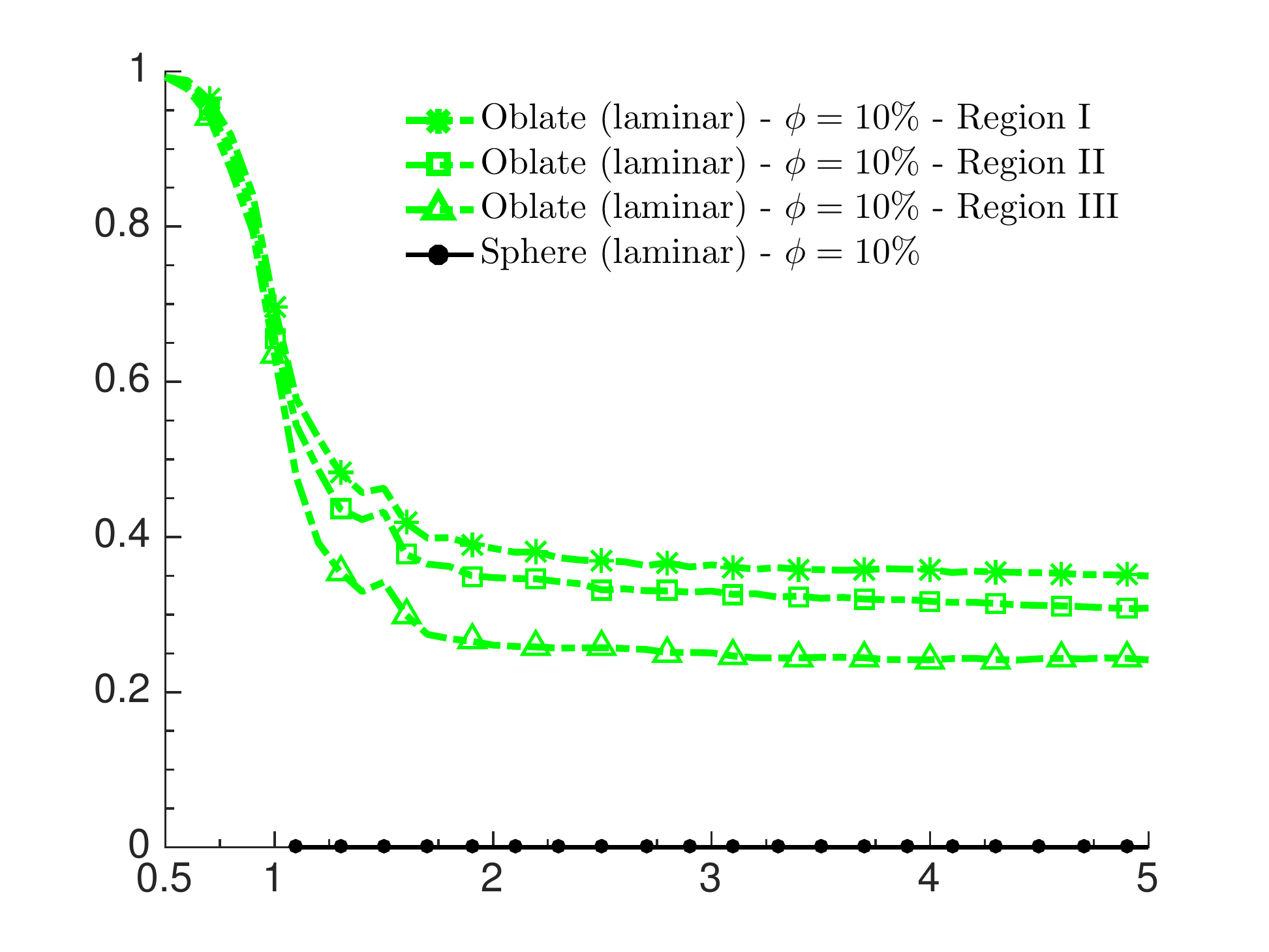}
   \put(-233,63){\rotatebox{90}{ \large $O.C.F\,(r)$}}
   \put(-128,-3){{ \large $r/D_{eq}$}}
  \caption{Orientational correlation function $O.C.F\,(r)$ versus center separation $r/D_{eq}$ for the laminar cases in $3$ regions, where regions I to III refer to $y/h < 1/3$, $1/3 < y/h < 2/3$ and $2/3 < y/h < 1$ respectively.}
\label{fig:OCF_Lam}
\end{figure}

The mean particle orientation and rotation  in laminar flow are compared in figure~\ref{fig:ParticleOrient_Lam}. The tendency of oblate particles to align with the gradient direction in the regions far from the wall is observed to be more pronounced in the laminar regime than in turbulent flow. 
The mean absolute value of the particle angular velocities, depicted in figure~\ref{fig:ParticleOrient_Lam}$(b)$ to $(d)$, show higher values for oblate particles in the streamwise and the wall-normal directions close to the wall and smaller for rotation rates in the spanwise direction.        
 
 The nematic order parameter $\lambda$ and the biaxial parameter $\zeta$ 
 are depicted in figure~\ref{fig:NemLam} versus the channel height. The profile of $\lambda$ in figure~\ref{fig:NemLam}$(a)$ is similar to the mean orientation profile of oblate particles in figure~\ref{fig:ParticleOrient_Lam}$(a)$; Values close to 1 in both figures show the tendency of the particles to be aligned with the symmetry axis normal to the wall. 
 Figure~\ref{fig:NemLam}$(b)$ displays two local maxima of the biaxial parameter $\zeta$: one close to the wall and the other one close to the channel center. Interestingly, these local maxima appear  in correspondence to  the local maxima of the local volume fraction profiles  in figure~\ref{fig:phi_Lam}, revealing that the tendency to be oriented also in the spanwise direction increases in those locations with higher local volume fractions. 
 
Finally, figure~\ref{fig:OCF_Lam} reports the orientational correlation function, $O.C.F.$, versus the particle centre separation $r/D_{eq}$ in $3$ regions, where regions I to III refer to $y/h < 1/3$, $1/3 < y/h < 2/3$ and $2/3 < y/h < 1$. The results for laminar flow show that the particle orientations are more correlated with respect to the turbulent cases, in other words that the order is disrupted by turbulent mixing.

\bibliographystyle{jfm}
\bibliography{jfm_Channel}

\end{document}